\documentclass{SciPost}

\hypersetup{
    colorlinks,
    linkcolor={red!50!black},
    citecolor={blue!50!black},
    urlcolor={blue!80!black}
}

\usepackage[bitstream-charter]{mathdesign}
\usepackage{subcaption}
\usepackage{epstopdf}

\DeclareSymbolFont{usualmathcal}{OMS}{cmsy}{m}{n}
\DeclareSymbolFontAlphabet{\mathcal}{usualmathcal}

\fancypagestyle{SPstyle}{
\fancyhf{}
\lhead{\colorbox{scipostblue}{\bf \color{white} ~SciPost Physics }}
\rhead{{\bf \color{scipostdeepblue} ~Submission }}

\fancyfoot[C]{\textbf{\thepage}}
}

\usepackage{amsmath}
\usepackage[makeroom]{cancel}
\usepackage{subcaption}
\usepackage{amsthm}
\usepackage{float}
\usepackage{graphicx}

\allowdisplaybreaks

\newcommand{\tr}{\text{tr}}

\begin{document}

\pagestyle{SPstyle}

\begin{center}{\Large \textbf{\color{scipostdeepblue}{
Open-boundary integrable quantum circuits with different geometries
}}}\end{center}

\begin{center}\textbf{
Miguel García Fernández\textsuperscript{1},
Chiara Paletta\textsuperscript{2},
Ana L. Retore\textsuperscript{3,4},
}\end{center}

\begin{center}
{\bf 1} Instituto Galego de F\'isica de Altas Enerx\'ias (IGFAE),
\\and Departamento de F\'\i sica de Part\'\i culas,\\
Universidade de Santiago de Compostela, 15705 Santiago de Compostela, Spain\\
{\bf 2} Department of Physics, Faculty of Mathematics and Physics,\\
 University of Ljubljana, Jadranska 19, SI-1000 Ljubljana, Slovenia\\
{\bf 3} Department of Mathematical Sciences, Durham University, Durham DH1 3LE, UK\\
{\bf 4} Deutsches Elektronen-Synchrotron DESY, Notkestr. 85, 22607 Hamburg, Germany\footnote{DESY-26-081}\\
\vspace{0.3cm}
{\bf 1} miguelg.fernandez@usc.es
{\bf 2} chiara.paletta@fmf.uni-lj.si
{\bf 3} ana.retore@desy.de
\end{center}

\section*{\color{scipostdeepblue}{Abstract}}

\textbf{We present a complete classification of integrable Yang–Baxter quantum circuits with open boundary conditions and arbitrary circuit geometries. Starting from the standard transfer-matrix construction with two types of staggered inhomogeneities, we derive a general mapping that determines the arrangement of circuit gates in terms of the inhomogeneities and the system size. We conjecture that time-periodic quantum circuits are integrable whenever the local bulk and boundary gates satisfy the Yang-Baxter equation and the same bulk gate is applied exactly once per period to every nearest-neighbor pair of spins. Our construction also provides an algorithm to detect Yang–Baxter integrability for circuits with arbitrary geometries. Furthermore, we introduce a third type of inhomogeneity, denoted by $\rho$, and demonstrate that the minimum possible circuit depth is four. We show that when these $\rho$-inhomogeneities are placed at the endpoints and in their immediate neighborhood, the resulting boundary gates can be interpreted as single gates acting on multiple sites. Our construction is fully general and applies to regular $R$-matrices, both of difference and non-difference type, together with their associated boundary matrices. As an application, we consider two-qubit gates corresponding to 6- and 8-vertex $R$-matrices of non-difference form satisfying the Yang–Baxter equation, and we construct the associated reflection matrices that generate integrable quantum circuits.}

\vspace{\baselineskip}





\noindent\rule{\textwidth}{1pt}
\tableofcontents
\noindent\rule{\textwidth}{1pt}
\vspace{10pt}

\section{Introduction}

Determining the spectrum of the operator that governs a system's dynamics is of broad interest. In lattice systems with a local finite-dimensional Hilbert space, the dimension of the Hamiltonian scales exponentially with the system size, making direct computation of the spectrum quickly unfeasible. When present, integrability is a powerful tool to achieve this goal, as it provides a robust toolkit to solve these models.

Thanks to the Yang-Baxter equation, several methods are available to obtain the spectrum of the dynamical evolution of integrable models. One of these is the algebraic Bethe ansatz, \cite{Takhtajan:1979iv,faddeev1996algebraic} (see \cite{analecture} for a recent review), where the problem of exact diagonalization of operators in a large Hilbert space is mapped to solving a coupled set of polynomial equations. This brings the additional advantage of being able to take the thermodynamic limit, see \cite{van2016introduction},  in which both the system size and the number of excitations are taken to infinity while keeping their ratio finite.  Other techniques include separation of variables \cite{SklyaninSoV}, the ODE/IM correspondence \cite{dorey1999anharmonic,fioravanti2023origin}, the Quantum Spectral Curve \cite{gromov2014quantum} (for a review see  \cite{levkovich2020review}), and various other forms of the Bethe ansatz \cite{frassek2017boundary,belliard2018modified}.

If we consider a one-dimensional continuous-time lattice model characterized by a many-body Hamiltonian $H$, the evolution of the density matrix $\rho(t)$ is governed by the Liouville-von Neumann equation 
\begin{equation}
    i \frac{d \rho(t)}{d t}=[H,\rho(t)].
\end{equation}

This system is integrable if $H$ belongs to a tower of commuting conserved charges. A well-known example is the Heisenberg (XXX) spin chain, where the Hamiltonian can be written as a sum of local terms $H=\sum_i \vec{\sigma}_i \cdot  \vec{\sigma}_{i+1}$ and the framework of Yang-Baxter integrability is well established, \cite{faddeev1996algebraic}. Similarly, one can consider a discrete-time evolution given by 
\begin{equation}
\rho_{t+1}=M\rho_t M^\dagger,
\end{equation} where $M$ is a unitary propagator, typically realized as a sequence of local operators acting on the spin chain. In the appropriate limit, this evolution corresponds to the continuous time evolution given by the Liouville-von Neumann equation. In this setting, integrability refers to the property that the propagator $M$ commutes with a set of local conserved charges. A clear example is the brickwork type circuit with periodic boundary conditions built from the XXX model in \cite{vanicat2018integrable2}: the $R$-matrix plays the role of a quantum gate and $M$ can be built from a transfer matrix with staggered inhomogeneities. 
Expanding the transfer matrix around the staggered values of the inhomogeneities provides the construction of two sets of local conserved charges. 

Although quantum circuits are often associated with unitary time evolution, the formalism naturally extends to non-unitary local gates. This broader setting may allow one to describe dissipative dynamics, and has recently attracted significant interest in the context of integrable systems, see for example, \cite{sa2021integrable}.

Mathematically, this approach allows one to interpret the matrix elements of the propagator of the model $ \langle{\psi'}|M^n |\psi \rangle$, where $|\psi\rangle$  and $|\psi'\rangle$ denote the initial and final states, respectively, as the equilibrium partition function of a corresponding vertex model, for which many computational techniques are readily available.

An additional advantage of using these discrete space-time models is that, in classical simulations, they avoid the computationally expensive Trotter-Suzuki limit \cite{suzuki1976generalized}, which requires a large number of small time steps to reproduce Hamiltonian evolution. Moreover, with the recent development of quantum computing, these dynamical systems naturally fit into modern quantum architectures. 

For the reasons above, integrable quantum circuits have, in recent years, become useful for calibration and error mitigation in modern engineered quantum platforms \cite{aleiner2021bethe,morvan2022formation, maruyoshi2023conserved,zhao2026effective}. They have also provided a framework for exploring universal superdiffusive scaling in quantum dynamics \cite{circuitBA3}. Moreover, in certain limits, a connection between the Floquet construction and non-rational conformal field theories has been established \cite{miao2024floquet}. It has further been demonstrated that, for quantum circuits obtained via the Trotterization of the XXZ spin chain, strong zero modes can emerge in specific regions of parameter space, \cite{vernier2024strong,Gehrmann:2025wnb}. More recently, within the integrable quantum circuit framework, correlation functions of spin operator strings have been explicitly computed, \cite{circuitBA4}. Another direction where these circuits are protagonists is the development of quantum algorithms used to prepare eigenstates of integrable Hamiltonians, for example, the adiabatic algorithm \cite{lutz2025adiabatic}, the variational algorithm \cite {nepomechie2020bethe},  or algorithms based on the algebraic Bethe circuit \cite{sopena2022algebraic, ruiz2025bethe}. A new type of circuit recently considered and whose dynamics can be simulated efficiently on classical computers hosts free fermions in disguise, see for example \cite{Szasz-Schagrin:2025xyq, fukai2025quantum}.

Over the years, the integrable Trotterization procedure has been generalized to several types of integrable quantum circuits, either by changing the model used as quantum gate (see for example \cite{vanicat2018integrable, vanicat2018integrable2} or \cite{sa2021integrable}, where the method has been generalized to build quantum circuits where the gate is related to a non-difference form $R$-matrix) or by changing the geometry, the spatial and temporal distribution of the gates. In \cite{miao2024floquet, Paletta:2025sap}, integrability of quantum circuits with periodic boundary conditions, in which gates are not arranged in the standard brickwork fashion, has been analysed. This mechanism takes inspiration from the light cone discretization of \cite{Baxter:1978xr, Destri:1987ze,Destri:1991zm}. One motivation for doing this is that, similarly to the continuous time model, the framework of generalized hydrodynamics can be used to describe the large-scale dynamics of Yang-Baxter integrable quantum circuits, and it can reveal features that do not necessarily appear in their continuous time counterparts. For example, spatially asymmetric dynamical correlation functions may emerge either by using gates that explicitly break the space-reflection symmetry by acting on distinct degrees of freedom or by changing their arrangement inside the circuit. 

Another possible construction corresponds to quantum circuits with open boundary conditions. These can be used, for example, to describe boundary-driven systems, \cite{Paletta:2024uzj}. In \cite{vanicat2018integrable}, the integrable Trotterization approach was developed for quantum circuits with open boundary conditions and the $R$-matrix of difference form. For these models, the periodic inhomogeneous transfer matrix is replaced by the double row transfer matrix of Sklyanin \cite{Sklyanin:1988yz}. The boundary gates are then taken to be the left and right $K$-matrices, which are solutions of the Sklyanin reflection algebra. In some cases, the system may still exhibit solvability of the non-equilibrium steady states \cite{Prosen:2011veg, prosen-exterior-lindblad} and, as shown by numerical techniques, may also display full-spectrum integrability, even when the boundary operators do not satisfy the Sklyanin algebra \cite{popkov2025exact,paletta2026integrability,Zhang:2026avw}.

\subsection{Summary of the paper}

In Sec. \ref{sec:basics}, we start by reviewing known concepts about quantum integrability. We define a Yang-Baxter integrable model with open boundary conditions as a model characterized by an $R$-matrix solution of the Yang-Baxter equation and $K$-matrices solutions of the reflection equations. In Sec. \ref{integrablequantumcircuitbuilding}, the original results begin. We consider quantum circuits with open boundary conditions and make further generalizations compared to what is available in the literature; see Table \ref{literature}. 
\begin{table}[H]
\hspace{1cm}\begin{tabular}{|c|c|c|c|}
\cline{2-4}
\multicolumn{1}{c|}{} & R difference form  & R non-difference form  & arbitrary geometry \\ \hline
periodic b.c & \cite{vanicat2018integrable2,vanicat2018integrable} &  \cite{sa2021integrable} &  \cite{miao2024floquet,Paletta:2025sap} \\ \hline
open b.c  &  \cite{vanicat2018integrable} & x & x \\ \hline
\end{tabular}
\caption{In this table, we summarize the different classes of models for which the integrable Trotterization procedure can be performed. The references indicate previously studied models, while the entries marked by a cross correspond to cases that had not been investigated before and are addressed in the present work.}
\label{literature}
\end{table}
In particular, we extend the Trotterization procedure to the case where the $R$-matrix has a non-difference form. This generalisation has important consequences, as it can be used to study the connections between the solvability of the non-equilibrium steady state of the Hubbard model circuits \cite{prosen2014exact,popkov2015infinitely} and the full integrable spectrum (in the same spirit as \cite{Paletta:2024uzj}).

In Sec. \ref{sec:result}, we present most of our new results. We begin in Sec. \ref{sec:gatespositions} where we construct all possible integrable quantum circuits with open boundary conditions and different geometries that can be generated by only two types of inhomogeneities, $\kappa$ and $-\kappa$. Each circuit is then fully determined by the positions of the $-\kappa$ inhomogeneities and the system size.

Analogously to the periodic case \cite{Paletta:2025sap}, we conjecture that, for open boundaries, every circuit whose gates satisfy the Yang--Baxter equation (YBE) and the boundary Yang-Baxter equation (BYBE), and which acts exactly once on each pair of spins per period, is integrable. We also provide a systematic algorithm for the inverse problem: starting from a given circuit, we determine the positions of the $-\kappa$ inhomogeneities that generate it.

We then examine the constructed circuits, generalising the proof from \cite{bensa2021fastest} to show that they share the same spectrum despite having distinct eigenvectors. We characterise the equivalence classes of these circuits using the number of $-\kappa$ inhomogeneities; since eigenvectors within the same class are related by similarity transformations only involving bulk operators, this parameter effectively groups them. In Sec. \ref{sec:minimumdept}, we identify among each class, the configurations that minimize the circuit depth, providing the most efficient realisations of the corresponding integrable dynamics. In particular, we find that the circuits with minimum depth are those of the staircase form (with $d$ steps) on the left- and on the right-hand sides, and a brickwork circuit in the middle.  Finally, we remark that if one is interested only in the eigenvalues, rather than in the equivalence classes introduced here, the minimum circuit depth is always $d=2$.

In Sec. \ref{sec:beyondkappa}, we discuss what happens when a third type of inhomogeneity is introduced. In Sec. \ref{sec:actualdepth}, we identify the configurations that generate the minimum possible depth in this case. This depth is equal to $d=4$, in contrast to the usual value $d=2$ obtained in the case with only two types of inhomogeneities. These are the analogs of the brickwork case, but for more types of inhomogeneities. In Sec. \ref{sec:effectivedepth},  we investigate the effect of placing the new inhomogeneity at the endpoints of the chain. This leads to the concept of \textit{effective} minimum depth. Finally, we provide a few insights into what happens as more types of inhomogeneities are introduced.

As an example, we consider quantum circuits where the gates act on two qubits and the boundary gates are a solution of the reflection algebra of Sklyanin.  We solve the reflection algebras for all known six- and eight-vertex $R$-matrices. For difference form, the $K$-matrices were classified in  \cite{Sklyanin:1988yz,Cherednik:1984vvp,Destri:1991zm,deVega:1993xi}. For non-difference form, new $R$-matrices were found in \cite{deLeeuw:2020ahe}, and in Sec. \ref{Kmatrices} we classify their corresponding boundary matrices. Focusing on  six- and eight-vertex models is enough here because we can map any Hermitian $4\times 4$ Hamiltonian (by using integrability preserving transformations such as the ones in Sec. \ref{subsec:symmetries}) to a six- or eight-vertex model (not necessarily Hermitian), see \cite{classificationybandboost} for more details. These newly built circuits may serve as a setup to study hydrodynamical quantities, such as dynamical correlation functions. We remark that the newly found $K$-matrices may be of interest for several different applications, not necessarily related to quantum circuits, for instance, studying the non-equilibrium steady states, or the continuous time dynamics. 

Finally, in Sec. \ref{sec:conclusions}, we discuss several open questions and outline potential avenues for future research.

We provide two Mathematica notebooks as supplementary material in the Zenodo repository \cite{zenodo}. The first, \texttt{OpenQCforDiffGeom.nb}, implements the construction of quantum circuits with arbitrary geometries for up to four types of inhomogeneities, while the second, \texttt{KL.nb}, contains a summary of the $K^R$ and $K^L$ matrices. Both notebooks include self-contained instructions on how to use them.

\section{Building an integrable open boundary quantum circuits}
\label{recap}
\subsection{The basics: integrability in a nutshell}\label{sec:basics}

\subsubsection{Constructing an open spin chain}

\paragraph{Yang-Baxter equation (YBE):} Integrable spin chains are discrete systems whose commuting conserved charges can be constructed from fundamental building blocks known as $R$-matrices, which are solutions of the Yang-Baxter equation
\begin{equation}
    R_{12}(u,v)R_{13}(u,w)R_{23}(v,w)=R_{23}(v,w)R_{13}(u,w)R_{12}(u,v).
    \label{eq:ybe}
\end{equation}
Here $u,v,w\in \mathbb{C}$ are called spectral parameters. Non-constant $R$-matrices are divided into those of difference form $R(u,v)=R(u-v)$ and non-difference form $R(u,v)\neq R(u-v)$. The difference form $R$-matrix can be obtained from \eqref{eq:ybe} upon setting $R(u,v)=R(u-v)$. 

For a local Hilbert space $V$ of dimension $D$, an $R$-matrix is such that $R: V\otimes V \mapsto V\otimes V$, where $V=\mathbb{C}^D$. Additionally, in Eq. \eqref{eq:ybe}, $R_{i,j}:V\otimes V\otimes V \mapsto V\otimes V\otimes V$, with $R_{12}=R\otimes \mathbb{I}$, $R_{23}=\mathbb{I}\otimes R$ and $R_{13}=P_{12}R_{23}P_{12}$, where $P$ is the permutation operator
\begin{equation}
    P=\sum_{i,j=1}^D e_{i,j}\otimes e_{j,i}, \quad (e_{i,j})_{k,l}=\delta_{i,k}\delta_{j,l}. 
\end{equation}
\paragraph{Boundary Yang-Baxter equation (BYBE):} For integrable \textbf{open} spin chains, two more ingredients are required: the ``right'' and ``left'' reflection matrices $K^R$ and $K^L$, respectively. Each of them satisfies a so-called Boundary Yang-Baxter equation \cite{Sklyanin:1988yz}, in particular, $K^R(u)$ satisfies
\begin{equation}
    R_{12}(u,v)K_1^R(u)R_{21}(v,-u)K_2^R(v)=K_2^R(v)R_{12}(u,-v)K_1^R(u)R_{21}(-v,-u).
    \label{eq:bybe}
\end{equation}

Once $R(u,v)$ and $K^R(u)$ are fixed, the construction of the other boundary matrix can vary depending on the symmetries of the $R$-matrix (see \cite{Sklyanin:1988yz,Mezincescu:1990hda,bajnok2006equivalences,Murgan:2008fs,thesisdelaRosa,Bielli:2025abu}).

By following \cite{vanicat2018integrable}, for $R$-matrices that do not necessarily exhibit any manifest symmetry, we first define the dual reflection matrix $\bar{K}(u)$, that satisfies the dual equation

\begin{equation}
    R_{12}^{-1}(u,v)\bar{K}_1(u)R_{21}^{-1}(v,-u)\bar{K}_2(v)=\bar{K}_2(v)R_{12}^{-1}(u,-v)\bar{K}_1(u)R_{21}^{-1}(-v,-u).
    \label{eq:dualbybediff}
\end{equation}

The $K^L(u)$ is related to $\bar{K}(u)$ by the automorphism
\begin{equation}
    K_1^L(u)=\text{tr}_0 \Big( \bar{K}_0(-u) \Big(\Big(\big(R_{01}(u,-u)\big)^{t_1}\Big)^{-1}\Big)^{t_1}P_{01}\Big),
    \label{KLandKbarintro}
\end{equation}
where $t_1$ is the transpose with respect to the first space: $0$ in $R_{01}$. $K^L(u)$ obeys the following reflection equation 
\begin{equation}
    R_{12}(-u,-v)K_1^L(u)\Big(\Big(\big(R_{12}(u,-v)\big)^{t_1}\Big)^{-1}\Big)^{t_1}K_2^L(v)=K_2^L(v)\Big(\Big(\big(R_{21}(v,-u)\big)^{t_2}\Big)^{-1}\Big)^{t_2}K_1^L(u)R_{21}(v,u).
    \label{eq:bybel}
\end{equation}

When the $R$-matrix satisfies additional symmetries, the automorphism \eqref{KLandKbarintro} and the reflection algebra \eqref{eq:bybel} take a simpler form, as discussed in the original work \cite{Sklyanin:1988yz}.
We also remark that, in some particular models, for example  some of the models in \cite{deLeeuw:2019zsi}, the operator $R_{ij}(u,-u)^{t_1}$ is singular, so the automorphism \eqref{KLandKbarintro} cannot be applied directly. We analyse this case in an upcoming work.

\paragraph{Transfer-matrix:} With the $R$-matrix and the corresponding reflection $K$-matrices, as in \cite{Sklyanin:1988yz}, we can define the double-row transfer matrix for an $N$-site spin chain as
\begin{equation}
    t(u,\{\theta_j\})=\text{tr}_a(K_a^L(u)T_a(u,\{\theta_j\})K_a^R(u)\hat{T}_a(u,\{\theta_j\})),\label{eq:transfer}
\end{equation}
where the monodromy matrices $T(u,\{\theta_j\})$ and $\hat{T}(u,\{\theta_j\})$ are given by
\begin{align}
    & T_a(u,\{\theta_j\})=R_{aN}(u,\theta_N)\cdots R_{a2}(u,\theta_2)R_{a1}(u,\theta_1),\label{eq:monodromy1}\\
    & \hat{T}_a(u,\{\theta_j\})\equiv T_a^{-1}(-u,\{\theta_j\})=R_{1a}(\theta_1,-u)R_{2a}(\theta_2,-u)\cdots R_{Na}(\theta_N,-u).\label{eq:monodromy2}
\end{align}
We use the notation $\{\theta_j\}=\{\theta_1,\theta_2,\dots,\theta_N\}$ to identify the inhomogeneities. In \eqref{eq:monodromy2}, we have used the unitarity\footnote{This property holds for any regular $R$-matrix ($R_{ij}(u,u)\propto P_{ij}$) solution of the Yang-Baxter equation.} property $R_{12}(u,v)R_{21}(v,u)=\mathbb{I}$.

The transfer matrix, as defined above, satisfies
\begin{equation}
    [t(u,\{\theta_j\}),t(v,\{\theta_j\})]=0,\label{eq:commutingtransfermatrices}
\end{equation}
as long as the set of  inhomogeneities $\{\theta_i\}$ are the same in both transfer matrices.

For simplicity, we will sometimes write $t(u)$ instead of  $t(u,\{\theta_j\})$. Unless otherwise specified, we always consider the inhomogeneous case, since the inhomogeneities are essential for constructing the integrable quantum circuit.

\paragraph{Conserved charges:} The transfer matrix is the generating function of the conserved\footnote{To be precise, for homogeneous models, we can define the Hamiltonian as $Q_2$ and hence, since all the charges commute with the Hamiltonian, they are conserved. For inhomogeneous models, we can define the dynamical evolution $\mathbb{U}=g(t(u))$. The quantum circuits we are considering belong to this class.} charges

 \begin{equation}
    Q_{n+1}=\frac{\partial^n f(t(u,\{\theta_i\}))}{\partial u^n}\Big|_{u=u_0},
\end{equation}
where $f$ is a function of the transfer matrix chosen to guarantee the locality of the charges\footnote{For a spin chain with all $\theta_i$ different, the charges will be non local, hence, for simplicity $f(t(u))=t(u)$.}, and $n=0,1,2\cdots$. 
For a homogeneous chain ($\theta_j=0$), $f(t(u))=t(u)$ and $u_0=0$, while for staggered inhomogeneities ($\theta_{odd}=\kappa,\,\theta_{even}=-\kappa$) $f(t(u))=\log t(u)$ and $u_0=\pm \kappa$. Due to \eqref{eq:commutingtransfermatrices}, the charges mutually commute $[Q_n,Q_m]=0$, for all $m,n=1,2,\cdots$.

For example, if we compute the first charge (corresponding to $n=1$) for the homogeneous case and a regular $R$-matrix, we obtain\footnote{In the following derivation, we assume that the limit of the product of two operators is equal to the product of the limits. For a more detailed explanation, we refer to \cite{Paletta:2024uzj}, Sec. D.1.} 
\begin{align}
    t'(0)=\Big(\tr \,K^L(0) \Big)\left[\left\{\sum_{j=1}^{N-1}h_{j,j+1},K_1^{R}(0)\right\}+K_1^{R,\prime}(0)+\frac{2\,\tr_a\left(K_a^L(0)h_{N,a}\right)}{\tr\, K^{L}(0)}K_1^{R}(0)+\frac{\tr\, K^{L,\prime}(0)}{\tr\, K^L(0)}K_1^{R}(0)\right],
\end{align}
where $h_{j,j+1}=h_{j,j+1}(u)=\partial_u\check{R}(u,v)|_{v=u}$ and $\check{R}(u,v)=P R(u,v)$. Assuming $K^{R}(0)$ is invertible and $\tr\,K^{L}(0)\neq 0$, we can define the Hamiltonian as 

\begin{align}
    &\mathbb{H} = \frac{t'(0)\,K_1^{R}(0)^{-1}}{2\tr\, K^{L}(0)}\nonumber\\
    &=\sum_{j=1}^{N-1}h_{j,j+1}-\frac{1}{2}h_{12}+\frac{1}{2}K_1^{R}(0)h_{12}K_1^{R}(0)^{-1}+\frac{1}{2}K_1^{R\,\prime}(0)K_1^{R}(0)^{-1}+\frac{\tr_a\left(K_a^{L}(0)h_{Na}\right)}{\tr\,K^{L}(0)}+\frac{\tr\,K^{L\,\prime}(0)}{2\tr\,K^{L}(0)},
\end{align}
where we can recognize the nearest-neighbor interactions in the bulk,  a one-site boundary term at the left boundary, and two-site terms on the right. For regular $K$-matrices\footnote{For a discussion on cases where $K^R$ is not regular and/or $\det(K^L(0))=0$ see the part following equation (38) in \cite{deVega:1993xi}.} $K^R(0)=\mathbb{I}$, the expression reduces to the known one of \cite{Sklyanin:1988yz}. 

\subsubsection{Symmetries of the Yang-Baxter equation (YBE) and the Boundary YBE}\label{subsec:symmetries}

There are several transformations of the $R$-matrix known to preserve the Yang--Baxter equation \eqref{eq:ybe}. To maintain the commutativity property \eqref{eq:commutingtransfermatrices}, the boundary matrices must remain solutions of the boundary Yang--Baxter equation \eqref{eq:bybe}. In this subsection, we briefly review the consequences of each $R$-matrix transformations for the right boundary matrix\footnote{Since the $K^L$ is constructed from the $K^R$ by using the automorphism \eqref{KLandKbarintro}, we do not discuss the consequences of these transformations on the $K^L$.}. For completeness, the corresponding derivations are presented in Appendix \ref{App:transformations}.

\paragraph{Local basis transformations}

\begin{equation}
    \tilde{R}_{12}(u,v)=V_1(u)V_2(v)R_{12}(u,v)V_1(u)^{-1}V_2(v)^{-1}\quad \Rightarrow\quad \tilde{K}^{R}(u)=V(u)K^R(u)V(-u)^{-1}. \label{eq:lbt}
\end{equation}

\paragraph{Normalization}

\begin{align}
    \tilde{R}(u,v)=&f(u,v)R(u,v)\quad \Rightarrow\quad \tilde{K}^R(u)=K^R(u)\nonumber\\
    & \text{iff}\quad f(u,v)f(v,-u)=f(u,-v)f(-v,-u).\label{eq:normalizationcondition}
\end{align}
For $R$-matrices of difference form, the condition \eqref{eq:normalizationcondition} is automatically satisfied, since in that case $f(v,-u)=f(u,-v)$ and $f(-v,-u)=f(u,v)$. For non-difference form $R$-matrices, however, the situation is different: if a given $K^R(u)$ satisfies the BYBE and the normalization of $R(u,v)$ is modified in such a way that \eqref{eq:normalizationcondition} no longer holds, then the BYBE must be solved again in order to determine the appropriate $K^R(u)$.

On the other hand, one may always perform an independent normalization directly at the level of the boundary matrix, namely $K^R(u)\to g(u) K^R (u)$.

\paragraph{Reparameterization}

\begin{align}
    \tilde{R}(u,v)&=R(g(u),g(v))\quad \Rightarrow \quad \tilde{K}^R(u)=K^R(g(u)),\nonumber\\
    & \text{iff} \quad g(-u)=-\,g(u).\label{eq:reparameterizationcondition}
\end{align}
If one applies a reparameterization that does not satisfy the condition $g(-u)=-\,g(u)$, there is no general transformation that can be applied directly to $K^R(u)$. In such a case, one must instead begin with the new $R$-matrix and solve the BYBE again in order to obtain the correct reflection matrices.

\paragraph{Twists}

\begin{align}
    \tilde{R}_{12}(u,v)&=W_1(u)R_{12}(u,v)W_2(v)^{-1}, \quad \text{ with } \quad [R_{12}(u,v),W_1(u)W_2(v)]=0\quad\nonumber\\
    &\Rightarrow \quad \tilde{K}^R(u)=W(u)K^R(u)W(-u)^{-1}.
\end{align}

We remark that the first three transformations preserve the symmetries of the models, while the twists can also change the degeneracies of the eigenvalues. For this reason, we will not use the twists in Sec. \ref{Kmatrices}.

\subsubsection{Symmetries of the spectrum}
\label{subsec:symmetriesspectrum}

The spectrum of a transfer matrix whose inhomogeneities are swapped $t(u;\sigma(\{\theta_i,\theta_j\}))$ is the same as the one for $t(u)$ before the swapping. This is a consequence of the fact that the two operators are related by a similarity transformation (see \cite{Ferrando:2023lrx} for a simple and clear proof for the periodic case). Since we were unable to find an explicit proof in the literature for the open case, especially for non-difference form $R$-matrices, we have provided a proof of this fact in Appendix \ref{App:swappinginhomogeneities} (heavily inspired by the proof presented in \cite{Ferrando:2023lrx}).

\subsection{Set up: building an integrable quantum circuit}
\label{integrablequantumcircuitbuilding}

\subsubsection{Basic idea: the integrable trotterization}

A many-body operator $M$ is defined as a quantum circuit if it can be decomposed into a finite sequence of two-particle operators, commonly known as gates. An integrable quantum circuit arises when $M$ commutes with the transfer matrix of an integrable model. This observation led to the development of the integrable Trotterization procedure. The underlying idea traces back to Baxter’s work \cite{baxter2016exactly}, where it was shown that the inhomogeneity parameters of the six-vertex model transfer matrix can be chosen to generate a discrete-time parallel updated dynamics on a periodic lattice. This construction was later extended in \cite{vanicat2018integrable,vanicat2018integrable2}, where an $R$-matrix-independent framework was introduced to define integrable unitary discrete-time dynamics. We remark that the quantum circuit framework can also be extended beyond unitary dynamics to generate dissipative evolution, for instance as a discretized version of Lindblad dynamics \cite{sa2021integrable, su2022integrable}. This is possible because the construction does not depend on the specific choice of $R$- and $K$-matrices, provided they satisfy the Yang–Baxter and boundary Yang–Baxter equations, Eqs.~\eqref{eq:ybe} and \eqref{eq:bybe}. The Trotterization scheme applies to both periodic and open boundary conditions. Here we extend the proof to systems with open boundaries and non-difference-form $R$-matrices. We first consider alternating inhomogeneities and show that, as in the difference-form case, they produce a two-step Floquet evolution. In subsequent sections, we discuss the construction for all other possible geometries.

\subsubsection{Brickwork construction}

In this section, for pedagogical reasons, we begin by considering the explicit construction of a circuit of fixed dimension. This framework is general enough to allow us to easily reconstruct the corresponding expression for arbitrary $N$.

We consider the transfer matrix \eqref{eq:transfer} with $N$ odd, specifically $N=5$, and alternating inhomogeneities 

\begin{equation}
    \theta_{\text{odd}}=\kappa, \quad \theta_{\text{even}}=-\kappa.
\end{equation}
This leads to
\begin{align}
    t(u)&=\text{tr}_a\left(K_a^L(u)R_{a5}(u,\kappa)R_{a4}(u,-\kappa)R_{a3}(u,\kappa)R_{a2}(u,-\kappa)R_{a1}(u,\kappa)\times\right.\nonumber\\
    & \hspace{1cm}\left.\times K_a^R(u)R_{1a}(\kappa,-u)R_{2a}(-\kappa,-u)R_{3a}(\kappa,-u)R_{4a}(-\kappa,-u)R_{5a}(\kappa,-u)\right).\label{eq:transferN5}
\end{align}

We now compute $t(\kappa)$
\begin{align}
    t(\kappa)&=\text{tr}_a\left(K_a^L(\kappa)R_{a5}(\kappa,\kappa)R_{a4}(\kappa,-\kappa)R_{a3}(\kappa,\kappa)R_{a2}(\kappa,-\kappa)R_{a1}(\kappa,\kappa)\times\right.\nonumber\\
    & \hspace{1cm}\left.\times K_a^R(\kappa)R_{1a}(\kappa,-\kappa)R_{2a}(-\kappa,-\kappa)R_{3a}(\kappa,-\kappa)R_{4a}(-\kappa,-\kappa)R_{5a}(\kappa,-\kappa)\right)\label{eq:tN5kappa1}\\
    &=g(\kappa)^3g(-\kappa)^2\text{tr}_a\left(K_a^L(\kappa)P_{a5}R_{a4}(\kappa,-\kappa)P_{a3}R_{a2}(\kappa,-\kappa)P_{a1}\times\right.\nonumber\\
    & \hspace{1cm}\left.\times K_a^R(\kappa)R_{1a}(\kappa,-\kappa)P_{2a}R_{3a}(\kappa,-\kappa)P_{4a}R_{5a}(\kappa,-\kappa)\right)\label{eq:tN5kappa2}\\
    &=g(\kappa)^3g(-\kappa)^2\text{tr}_a\left(P_{a5}P_{a4}\check{R}_{a4}(\kappa,-\kappa)P_{a3}P_{a2}\check{R}_{a2}(\kappa,-\kappa)P_{a1}K_a^R(\kappa)\times\right.\nonumber\\
    & \hspace{1cm}\left.\times P_{1a}\check{R}_{1a}(\kappa,-\kappa)P_{2a}P_{3a}\check{R}_{3a}(\kappa,-\kappa)P_{4a}P_{5a}\check{R}_{5a}(\kappa,-\kappa)K_a^L(\kappa)\right)\label{eq:tN5kappa3}\\
    &=g(\kappa)^3g(-\kappa)^2\check{R}_{45}(\kappa,-\kappa)\check{R}_{23}(\kappa,-\kappa) K_1^R(\kappa)\check{R}_{12}(\kappa,-\kappa)\check{R}_{34}(\kappa,-\kappa)\text{tr}_a\left(\check{R}_{5a}(\kappa,-\kappa)K_a^L(\kappa)\right),\label{eq:tN5kappa4}
\end{align}
where from Eq. \eqref{eq:tN5kappa1} to Eq. \eqref{eq:tN5kappa2} we use the regularity condition  $R(u,u)=g(u)P$. From Eq. \eqref{eq:tN5kappa2} to Eq. \eqref{eq:tN5kappa3}, we rewrite $R=P\check{R}$ and cyclically permute $K_a^L(\kappa)$ to the end using the cyclicity of the trace in the auxiliary space $a$. 
Finally, from Eq. \eqref{eq:tN5kappa3} to Eq. \eqref{eq:tN5kappa4}, we use the relations $P_{ak}\mathcal{O}_{aj}=\mathcal{O}_{kj}P_{ak}$ and $P_{ak}\mathcal{O}_{ja}=\mathcal{O}_{jk}P_{ak}$ to move all permutation operators $P_{ij}$ in the first line to their corresponding positions in the second line, and then apply $P^2=\mathbb{I}$. We also observe that all operators except $\check{R}_{5a}(\kappa,-\kappa)K_a^L(\kappa)$ are independent of the auxiliary space index $a$ and can therefore be taken out of the trace.

We can then define the circuit evolution operator as $M=t(\kappa)/\left(g(\kappa)^3g(-\kappa)^2\right)$, 

\begin{equation}
    M=K_1^R(\kappa)U_{23} U_{45} U_{12}U_{34} \tilde{K}^L_5 (\kappa)
\end{equation}
where 

\begin{equation}
    U=\check{R}(\kappa,-\kappa) \quad \text{and} \quad \tilde{K}^L_5(\kappa)=\text{tr}_a\left(\check{R}_{5a}(\kappa,-\kappa)K_a^L(\kappa)\right).
    \label{eq:UandtildeKN5}
\end{equation}

By construction, since $[t(u),t(v)]=0$ for any choice of inhomogeneities, $M$ satisfies
\begin{equation}
    [M,t(u)]=0
\end{equation}
\noindent and is therefore integrable.

\begin{figure}[H]
    \centering
    \includegraphics[scale=1]{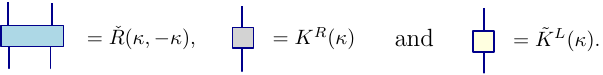}
    \caption{Graphical representation of the basic gates}
    \label{fig:basicgates}
\end{figure}

Graphically, in Figure \ref{fig:basicgates}, we represent the gates $U$, $K^R(\kappa)$ and $\tilde{K}^L(\kappa)$, respectively, as

\begin{figure}[H]
    \centering
        \includegraphics[scale=1]{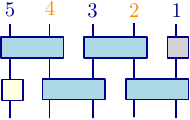}
        \caption{Brickwork circuit of length $N=5$, with time flowing from bottom to top. The orange colour identifies the positions of the $-\kappa$ inhomogeneities in the transfer matrix \eqref{eq:transfer}. All remaining inhomogeneities are set equal to $\kappa$.}
        \label{fig1brickworkN5}
\end{figure}

\noindent which leads to the quantum circuit $M$ represented in Fig  \ref{fig1brickworkN5}.

By generalising the construction above, we obtain the following expression for an odd number of sites $N$ 
\begin{equation}
M=\left(\prod_{i=1}^{\frac{N-1}{2}}U_{2i,2i+1}\right)K_1^R(\kappa)\left(\prod_{j=1}^{\frac{N-1}{2}}U_{2j-1,2j}\right)\tilde{K}_N^L(\kappa),
\label{brickworkodd}
\end{equation}
while for even $N$ we find
\begin{equation}
M=\tilde{K}_N^L(\kappa)\left(\prod_{i=1}^{\frac{N}{2}-1}U_{2i,2i+1}\right)K_1^R(\kappa)\left(\prod_{j=1}^{\frac{N}{2}}U_{2j-1,2j}\right),\label{brickworkeven}
\end{equation}
where $\tilde{K}^L_N(\kappa)=\text{tr}_a\left(\check{R}_{Na}(\kappa,-\kappa)K_a^L(\kappa)\right)$. Above, in equations \eqref{brickworkodd} and \eqref{brickworkeven}, the $U_{j,j+1}$ inside each product commute. This will not be the case for most expressions in the remainder of this paper. Therefore, it is convenient to define \textit{ordered} products as follows

\begin{align}
    &\mathop{\overset{\leftarrow}{\prod}}\limits_{1 \le i \le b}A_i=A_b\cdots A_2\,A_1,\\
    &\mathop{\overset{\rightarrow}{\prod}}\limits_{1 \le i \le b}A_i=A_1\,A_2\cdots A_b. 
\end{align}

\begin{figure}[H]
    \centering
        \includegraphics[scale=0.4]{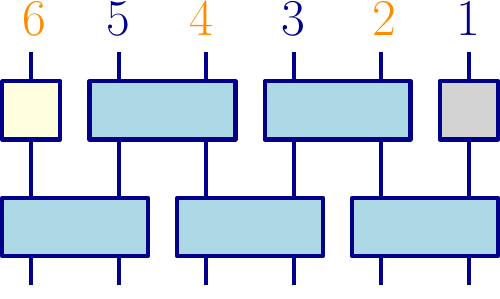}
        \caption{Brickwork circuit of length $N=6$, with time flowing from bottom to top. }
        \label{fig1brickworkN6}
\end{figure}

The graphical representation of $M$ for even $N$ differs slightly from the odd case. In fact, when compared with Fig. \ref{fig1brickworkN5}, the bottom layer will consist only of $N/2$ $U$-gates, while the upper layer will contain $K_1^R(\kappa)$, $\tilde{K}_N^L(\kappa)$ and $N/2-1$ $U$-gates. As an example, see Fig. \ref{fig1brickworkN6} for $N=6$.

This generalises the construction of \cite{vanicat2018integrable} for $R$-matrices of non-difference form. It is easy to check that upon imposing the restriction $R(u,v)\to R(u-v)$, our expressions reduce to those found in the literature, \cite{vanicat2018integrable}. In the next section, we generalise this construction to circuits characterized by different gate configurations (different geometries).

\section{Integrable open boundary quantum circuits with different geometries}\label{sec:result}

\subsection{Positions of the gates as a function of their inhomogeneities}\label{sec:gatespositions}

First, we construct quantum circuits with different geometries. 

\subsubsection{General construction }\label{sec:generalconstruction}

Different geometries are characterized by the specific placement of the gates. In this section, we provide a closed-form expression for the circuit evolution, derived from the positions of the $-\kappa$ inhomogeneities, valid for cases where exactly two types of inhomogeneities are present.

To do so, we consider the transfer matrix of an open spin chain with $N$ sites, where some inhomogeneities are set to $-\kappa$ and others to $\kappa$. We denote the number of $-\kappa$ inhomogeneities by $\kappa_{-}$ and the number of $\kappa$ inhomogeneities by $\kappa_{+}=N-\kappa_-$. The $-\kappa$ inhomogeneities are located at positions $n_1, n_2, \ldots, n_{\kappa_{-}}$, and we denote such a configuration by the notation

\begin{equation}
    \Vec{n}=({n_{\kappa_-},\cdots, n_2, n_1}), \quad \text{with} \quad 1\le n_1 < n_2 < \cdots < n_{\kappa_-}\le N.\label{eq:vecndef}
\end{equation}

In the following, we use as gates $K^{R}(\kappa)$, $U$ and $\tilde{K}^L(\kappa)$, with the last two defined by

\begin{equation}
    U=\check{R}(\kappa,-\kappa), \quad \text{and} \quad \tilde{K}^L_N(\kappa)=\text{tr}_a\left(K_a^L(\kappa)\check{R}_{Na}(\kappa,-\kappa)\right).
    \label{eq:UandtildeKdefinition}
\end{equation}
Their graphical representation is shown in Fig. \ref{fig:basicgates}.

In the following results (Lemma~1, Theorem~1, Lemma~2, and Theorem~2), we give general expressions that determine the corresponding quantum circuit from $\vec{n}$. They are valid for both even and odd $N$.

\paragraph{Lemma 1:} Consider a spin chain with double-row transfer matrix \eqref{eq:transfer}--\eqref{eq:monodromy2} and regular $R$-matrix ($R(u,u)=g(u)P$). If $n_{\kappa_-}<N$ (i.e.  $\theta_N=\kappa$), then

    \begin{equation}
    t(\kappa)=g(\kappa)\left(\mathop{\overset{\leftarrow}{\prod}}\limits_{1 \le i \le N-1}\check{R}_{i,i+1}(\kappa,\theta_{i})\right)K_1^R(\kappa)\left(\mathop{\overset{\rightarrow}{\prod}}\limits_{1 \le j \le N-1}\check{R}_{j,j+1}(\theta_j,-\kappa)\right)\tilde{K}_N^{L}(\kappa),\label{eq:lemma1}
\end{equation} 
where $\tilde{K}^L$ is defined in \eqref{eq:UandtildeKdefinition}.

\begin{proof}
Consider the transfer matrix \eqref{eq:transfer}--\eqref{eq:monodromy2} for a chain with $N$ sites and inhomogeneities $\theta_1,\ldots,\theta_N$. Let $\theta_N=\kappa$, while $\theta_j$ for $j=1,\ldots,N-1$ remain arbitrary. Then $t(\kappa)$ takes the form
\begin{align}
    t(\kappa)&=\tr_a\left(K_a^L(\kappa)R_{a,N}(\kappa,\kappa)R_{a,N-1}(\kappa,\theta_{N-1})\cdots R_{a,2}(\kappa,\theta_{2})R_{a,1}(\kappa,\theta_{1}) \times\right.\nonumber\\
    &\hspace{0.2cm} \left. \times \, K_a^R(\kappa)R_{1,a}(\theta_1,-\kappa)R_{2,a}(\theta_2,-\kappa)\cdots R_{N-1,a}(\theta_{N-1},-\kappa)R_{N,a}(\kappa,-\kappa)\right)\label{eq:lemma1p1}\\[0.2cm]
    &=g(\kappa)\tr_a\left(K_a^L(\kappa)P_{a,N}R_{a,N-1}(\kappa,\theta_{N-1})\cdots R_{a,2}(\kappa,\theta_{2})R_{a,1}(\kappa,\theta_{1}) \times\right.\nonumber\\
    &\hspace{0.2cm} \left. \times \, K_a^R(\kappa)R_{1,a}(\theta_1,-\kappa)R_{2,a}(\theta_2,-\kappa)\cdots R_{N-1,a}(\theta_{N-1},-\kappa)R_{N,a}(\kappa,-\kappa)\right) \label{eq:lemma1p2} \\[0.2cm]
    &=g(\kappa)\tr_a\left(K_a^L(\kappa)R_{N,N-1}(\kappa,\theta_{N-1})\cdots R_{N,2}(\kappa,\theta_{2})R_{N,1}(\kappa,\theta_{1}) \times\right.\nonumber\\
    &\hspace{0.2cm} \left. \times \, K_N^R(\kappa)R_{1,N}(\theta_1,-\kappa)R_{2,N}(\theta_2,-\kappa)\cdots R_{N-1,N}(\theta_{N-1},-\kappa)P_{a,N}R_{N,a}(\kappa,-\kappa)\right) \label{eq:lemma1p3}\\[0.2cm]
    &=g(\kappa)R_{N,N-1}(\kappa,\theta_{N-1})\cdots R_{N,2}(\kappa,\theta_{2})R_{N,1}(\kappa,\theta_{1})K_N^R(\kappa)R_{1,N}(\theta_1,-\kappa) \times\nonumber\\
    &\hspace{0.2cm} \times \, R_{2,N}(\theta_2,-\kappa)\cdots R_{N-1,N}(\theta_{N-1},-\kappa)\tr_a\left(K_a^L(\kappa)P_{a,N}R_{N,a}(\kappa,-\kappa)\right) \label{eq:lemma1p4}\\[0.2cm]
    &=g(\kappa)\check{R}_{N-1,N}(\kappa,\theta_{N-1})\cdots \check{R}_{2,3}(\kappa,\theta_{2})\check{R}_{1,2}(\kappa,\theta_{1})K_1^R(\kappa)\check{R}_{1,2}(\theta_1,-\kappa) \times\nonumber\\
    &\hspace{0.2cm} \times \, \check{R}_{2,3}(\theta_2,-\kappa)\cdots \check{R}_{N-1,N}(\theta_{N-1},-\kappa)\tilde{K}_N^L(\kappa) \label{eq:lemma1p5}\\[0.2cm]
    &=g(\kappa)\left(\mathop{\overset{\leftarrow}{\prod}}\limits_{1 \le i \le N-1}\check{R}_{i,i+1}(\kappa,\theta_{i})\right)K_1^R(\kappa)\left(\mathop{\overset{\rightarrow}{\prod}}\limits_{1 \le j \le N-1}\check{R}_{j,j+1}(\theta_j,-\kappa)\right)\tilde{K}_N^{L}(\kappa), \label{eq:lemma1p6}
\end{align}
\noindent where from Eq. \eqref{eq:lemma1p1} to Eq.  \eqref{eq:lemma1p2}  we use the regularity condition $R(u,u)=g(u)P$. From Eq. \eqref{eq:lemma1p2} to Eq. \eqref{eq:lemma1p3}, we use the relation $P_{a,N}\mathcal{O}_{a,j}=\mathcal{O}_{N,j}P_{a,N}$ (for $j\neq a$), to commute through the $R$-matrices and move it close to the end of the trace. We then notice that from $R_{N,N-1}$ up to $R_{N-1,N}$, no operator acts non-trivially on the auxiliary space $a$, and therefore all such factors can be taken outside the trace. The next step consists of two parts. First, we identify $\tr_a\left(K_a^L(\kappa)P_{a,N}R_{N,a}(\kappa,-\kappa)\right)$ with $\tilde{K}_N^L(\kappa)$, as defined in \eqref{eq:UandtildeKdefinition}. Second, we define $\check{R}=PR$ and again use permutation identities such as $P_{a,N}\mathcal{O}_{a,j}=\mathcal{O}_{N,j}P_{a,N}$ to rewrite the expression in the form \eqref{eq:lemma1p5}. Finally, we recognize that the resulting product of $R$-matrices coincides exactly with the products appearing in \eqref{eq:lemma1p6}.\end{proof}

\paragraph{Theorem 1:} A quantum circuit built for a configuration of inhomogeneities given by $\Vec{n}$ with  $n_{\kappa_-}<N$, is given by the following general formula
\begin{equation}
t(\kappa)=g(\kappa)^{N-\kappa_-}g(-\kappa)^{\kappa_-}\left(\mathop{\overset{\leftarrow}{\prod}}\limits_{1 \le r \le \kappa_-}U_{n_{r},n_{r}+1}\right)K_1^R(\kappa)\left(\mathop{\overset{\rightarrow}{\prod}}\limits_{\substack{
1 \le j \le N-1 \\[0.07cm]
j \neq n_1,n_2,\ldots
n_{\kappa_-}}}U_{j,j+1}\right)\tilde{K}^L_N(\kappa). 
\label{eq:generalformulaforopen1}
\end{equation}

\begin{proof}
    Since $\check{R}(u,u)=g(u)\mathbb{I}$, if we consider a system with inhomogeneities $-\kappa$ at positions $\vec{n}$ as in Eq. \eqref{eq:vecndef}, and with  $n_{\kappa_-}<N$, the $\check{R}_{i,i+1}(\theta_i,-\kappa) $ in the second product in Lemma 1 becomes proportional to $\mathbb{I}$ for all $i\in \vec{n}$. This means that the second product can be written as 
    \begin{align}
        \left(\mathop{\overset{\rightarrow}{\prod}}\limits_{1 \le i \le N-1}\check{R}_{i,i+1}(\theta_i,-\kappa)\right)&=g(-\kappa)^{\kappa_-}\left(\mathop{\overset{\rightarrow}{\prod}}\limits_{\substack{
        1 \le j \le N-1 \\[0.07cm]
        j \neq n_1,n_2,\ldots
        n_{\kappa_-}}}\check{R}_{j,j+1}(\kappa,-\kappa)\right)\\
        &=g(-\kappa)^{\kappa_-}\left(\mathop{\overset{\rightarrow}{\prod}}\limits_{\substack{
        1 \le j \le N-1 \\[0.07cm]
        j \neq n_1,n_2,\ldots
        n_{\kappa_-}}}U_{j,j+1}\right),\label{eq:thm1firstprod}
    \end{align}
\noindent where in the first line,  the factor $g(-\kappa)^{\kappa_-}$ arises from the fact that $\kappa_-$ of the $\check{R}_{i,i+1}(\theta_i,-\kappa)$ becomes $\check{R}_{i,i+1}(-\kappa,-\kappa)=g(-\kappa)\mathbb{I}$. In the second equality, we use the first definition in \eqref{eq:UandtildeKdefinition}.

We now observe that every $\check{R}_{i,i+1}$ depending on the pair $(\theta_i,-\kappa)$ in the second product of Lemma~1 also appears in the first product, where it depends on $(\kappa,\theta_{i})$. This implies that, for a given configuration $\vec{n}$, every $\check{R}_{i,i+1}$ that reduces to a multiple of $\mathbb{I}$ in the second product corresponds to a $U$ in the first product, and vice versa. Therefore, the first product in Lemma 1 can be rewritten as

\begin{align}
    \left(\mathop{\overset{\leftarrow}{\prod}}\limits_{1 \le i \le N-1}\check{R}_{i,i+1}(\kappa,\theta_{i})\right)&=g(\kappa)^{N-\kappa_-}U_{n_{\kappa_-},n_{\kappa_-}+1}\cdots U_{n_2,n_2+1}U_{n_1,n_1+1}\\
    &=g(\kappa)^{N-\kappa_-}\left(\mathop{\overset{\leftarrow}{\prod}}\limits_{1 \le r \le \kappa_-}U_{n_{r},n_{r}+1}\right)\label{eq:thm1secondprod}
\end{align}
\noindent where we used multiple times that $ \check{R}(\kappa,\kappa)=g(\kappa)\mathbb{I}$.

Finally, substituting Eqs. \eqref{eq:thm1firstprod} and \eqref{eq:thm1secondprod} into Lemma 1, we recover Eq. \eqref{eq:generalformulaforopen1}.
\end{proof}

\paragraph{Lemma 2: } Consider a spin chain with double-row transfer matrix \eqref{eq:transfer}--\eqref{eq:monodromy2} and regular $R$-matrix ($R(u,u)=g(u)P$). If $n_{\kappa_-}=N$, then
    \begin{equation}
    t(\kappa)=g(-\kappa)\tilde{K}_N^{L}(\kappa)\left(\mathop{\overset{\leftarrow}{\prod}}\limits_{1 \le i \le N-1}\check{R}_{i,i+1}(\kappa,\theta_{i})\right)K_1^R(\kappa)\left(\mathop{\overset{\rightarrow}{\prod}}\limits_{1 \le j \le N-1}\check{R}_{j,j+1}(\theta_j,-\kappa)\right),\label{eq:lemma2} 
\end{equation} 
where $\tilde{K}^L$ is defined in \eqref{eq:UandtildeKdefinition}. 

\begin{proof}

\begin{align}
    t(\kappa)&=\tr_a\left(K_a^L(\kappa)R_{a,N}(\kappa,-\kappa)R_{a,N-1}(\kappa,\theta_{N-1})\cdots R_{a,2}(\kappa,\theta_{2})R_{a,1}(\kappa,\theta_{1}) \times\right.\nonumber\\
    &\hspace{0.2cm} \left. \times \, K_a^R(\kappa)R_{1,a}(\theta_1,-\kappa)R_{2,a}(\theta_2,-\kappa)\cdots R_{N-1,a}(\theta_{N-1},-\kappa)R_{N,a}(-\kappa,-\kappa)\right) \label{eq:lemma2p1}\\[0.2cm]
    &=g(-\kappa)\tr_a\left(K_a^L(\kappa)R_{a,N}(\kappa,-\kappa)R_{a,N-1}(\kappa,\theta_{N-1})\cdots R_{a,2}(\kappa,\theta_{2})R_{a,1}(\kappa,\theta_{1}) \times\right.\nonumber\\
    &\hspace{0.2cm} \left. \times \, K_a^R(\kappa)R_{1,a}(\theta_1,-\kappa)R_{2,a}(\theta_2,-\kappa)\cdots R_{N-1,a}(\theta_{N-1},-\kappa)P_{N,a}\right) \label{eq:lemma2p2} \\[0.2cm]
    &=g(-\kappa)\tr_a\left(K_a^L(\kappa)P_{a,N}R_{N,a}(\kappa,-\kappa)\right)R_{N,N-1}(\kappa,\theta_{N-1})\cdots R_{N,2}(\kappa,\theta_{2})R_{N,1}(\kappa,\theta_{1}) \times\nonumber\\
    &\hspace{0.2cm} \times \, K_N^R(\kappa)R_{1,N}(\theta_1,-\kappa)R_{2,N}(\theta_2,-\kappa)\cdots R_{N-1,N}(\theta_{N-1},-\kappa) \label{eq:lemma2p4}\\[0.2cm]
    &=g(-\kappa)\tilde{K}_N^{L}(\kappa)\check{R}_{N-1,N}(\kappa,\theta_{N-1})\cdots \check{R}_{2,3}(\kappa,\theta_{2})\check{R}_{1,2}(\kappa,\theta_{1})K_1^R(\kappa)\check{R}_{1,2}(\theta_1,-\kappa) \times\nonumber\\
    &\hspace{0.2cm} \times \, \check{R}_{2,3}(\theta_2,-\kappa)\cdots \check{R}_{N-1,N}(\theta_{N-1},-\kappa) \label{eq:lemma2p5}\\[0.2cm]
    &=g(-\kappa)\tilde{K}_N^{L}(\kappa)\left(\mathop{\overset{\leftarrow}{\prod}}\limits_{1 \le i \le N-1}\check{R}_{i,i+1}(\kappa,\theta_{i})\right)K_1^R(\kappa)\left(\mathop{\overset{\rightarrow}{\prod}}\limits_{1 \le j \le N-1}\check{R}_{j,j+1}(\theta_j,-\kappa)\right). \label{eq:lemma2p6}
\end{align}

The steps of this proof are almost exactly the same as for Lemma 1, the only difference is that now $P_{N,a}$ appears at the end of the expression instead of at the beginning. As a consequence, we move $P_{N,a}$ back to the left in order to identify  $\tr_a\left(K_a^L(\kappa)P_{a,N}R_{N,a}(\kappa,-\kappa)\right)$ with $\tilde{K}_N^L(\kappa)$ and to take all remaining factors out of the trace. As a result, $\tilde{K}^L_N(\kappa)$ appears on the left-hand side of the expression rather than on the right.
\end{proof}

\paragraph{Theorem 2:}A quantum circuit built for a given inhomogeneity configuration  $\Vec{n}$, with  $n_{\kappa_-}=N$, is given by the following general formula\vspace{-0.5cm}
\begin{equation}
t(\kappa)=g(\kappa)^{N-\kappa_-}g(-\kappa)^{\kappa_-}\tilde{K}^L_N(\kappa)\left(\mathop{\overset{\leftarrow}{\prod}}\limits_{1 \le r \le \kappa_- -1}U_{n_{r},n_{r}+1}\right)K_1^R(\kappa)\left(\mathop{\overset{\rightarrow}{\prod}}\limits_{\substack{
1 \le j \le N-1 \\[0.07cm]
j \neq n_1,n_2,\ldots
n_{\kappa_-}}}U_{j,j+1}\right).
        \label{eq:generalformulaforopen2}
\end{equation}

\begin{proof}

    Lemma~1 and Lemma~2 are structurally similar. There are two differences: the sign in $g(\kappa)$ and the position of $\tilde{K}^L(\kappa)$. 
    As a consequence, the proof of Theorem 2 is basically the same as for Theorem 1. The only difference is that since  now $n_{\kappa_-}=N$, the gate $U_{n_{\kappa},n_{\kappa}+1}$ is no longer included in the first product but is instead moved to the second one.
\end{proof}

From now on, we will always define the quantum circuit as 

\begin{equation}
    M=\frac{t(\kappa)}{g(\kappa)^{N-\kappa_-}g(-\kappa)^{\kappa_-}}.
\end{equation}

Using the general expressions \eqref{eq:generalformulaforopen1}-\eqref{eq:generalformulaforopen2} of Theorems~1 and~2, we can reproduce the brickwork-type circuit expressions \eqref{brickworkodd}--\eqref{brickworkeven}. In particular, to obtain \eqref{brickworkodd} for $N$ odd, we set $(N-1)/2$ inhomogeneities to $-\kappa$, with $\vec{n}=(N-1,N-3,\dots, 2)$. Since in this case $n_{\kappa_-} < N$, we apply Theorem 1, and Eq. \eqref{eq:generalformulaforopen1} reproduces Eq. \eqref{brickworkodd}. To obtain Eq. \eqref{brickworkeven}, for $N$ even, we set $N/2$ inhomogeneities to $-\kappa$ with $\vec{n}=(N,N-2,\dots, 2)$. In this case we apply Theorem~2, and Eq.~\eqref{eq:generalformulaforopen2} reproduces Eq.~\eqref{brickworkeven}.

We have written a Mathematica notebook named \texttt{OpenQCforDiffGeom.nb}, see \cite{zenodo}, that implements Lemmas 1 and 2.

We note that, unlike in the periodic case, for spin chains with open boundaries the choice of inhomogeneities $(\kappa,-\kappa)$ is not equivalent to a generic choice $(\kappa_1,\kappa_2)$.\footnote{Starting from a transfer matrix with inhomogeneities $(\kappa_1,\kappa_2)$, one might try to shift $u$, $\kappa_1$, and $\kappa_2$ by $\alpha=-(\kappa_1+\kappa_2)/2$ and then define $\kappa=(\kappa_1-\kappa_2)/2$, suggesting that the two models are equivalent. However, this is not the case: such a shift of the spectral parameters does not satisfy the property \eqref{eq:reparameterizationcondition}, and therefore gives rise to a different model.} We focus on the choice $(\kappa,-\kappa)$ for two reasons: it yields quantum circuits of smaller depth, and it produces circuits in which each gate acts on a given pair of spins only once. 

\subsubsection{From the circuit to $\vec{n}$}\label{sec:fromcircuittoinhom}

For the open-boundary case, similarly to the periodic setting \cite{Paletta:2025sap}, we conjecture that any circuit in which each gate $U_{i,i+1}$ (constructed from an $\check{R}$-matrix) appears exactly once per period to every nearest-neighbor pair of spins, and where each boundary gate is constructed from a $K$-matrix, is integrable. 

In this section, we go one step further by providing a procedure that, given the operator that describes the evolution of a quantum circuit with open boundaries, identifies the corresponding $\vec{n}$-configuration that generates it. The procedure consists of manipulating the circuit expression using only the allowed commutation relations until it is brought into one of the forms stated in Theorem 1 or Theorem 2.

The procedure is described below, followed by two explicit examples.

\paragraph{Step 1:} Write down the expression corresponding to the circuit by reading off the position of each gate. Recall that time runs from bottom to top.

\paragraph{Step 2:} Verify if the left boundary gate $\tilde{K}^L_{N}$ lies above or below  $U_{N-1,N}$. If  $\tilde{K}^L_{N}$ lies below $U_{N-1,N}$, then $n_{\kappa_-}\neq N$ and the circuit falls under Theorem 1. If $\tilde{K}^L_{N}$ lies above    $U_{N-1,N}$ instead, $n_{\kappa_-}= N$ and the circuit falls under Theorem 2.

\paragraph{Step 3:} $U_{i,i+1}$ commutes with any gate that does not act on sites $i$ or $i+1$. Respecting the allowed commutation relations, use this fact to move $U$-gates through each other and through $K_1^{R}$ until the following two configurations are simultaneously obtained: (1) the  gates to the left of $K_1^R$ are in \textit{decreasing} order from left to right; (2) the remaining gates appear in \textit{increasing} order from left to right, starting with $K_1^R$ itself. 

\paragraph{Step 4:} Count the number of $U$ gates to the left of $K_1^R$ and call that $\tilde{\kappa}_-$. If in step 2, we had $n_{\kappa_-}\neq N$ , then $\kappa_-=\tilde{\kappa}_-$, if instead we had $n_{\kappa_-}= N$, then $\kappa_-=\tilde{\kappa}_-+1$.

\paragraph{Step 5:} Following these steps, the components of $\vec{n}$ not fixed by Step 2 are determined by the first index of each $U$ to the left of $K_1^R$ written.

\paragraph{Step 6:} Use Theorem 1 or 2 to check that this $\vec{n}$ gives the correct circuit.

\

We now illustrate the procedure through two examples.

\begin{figure}[H]
    \centering
        \includegraphics[scale=0.9]{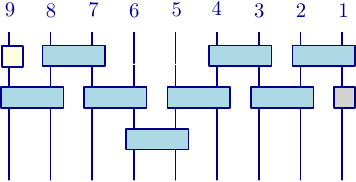}
        \caption{A quantum circuit for $N=9$. Applying our procedure leads to the inhomogeneities configuration in equation \eqref{eq:example1} and it falls under Theorem 2. }
        \label{fig:example1}
\end{figure}

\paragraph{Example 1:} Assume we are given the circuit in figure \ref{fig:example1} and asked to find the corresponding $\vec{n}$ that leads to it. 

To do so, we apply the steps (1)-(6) in order.

\

\noindent Step 1: Write $M$ by looking at Figure \ref{fig:example1} and identifying the position of each gate. This leads to\footnote{Notice that gates in the same row of the circuit commute with each other. As a consequence, at this stage, you can write them in your preferred order.}
\begin{equation}
    M=\tilde{K}_9^L(\kappa)U_{78}U_{34}U_{12}K_1^R(\kappa)U_{23}U_{45}U_{67}U_{89}U_{56}.
\end{equation}

\noindent Step 2: $\tilde{K}_9^L$ is above $U_{89}$, therefore, we know that $n_{\kappa_-}=9$.

\

\noindent Step 3: Following the instructions in step 3, we write 
\begin{equation}
    M=\tilde{K}_9^L(\kappa)U_{78}U_{67}U_{34}U_{12}K_1^R(\kappa)U_{23}U_{45}U_{56}U_{89}.
\end{equation}
Notice that to the left of $K_1^R$ we now have decreasing indices from left to right  and every gate is in its place. To the right of $K_1^R$ we only have increasing indices from left to right, starting with $K_1^R$ itself. Very importantly, only allowed commutations were performed.

\

\noindent Step 4: We know from step 2 that $n_{\kappa_-}=9$, and from step 3, we see that we have four $U$ gates to the left of $K_1^R$. So, $\kappa_-=1+4=5$.

\ 
\noindent Step 5: The first indices in the $U$ gates to the left of $K_1^R$ are $(7,6,3,1)$. Together with the information from step 2, we know that 
\begin{equation}
    \vec{n}=(9,7,6,3,1).\label{eq:example1}
\end{equation}

\noindent Step 6: The $\vec{n}$ written in step 5 automatically leads to the circuit in Figure \ref{fig:example1} if we use Theorem 2, and is therefore correct. 

\paragraph{Example 2:} If we are given the following circuit:

\begin{figure}[H]
    \centering
        \includegraphics[scale=0.9]{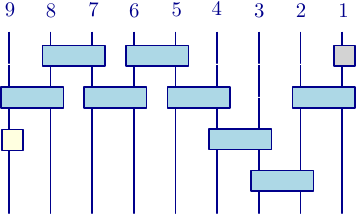}
        \caption{A quantum circuit for $N=9$. This falls under Theorem 1. }
        \label{fig:example2}
\end{figure}

\noindent Step 1: $M=U_{78}U_{56}K_1^R(\kappa)U_{12}U_{45}U_{67}U_{89}U_{34}\tilde{K}_9^L(\kappa)U_{23}$.

\

\noindent Step 2: $\tilde{K}_9^L$ is below $U_{89}$, therefore, we know that $n_{\kappa_-}\neq 9$. This implies $\theta_{N}=+\kappa$, consistent with the framework of Theorem 1.

\

\noindent Step 3: $M=U_{78}U_{56}U_{45}U_{34}K_1^R(\kappa)U_{12}U_{23}U_{67}U_{89}\tilde{K}_9^L(\kappa)$. 

\

\noindent Step 4: We know from step 2 that $n_{\kappa_-}\neq 9$, and from step 3 that there are four $U$ gates to the left of $K_1^R$. So, $\kappa_-=4$.

\ 

\noindent Step 5: The first indices in the $U$ gates to the left of $K_1^R$ are $(7,5,4,3)$. Given that from step 2 we learned that there is  a $+\kappa$ at site $N$, we have that $\vec{n}$ is completely fixed by the number of $U$ gates to the left of $K_1^R$: $\vec{n}=(7,5,4,3)$.

\

\noindent Step 6: The $\vec{n}$ written in step 5 directly leads to the circuit in Figure \ref{fig:example2} if we use Theorem 1.
\

\paragraph{Numerics:} The discussion above is also useful from a numerical perspective. For example, consider an open quantum circuit in which the gates are known only numerically, the gate $U$ acts on each pair ${i,i+1}$ exactly once per period, $K^R(\kappa)$ appears once and only at site 1, and similarly for $\tilde{K}^{L}$, but  at site $N$. To check whether this circuit is integrable, it is sufficient to perform the following three steps:
\begin{itemize}
    \item check that $U$ is made of an $\check{R}$-matrix satisfying the Yang-Baxter equation. A discussion is presented in Appendix \ref{app:numericBYBE}.
    \item check that $K^R(\kappa)$ and $U$ together, satisfy the Boundary Yang-Baxter equation. A procedure for both difference and non-difference form is proposed in Appendix \ref{app:numericBYBE}.
    \item usually, $K^L$ is obtained from $K^R(\kappa)$ via an automorphism. Therefore, certain basic properties are preserved. So, once the previous point is concluded, it is enough to check that $\tilde{K}^{L}(\kappa)$ shares the same rank and number of non-zero elements as $K^R(\kappa)$. 
\end{itemize} 

\subsubsection{Equivalent diagrams}\label{sec:equivalent}

The quantum circuits introduced above can be characterised by their spectra and eigenvectors. The following theorem generalizes the result of \cite{bensa2021fastest}, where it was shown that, for open boundary conditions, all circuits in which each gate acts exactly once per period are isospectral. Here we extend this result to circuits containing the two different boundary gates. Although all these circuits turn out to have the same spectrum, we will distinguish equivalence classes according to the number of $-\kappa$ inhomogeneities. Indeed, circuits with the same value of $\kappa_-$ are related by similarity transformations involving only bulk gates, whereas relating circuits with different values of $\kappa_-$ necessarily require the boundary operators.

\vspace{0.4cm}

\textbf{Theorem 3:} All integrable quantum circuits constructed from the bulk gate $U$ and the boundary gates $K^R(u)$ and $K^L(u)$, satisfying the Yang--Baxter and reflection equations, are isospectral.

\begin{proof}
    From the proof presented in Appendix~\ref{App:swappinginhomogeneities}, it follows that circuits with the same number of $-\kappa$ inhomogeneities are related by similarity transformations and therefore have identical spectra\footnote{The same conclusion also follows from the invariance of the spectrum under cyclic permutation.}. It therefore remains to show that the spectrum is independent of the value of $\kappa_-$.

We illustrate the argument for $N=5$; the extension to arbitrary $N$ follows by exactly the same reasoning. Without loss of generality, we place all the $-\kappa$ inhomogeneities at the beginning of the chain, so that $\vec{n}=\{n_{\kappa_-},\dots,2,1\}$.

    For $N=5$, Theorem~1 (for $\kappa_-=0,1,2,3,4$) and Theorem~2 (for ${\kappa_-}=N=5$) give\footnote{The reader can easily reproduce this result with the Mathematica notebook \texttt{OpenQCforDiffGeom.nb}, \cite{zenodo}.}
    \begin{itemize}
        \item $\kappa_-=0$, $M=K_1^R(\kappa)U_{12}U_{23}U_{34}U_{45}\tilde{K}_5^L(\kappa)$,
        \item $\kappa_-=1$, $M=U_{12}K_1^R(\kappa)U_{23}U_{34}U_{45}\tilde{K}_5^L(\kappa)$,
        \item $\kappa_-=2$, $M=U_{23}U_{12}K_1^R(\kappa)U_{34}U_{45}\tilde{K}_5^L(\kappa)$,
        \item $\kappa_-=3$, $M=U_{34}U_{23}U_{12}K_1^R(\kappa)U_{45}\tilde{K}_5^L(\kappa)$,
        \item $\kappa_-=4$, $M=U_{45}U_{34}U_{23}U_{12}K_1^R(\kappa)\tilde{K}_5^L(\kappa)$ ,
        \item $\kappa_-=5$, $M=\tilde{K}_5^L(\kappa)U_{45}U_{34}U_{23}U_{12}K_1^R(\kappa)$.       
    \end{itemize}
We proceed to prove that all these operators are isospectral. To this end, we use that for all square matrices $A$ and $B$ of the same size, the products $AB$ and $BA$ have identical spectra.
        
    Now, in order to relate the circuits with $\kappa_-=0$ and $\kappa_-=1$, we identify $A=K_1^R(\kappa)$ and $B=U_{12}U_{23}U_{34}U_{45}\tilde{K}_5^L(\kappa)$. Then, it is obvious that the product $A B$ gives precisely the circuit with $\kappa_-=0$, while the product $BA$ gives the circuit with $\kappa_-=1$. In fact,
\begin{align}
    & BA=U_{12}U_{23}U_{34}U_{45}\tilde{K}_5^L(\kappa)K_1^R(\kappa)=U_{12}K_1^R(\kappa)U_{23}U_{34}U_{45}\tilde{K}_5^L(\kappa),
\end{align}
where in the last step we commuted $K_1^R(\kappa)$ past the remaining operators, which act trivially on the first site.

Similarly, to relate the circuits with $\kappa_-=1$ and $\kappa_-=2$, we choose: $A=U_{12}K_1^R(\kappa)$ and $B=U_{23}U_{34}U_{45}\tilde{K}_5^L(\kappa)$ and repeat the same argument.

Proceeding inductively, at each step the operator $A$ is enlarged to include $K_1^R(\kappa)$ together with all bulk gates to its left. Iterating this construction establishes the result for arbitrary chain length $N$.
\end{proof}

Consequently, for fixed $N$, all integrable quantum circuits with open boundary conditions possess the same spectrum. Their eigenvectors, however, are generally different and are related by similarity transformations.

For circuits with the same number of $\kappa_-$ inhomogeneities, the corresponding similarity transformation is given in Appendix~\ref{App:swappinginhomogeneities}, Eq.~\eqref{eq:theorem3}, and involves only bulk gates. By contrast, when comparing circuits with different values of $\kappa_-$, the proof of Theorem~3 shows that the similarity transformation necessarily also involves the boundary operators\footnote{If $AB \vec{v}=\lambda \vec{v}$ with $\lambda\neq0$, then $BA\,B\vec{v}=\lambda B \vec{v}$, showing that the corresponding eigenvectors are related by the action of $B$.}.

For this reason, we consider all quantum circuits with the same $N$ and $\kappa_-$ as belonging to the same equivalence class. For fixed $N$ and $\kappa_-$, there are

\begin{equation}
    \#_\text{equiv}=\binom{N}{\kappa_-}=\frac{N!}{(N-\kappa_-)!\kappa_-!}
\end{equation}
\noindent equivalent circuits. For the same reason, for a given $N$ there are only $N+1$ inequivalent quantum circuits, corresponding to the choice of $\kappa_-=0,1,2,\dots N$.

\begin{figure}[H]
    \centering
    \begin{subfigure}[t]{0.32\textwidth}
        \centering
        \includegraphics[scale=1.1]{N5vecn42.pdf}
        \caption{$\vec{n}=(4,2)$}
        \label{fig:N5vecn42}
    \end{subfigure}\hfill
    \begin{subfigure}[t]{0.32\textwidth}
        \centering
        \includegraphics[scale=1.1]{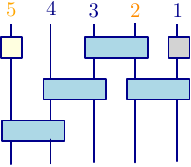}
        \caption{$\vec{n}=(5,2)$}
        \label{fig:N5vecn52}
    \end{subfigure}\hfill
    \begin{subfigure}[t]{0.32\textwidth}
        \centering
        \includegraphics[scale=1.1]{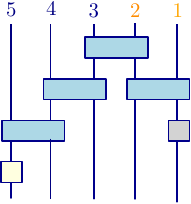}
        \caption{$\vec{n}=(2,1)$}
        \label{fig:N5vecn21}
    \end{subfigure}
    \caption{Quantum circuits for $N=5$ for three different choices of $\vec{n}$. The orange colour indicates the position of the $-\kappa$'s.}
    \label{fig:N=5quantumcircuits}
\end{figure}

As an example, we consider three circuits belonging to the same equivalence class. Using Eqs.~\eqref{eq:generalformulaforopen1} and \eqref{eq:generalformulaforopen2}, we evaluate $t(\kappa)$ for $N=5$ with $\kappa_-=2$ and three different choices of $\vec{n}$. The corresponding circuit diagrams are shown in Fig.~\ref{fig:N=5quantumcircuits}. \vspace{-0.4cm}

In this example and in Eqs.~\eqref{eq:generalformulaforopen1} and \eqref{eq:generalformulaforopen2}, we observe that circuits with the same $\kappa_-$ but different configurations $\vec{n}$ can differ significantly in structure, and in particular may have different depths, i.e., different numbers of sequential gate layers required to implement the circuit (respectively 2, 3, and 4 from left to right in Fig. \ref{fig:N=5quantumcircuits}). 

\subsection{Circuits with minimum depth $d$ and independent geometries}\label{sec:minimumdept}

In Sec. \ref{sec:generalconstruction}, we presented a procedure to construct circuits for all possible geometries. However, as discussed in Sec. \ref{sec:equivalent}, we grouped the circuits into equivalence classes depending on the number of inhomogeneities $-\kappa$. We identify a canonical representative for each equivalence class by minimizing the circuit depth. For example, in Fig. \ref{fig:N=5quantumcircuits}, the configuration $\vec{n}=\{4,2\}$ (Fig. \ref{fig:N5vecn42})  is the optimal representative, as it possesses the minimum depth compared to the other two configurations.

\subsubsection{Position of the inhomogeneities}

We identify the configurations $\vec{n}$ that minimize the circuit depth among circuits with the same number of $-\kappa$ inhomogeneities. We remark that, if one is interested only in the spectrum and not in the equivalence classes introduced above, the minimum circuit depth is always $d=2$. Nevertheless, the notion of minimum depth within a fixed equivalence class will prove particularly useful in Sec.~\ref{sec:beyondkappa}, where we introduce circuits containing inhomogeneities beyond $\kappa$ and $-\kappa$. We explicitly constructed quantum circuits for chain lengths $N=2, \dots, 10$ using Theorems 1 and 2, and subsequently verified these results for system sizes up to $N=51$. We find that for the case $\kappa_-=0$, the circuit is a staircase of depth $d=N+1$ ($N-1$ bulk gates and two boundary gates). For all other values of $\kappa_-$, our analysis points to the following two conjectures. 

\paragraph{Conjecture 1} For odd $N$ and $0<\kappa_-\le \frac{N-1}{2}$, the minimum depth is given by

\begin{equation}
    d=\frac{1}{2}(N+3)-\kappa_-.
\end{equation}
In particular, this means that for the staggered case, where $ \kappa_-= \frac{N-1}{2} $, we obtain $d=2$. From there, each time we reduce $\kappa_-$ by one, the depth increases by one, until $\kappa_-=1$, leading to $d= \frac{N+1}{2}$. By investigating a number of cases ($N=3,\dots, 9$ and all possible positions for the inhomogeneities), we find that by placing the $-\kappa$'s in the following positions:
\begin{equation}
    \vec{n}=(N-(d-1),\,N-(d+1),\cdots, d+2,\,d), \quad \text{for} \quad N\ge 2d-1,\label{eq:vecnmindoddN}
\end{equation}
we obtain a circuit with a minimum depth $d$. In other words, for a target minimum depth $d$ and $N$ sites, expression \eqref{eq:vecnmindoddN} gives the positions where the $-\kappa$'s must be placed to achieve it. We checked this conjecture for odd $N$ up to $N=51$. The reader can easily reproduce this result and the one in the next sections by using the Mathematica notebook \texttt{OpenQCforDiffGeom.nb} stored in \cite{zenodo}.

\paragraph{Conjecture 2} For even $N$ and $0<\kappa_-\le \frac{N}{2}$, the minimum depth is given by
\begin{equation}
    d=\frac{1}{2}(N+4)-\kappa_-.
\end{equation}
By studying $N=2,\dots,10$ we found that, for a target minimum depth $d$ the $-\kappa$ inhomogeneities have to be placed in positions 
\begin{equation}
    \vec{n}=(N-(d-2),\,N-d,\cdots, d+2,\,d), \quad \text{for} \quad N\ge 2d-2,
\end{equation}
leads to minimum depth $d$. We checked this conjecture for even $N$ up to $N=50$.

Now, we investigate how, starting from a given system size $N$ and value of $\kappa_-$, such that $0\le\kappa_-\le\frac{N-1}{2}$ ($N$ odd) and $0\le\kappa_-\le\frac{N}{2}$ ($N$ even), one can represent among the equivalent circuits, those characterized by minimal depth $d$.

\vspace{0.5cm}

We remark that the condition on $\kappa_-$ is not restrictive since there is a duality between $\kappa_-$ and $N-\kappa_-$. See the comment after Fig. \ref{fig:allgeometriesN=13}. 

\subsubsection{Positions of the gates}
\label{sec:positiongatesoddN}

Plugging the inhomogeneity configurations from Conjectures 1 and 2 into Theorems 1 and 2 leads to closed expressions for the quantum circuit with minimum depth.

For odd $N$, we place the configurations given in Conjecture 1 \eqref{eq:vecnmindoddN} into Theorem 1 \eqref{eq:generalformulaforopen1}. In particular, for the first product, this leads to  
\begin{align}
    \left(\mathop{\overset{\leftarrow}{\prod}}\limits_{1 \le r \le \kappa_-}U_{n_{r},n_{r}+1}\right)&=U_{n_{\kappa_-},n_{\kappa_-}+1},\cdots U_{n_2,n_2+1}U_{n_1,n_1+1}\\
    &=U_{N-d+1,N-d+2}U_{N-d-1,N-d}\cdots U_{d+2,d+3}U_{d,d+1}\\[0.2cm]
    &=U_{d,d+1}U_{d+2,d+3}\cdots U_{N-d-1,N-d}U_{N-d+1,N-d+2},
\end{align}
where in the last step we use that in Conjecture 1, $|n_i-n_j|\ge 2$ and therefore, \\$\left[U_{n_i,n_{i}+1},U_{n_j,n_{j}+1}\right]=0$, $\forall \, n_i\in \vec{n}$. With this in mind and repeating the computation for even $N$ and $d>2$, we find the general expression
\begin{equation}
    M=\left(\mathop{\overset{\rightarrow}{\prod}}\limits_{\frac{d+\alpha}{2} \le i \le \frac{n_{\kappa_-}+\alpha}{2}}U_{2i-\alpha,2i-\alpha+1}\right)K_1^{R}(\kappa)\left(\mathop{\overset{\rightarrow}{\prod}}\limits_{\substack{
1 \le j \le N-1 \\[0.07cm]
j \notin \vec{n}}}U_{j,j+1}\right)\tilde{K}_N^{L},
\end{equation}
\noindent where
\begin{equation}
    \alpha=\begin{cases}
        0, & \text{ for even } d\\
        1, & \text{ for odd } d
    \end{cases}.\label{eq:alpha}
\end{equation}

\begin{figure}[H]
\centering
    \includegraphics[scale=0.95]{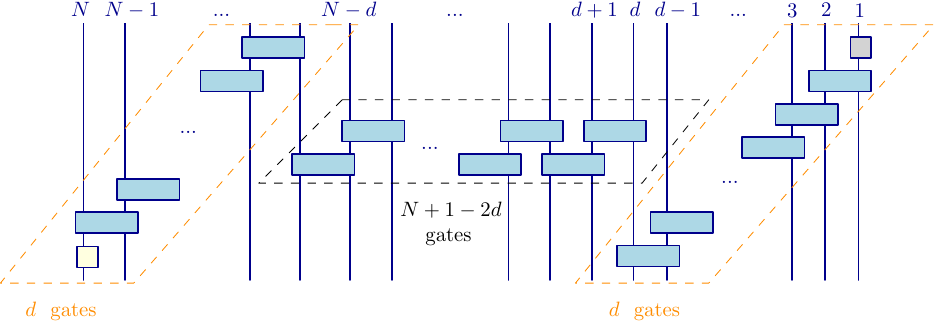}
    \caption{Representation of a quantum circuit with odd $N$ and minimal depth $d$}\label{fig:generaloddN}
\end{figure}

When visualizing these quantum circuits, for odd $N$, the operator $M$ decomposes into three regions: a "staircase" structure on the left containing $d$ gates (the boundary gate $\tilde{K}^L$ and $d-1$ bulk gates $U$), a mirrored staircase on the right containing $d$ gates (the boundary gate $K^R$ and $d-1$ bulk gates $U$), and a central brickwork circuit consisting of $N+1-2d$ bulk gates. This structure, valid for $0 \le \kappa_- \le \frac{N-1}{2}$, is illustrated in Fig. \ref{fig:generaloddN}.\

\paragraph{Example: $\mathbf{N=13}$}\

\begin{figure}[H]
\centering
\begin{subfigure}[t]{0.48\textwidth}
    \vspace{-3.4cm}
    \includegraphics[scale=0.8]{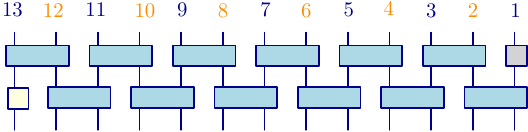}
    \label{fig:N13kappa_-=6}
    \caption{$\kappa_-=6,$ depth $=2$ and $\Vec{n}=(12,10,8,6,4,2)$}
    \vspace{1.2cm}
    \includegraphics[scale=0.8]{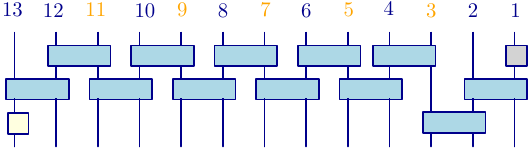}
    \label{fig:N13kappa_-=5}
    \caption{$\kappa_-=5,$ depth $=3$ and $\Vec{n}=(11,9,7,5,3)$}
    \vspace{1.2cm}
    \includegraphics[scale=0.8]{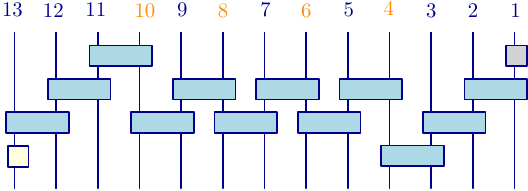}
    \label{fig:kappa_-=4}
    \caption{$\kappa_-=4,$ depth $=4$ and $\Vec{n}=(10,8,6,4)$}
    \vspace{1.2cm}
    \includegraphics[scale=0.8]{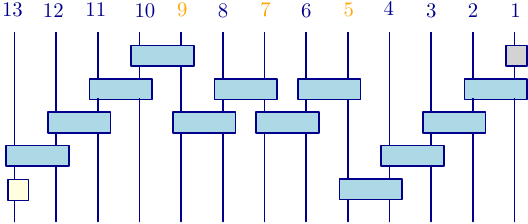}
    \label{fig:kappa_-=3}
    \caption{$\kappa_-=3,$ depth $=5$ and $\Vec{n}=(9,7,5)$}
\end{subfigure}\hfill
\begin{subfigure}[t]{0.48\textwidth}
    \includegraphics[scale=0.8]{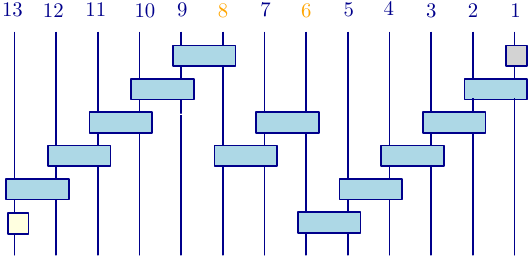}
    \label{fig:N13kappa_-=2}
    \caption{$\kappa_-=2,$ depth $=6$ and $\Vec{n}=(8,6)$}
    \medskip
    \includegraphics[scale=0.8]{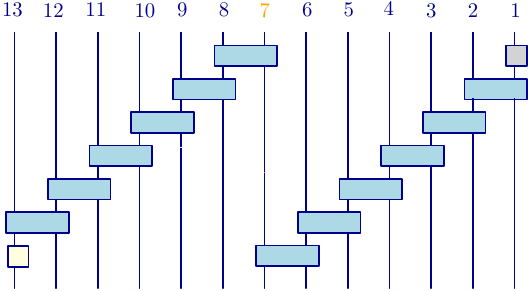}
    \label{fig:N13kappa_-=1}
    \caption{$\kappa_-=1,$ depth $=7$ and $\Vec{n}=(7)$}
    \medskip
    \includegraphics[scale=0.8]{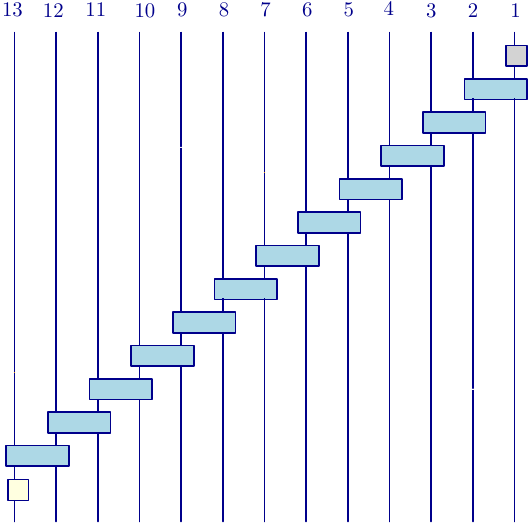}
    \label{fig:N13kappa_-=0}
    \caption{$\kappa_-=0,$ depth $=14$ and $\Vec{n}=(\emptyset)$}
\end{subfigure}
\caption{Quantum circuits with $N=13$ and all the possible choices of $\kappa_-$, $0\le \kappa_-\le \frac{N-1}{2}$ that correspond to a quantum circuit with minimum depth.}
\label{fig:allgeometriesN=13}
\end{figure}

\noindent In Fig. \ref{fig:allgeometriesN=13}, as an example, we explicitly show all the circuits with minimal depth for $N=13$ with $0\le\kappa_-\le\frac{N-1}{2}$.

\paragraph{What about $\frac{N-1}{2}<\kappa_-\le N$?}\

\

\begin{figure}[H]
\centering
    \includegraphics[scale=0.95]{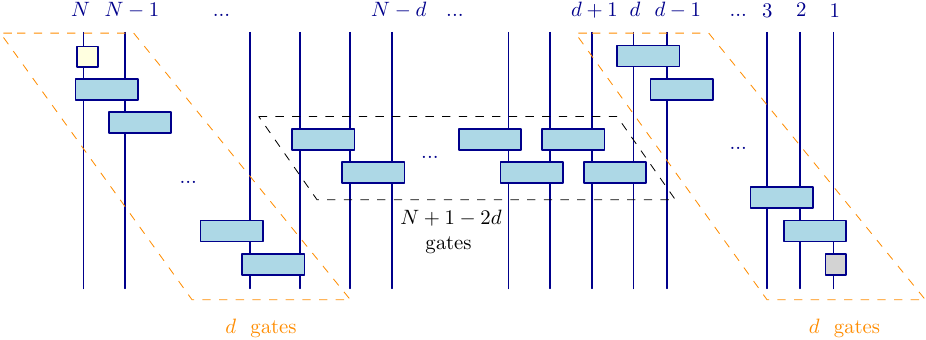}
    \caption{Circuit for odd $N$, depth $d$ and $\frac{N-1}{2}<\kappa_-\le N$.}
    \label{fig:generaloddNOpOrder}
\end{figure}

\noindent The cases analysed cover all  possible values of $\kappa_-$. In fact, there is a duality between $\kappa_-$ and $N-\kappa_-$. Therefore, the circuit for $\frac{N-1}{2}<\kappa_-\le N$ is obtained by reversing the order of the time layers of the one for $0\le \kappa_-\le \frac{N-1}{2}$, as represented in Fig. \ref{fig:generaloddNOpOrder}.

\begin{figure}[H]
\centering
    \includegraphics[scale=0.95]{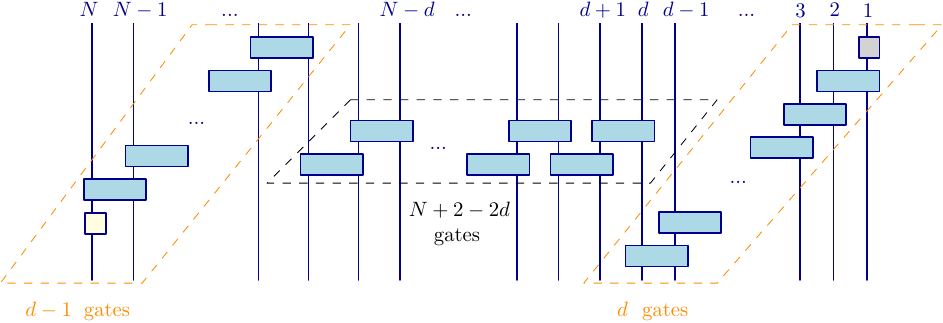}
    \caption{Circuit for even $N$ and depth $d>2$.}
    \label{fig:generalevenN}
\end{figure}

Similarly, for even $N$, $d>2$ we also obtain the same pattern, but now with $d-1$ gates on the left side. This is represented in Figure \ref{fig:generalevenN}.
As we discussed for odd $N$, there also exists a duality between $\kappa_-$ and $N-\kappa_-$ for even $N$. The corresponding quantum circuit can be represented as Fig. \ref{fig:generaloddNOpOrder} but with one less gate on the left and $N+2-2d$ gates in the brickwork part.

\begin{figure}[H]
\centering
    \includegraphics[scale=0.9]{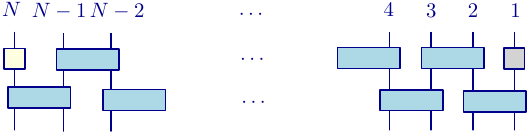}
    \caption{Circuit for even $N$ and depth $d=2$.}
    \label{fig:generalevenNd2}
\end{figure}

As observed above, Conjecture 1 for odd $N$ with $d\ge 2 $ and Conjecture 2 for even $N$ with $d>2$ lead to very similar quantum circuits. This is because both cases fall under Theorem 1. However, for even $N$ and $d=2$, Conjecture 2 puts a $-\kappa$ at position $N$, falling under Theorem 2. The corresponding quantum circuit is written as\footnote{Please notice that this is a choice, since we aim for a single Conjecture 2. We could have  chosen to have a single circuit instead (similar to Fig. \ref{fig:generalevenN}) representing the conjecture. The downside would have been to separate Conjecture 2 into two parts: $d=2$  and $d>2$.}
\begin{equation}
    M=\tilde{K}_N^{L}(\kappa)\left(\mathop{\overset{\rightarrow}{\prod}}\limits_{1 \le i \le \frac{N}{2}-1}U_{2i,2i+1}\right)K_1^{R}(\kappa)\left(\prod_{\substack{j=1\\j\notin \vec{n}}}^{N-1}U_{j,j+1}\right).\label{MNoddandd2}
\end{equation}
This is represented in Figure \ref{fig:generalevenNd2}.

\subsubsection{Discussion}

There is a clear intuition for why Conjectures 1 and 2 lead to the quantum circuits shown in Figures \ref{fig:generaloddN}--\ref{fig:generalevenNd2}. On the one hand, purely staggered chains naturally produce brickwork-type quantum circuits. On the other hand, taking all inhomogeneities to be equal gives rise to staircase-type circuits. A closer inspection of Conjectures 1 and 2 shows that the predicted arrangement of inhomogeneities yielding minimal depth consists precisely of three parts: an initial block of $d-1$ inhomogeneities equal to $\kappa$, followed by an alternating block of $(-\kappa,\kappa)$ values, and finally a second block containing only $\kappa$ values. This structure naturally gives rise to a staircase+brickwork+staircase circuit architecture, as illustrated in Figures \ref{fig:generaloddN}--\ref{fig:generalevenN}.

\subsection{Inhomogeneities beyond $(\kappa,-\kappa)$ }\label{sec:beyondkappa}

In the previous sections, we considered systems with inhomogeneities taking the values $\kappa$ or $-\kappa$. In this section, we investigate what happens when a third value, denoted by $\rho$, is introduced. In the periodic case, the introduction of a third value for the inhomogeneity was first presented in  \cite{Paletta:2025sap}.

For the discussion below, it is useful to introduce, in addition to gate $U=\check{R}(\kappa,-\kappa)$, two other gates
\begin{figure}[H]
    \centering
    \includegraphics[scale=0.9]{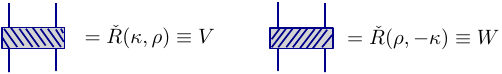}
    \caption{Quantum gates depending on $\kappa$ and the new inhomogeneity $\rho$.}
    \label{fig:newgates}
\end{figure}

\subsubsection{Circuits with minimum depth and different geometries}\label{sec:actualdepth}

As before, our goal is to minimise the depth. To this end, first observe that, if we set $M=t(\kappa)$, Lemmas 1 and 2 remain valid even after introducing a third type of inhomogeneity\footnote{The only restriction is that this third type of inhomogeneity should not be at position $N$.}, denoted by $\rho$.
 
When only two inhomogeneities, $\kappa$ and $-\kappa$, were present, the minimum achievable depth was $d=2$. This corresponds to the brickwork circuit obtained by staggering the inhomogeneities.

We determined that, in the presence of the additional inhomogeneities  $\rho$, the analog overall minimum depth is equal to $d=4$. Below, we present the configurations that lead to this minimum. To make the notation clearer, as before, we denote the number of  $-\kappa$ inhomogeneities as $\kappa_-$  and their location as $\vec{n}$,
\begin{equation}
    \Vec{n}=({n_{\kappa_-},\cdots, n_2, n_1}), \quad \text{with} \quad 1\le n_1 < n_2 < \cdots < n_{\kappa_-}\le N.\label{eq:vecnkappa}
\end{equation}
Additionally, we now define the number of $\rho$ inhomogeneities by $\rho_+$ and their position by $\vec{n}^{\rho}$ satisfying
\begin{equation}
    \Vec{n}^{\rho}=({n^{\rho}_{\rho_+},\cdots, n^{\rho}_2, n^{\rho}_1}), \quad \text{with} \quad 1\le n^{\rho}_1 < n^{\rho}_2 < \cdots < n^{\rho}_{\rho_+}\le N-1.\label{eq:vecnkappa2}
\end{equation}

All remaining sites have inhomogeneities $+\kappa$.\

\paragraph{With inhomogeneities of types $+\kappa$ and $\rho$}\

\

Let us assume that we have only inhomogeneities of types $+\kappa$ and $\rho$, but no $-\kappa$. Also, always consider a $+\kappa$ inhomogeneity at site $N$. As discussed before, contrary to the periodic case, for open spin chains, this case is very different from the one with only $\kappa$ and $-\kappa$. The first particularity is that, contrary to the case with $(\kappa,-\kappa)$, whose maximum depth is $d=N+1$, here $d=N+1$ is actually the minimum possible depth within this construction. 

For a system with odd $N$, $\rho_+$ inhomogeneities of type $\rho$ and $1\le \rho_+\le \frac{N-1}{2}$ the minimum depth is obtained by placing the  $\rho$  in positions\footnote{Please notice that $\vec{n}^\rho$ contains the positions of the $\rho$ inhomogeneities. Therefore, it does not make sense to have the same number appearing more than once. As soon as any of the smaller numbers becomes equal to $\frac{N+1}{2}+\rho_+-1$ we stop. For $\rho_+=1$ for example, both  $\frac{N+1}{2}+\rho_+-1$ and $\frac{N+1}{2}-\rho_++1$ become just $(N+1)/2$, so in that case we write $\vec{n}^{\rho}=\left(\frac{N+1}{2}\right)$.} 
\begin{equation}
    \vec{n}^\rho=\left(\frac{N+1}{2}+\rho_+-1,\cdots, \frac{N+1}{2}-\rho_++5,\frac{N+1}{2}-\rho_++3,\frac{N+1}{2}-\rho_++1\right)\label{eq:formindepthkr}.
\end{equation}
With this, we define $\rho_+$ different equivalence classes, one for each inequivalent geometry with minimum depth $d=N+1$. The corresponding quantum circuit $M\propto t(\kappa)$, is given by 
\begin{equation}
    M=\left(\mathop{\overset{\rightarrow}{\prod}}\limits_{1 \le r \le \rho_+}V_{n_{r}^{\rho},n_{r}^{\rho}+1}\right)K_1^R(\kappa)\left(\mathop{\overset{\rightarrow}{\prod}}\limits_{1 \le j \le N-1}Y_{j,j+1}\right)\tilde{K}_N^L(\kappa),
\end{equation}
with
\begin{equation}
    Y_{j,j+1}=\begin{cases}
        W_{j,j+1},& j\in \vec{n}^{\rho}\\
        U_{j,j+1},& j\notin \vec{n}^{\rho}
    \end{cases}.
\end{equation}

\begin{figure}[H]
    \centering

    \begin{subfigure}[t]{0.31\textwidth}
        \centering
        \includegraphics[width=\linewidth]{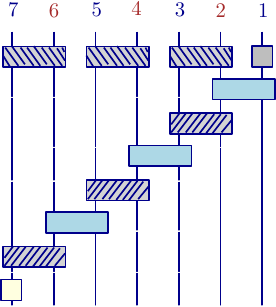}
        \caption{For $\vec{n}^{\rho}=(6,4,2)$.}
        \label{fig:N7kapparho3}
    \end{subfigure}
    \hfill
    \begin{subfigure}[t]{0.31\textwidth}
        \centering
        \includegraphics[width=\linewidth]{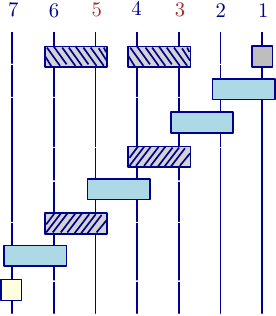}
        \caption{For $\vec{n}^{\rho}=(5,3)$.}
        \label{fig:N7kapparho2}
    \end{subfigure}
    \hfill
    \begin{subfigure}[t]{0.31\textwidth}
        \centering
        \includegraphics[width=\linewidth]{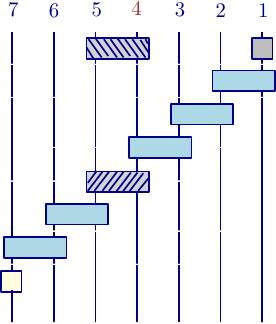}
        \caption{For $\vec{n}^{\rho}=(4)$.}
        \label{fig:N7kapparho1}
    \end{subfigure}

    \caption{Quantum circuits with minimum depth for $N=7$ and $\rho_+=3,2,1$. }
    \label{fig:N7kapparho}
\end{figure}

For $N=7$, they can be represented graphically as in Figure \ref{fig:N7kapparho}.

\begin{figure}[H]
    \centering
     \includegraphics[scale=0.7]{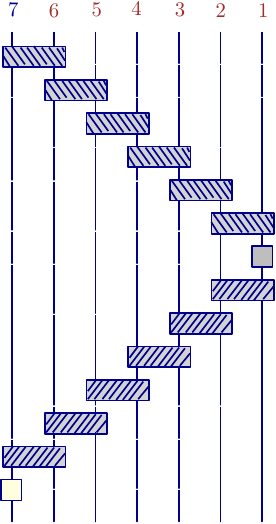}
     \caption{Quantum circuit with $N=7$ and $\vec{n}^{\rho}=(6,5,4,3,2,1)$.}
     \label{fig:rho+=N-1}
\end{figure}

Moreover, contrary to the $(\kappa,-\kappa)$, here there is no symmetry between $\rho_+$ and $N-\rho_+$. Additionally, we remark that deviating from the configurations in equation \eqref{eq:formindepthkr} can lead to an increase in depth. In particular, for $\vec{n}^{\rho}=(N-1,N-2,\cdots,3,2,1)$ the circuit has depth $d=2N$ (see for example, Figure \ref{fig:rho+=N-1}). Furthermore,  our Mathematica notebook \texttt{OpenQCforDiffGeom.nb} (available in \cite{zenodo}) is not limited to the cases with minimum depth. So, the reader can generate both the quantum circuits in Figure \ref{fig:N7kapparho} and in Figure \ref{fig:rho+=N-1}, and beyond.\vspace{-0.3cm}

We remark that in all cases we checked with odd $N$ and only $+\kappa$ and $\rho$ (but no $-\kappa$), the depth was always $d\ge N+1$. We checked this for all cases for $N=5,6,7,9$ and some randomly chosen cases for $N=11,13,15$.

Another difference is that, for this case with $+\kappa$ and $\rho$, there are three types of bulk gates ($V$, $W$ and $U$), and it is easy to see that Theorem 3 does not apply. Therefore, the spectrum will change for different values of $\rho_+$. But there is a subtlety associated to this. For a fixed $\rho_+$, we can replace some $+\kappa$'s by $-\kappa$'s because the spectrum is invariant\footnote{This statement can be proved analogously to Theorem 3, using the invariance of the spectrum under cyclic permutation of the matrix product.} under all possible choices of $\kappa_-$. Moreover, the eigenvectors corresponding to different values of $\kappa_-$ are related by a similarity transformation involving the bulk and boundary operators. An interesting consequence of this, is that we can use the introduction of $-\kappa$ inhomogeneities as a mechanism to reduce the minimum depth while preserving the spectrum. For example, in the next subsection, we will have circuits with minimum possible depth equal to four instead of $d=N+1$.

\paragraph{With inhomogeneities of types $\kappa,$ $-\kappa$ and $\rho$:}\

\

Minimising the depth in general is complicated in this case, but finding the analog of the brickwork case is straightforward. First, notice that for the brickwork we had two different inhomogeneities, and alternating between them led to the minimum possible depth in a circuit of that type, which is $d=2$. Additionally, circuits with an odd or even number of sites behaved slightly differently.

Now, we have three types of inhomogeneities so alternating between $\kappa$, $-\kappa$, and $\rho$ is a natural choice, and indeed the correct one for minimizing the depth. Moreover, circuits with $3m$ sites, $3m-1$ sites and $3m-2$ sites will behave slightly differently from each other.

With all this in mind, we make the following conjectures:

\paragraph{Conjecture 3a: } For a system with $N=3m$ sites, $m\in \mathbb{N}_{\ge 1}$, $\kappa_-=m$ and $1\le \rho_+\le m $, the configuration that minimises the depth is 
\begin{align}
    &\vec{n}=(N-1,N-4,\cdots, 5, 2),\nonumber\\
    &\vec{n}^{\rho}=(3\rho_+-2,3\rho_+-5,\cdots,4,1).
\end{align}
The cases described by this conjecture always lead to a depth $d=4$, and are the analog to the brickwork case. Notice, however, that for each $N$ we have $\rho_+$ different circuits of this type.

\begin{figure}[H]
    \centering

    \begin{subfigure}[t]{0.45\textwidth}
        \centering
        \includegraphics[width=\linewidth]{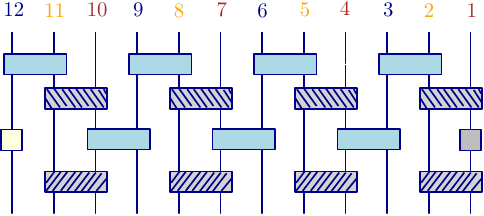}
        \caption{Quantum circuit for $\rho_+=4$, with $\rho$'s located at $\vec{n}^{\rho}=(10,7,4,1)$.}
        \label{fig:N12rho4}
    \end{subfigure}
    \hfill
    \begin{subfigure}[t]{0.45\textwidth}
        \centering
        \includegraphics[width=\linewidth]{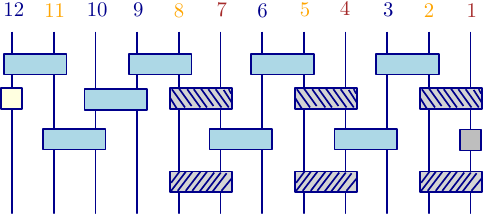}
        \caption{Quantum circuit for $\rho_+=3$, with $\rho$'s located at $\vec{n}^{\rho}=(7,4,1)$.}
        \label{fig:N12rho3}
    \end{subfigure}

    \medskip

    \begin{subfigure}[t]{0.45\textwidth}
        \centering
        \includegraphics[width=\linewidth]{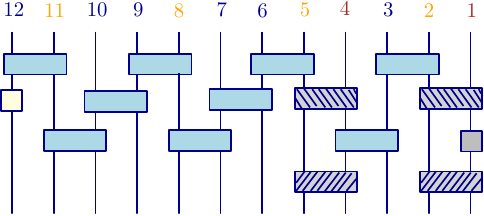}
        \caption{Quantum circuit for $\rho_+=2$, with $\rho$'s located at $\vec{n}^{\rho}=(4,1)$.}
        \label{fig:N12rho2}
    \end{subfigure}
    \hfill
    \begin{subfigure}[t]{0.45\textwidth}
        \centering
        \includegraphics[width=\linewidth]{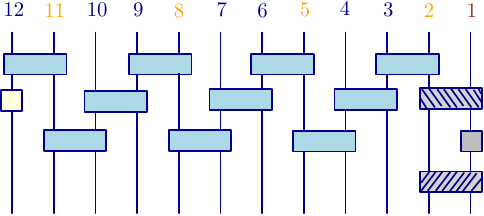}
        \caption{Quantum circuit for $\rho_+=1$, with $\rho$'s located at $\vec{n}^{\rho}=(1)$.}
        \label{fig:N12rho1}
    \end{subfigure}

    \caption{All quantum circuits for $N=12$ under Conjecture 3a. They correspond to $m=4$, with $\kappa_-=4$ and $1\le \rho_+\le 4 $. The $-\kappa$'s are located in the sites $\vec{n}=(11,8,5,2)$. The orange (brown) colour indicates the positions of inhomogeneities $-\kappa$ ($\rho$).}
    \label{fig:N12rho}
\end{figure}

Figure \ref{fig:N12rho} presents an example of all quantum circuits in this conjecture for $N=12$ (which corresponds to $m=4$).

\begin{figure}[H]
    \centering

    \begin{subfigure}[t]{0.45\textwidth}
        \centering
        \includegraphics[width=\linewidth]{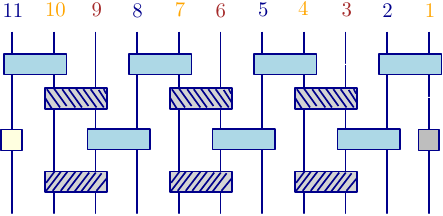}
        \caption{Quantum circuit for $\rho_+=3$, with $\rho$'s located at $\vec{n}^{\rho}=(9,6,3)$.}
        \label{fig:N11rho3}
    \end{subfigure}
    \hfill
    \begin{subfigure}[t]{0.45\textwidth}
        \centering
        \includegraphics[width=\linewidth]{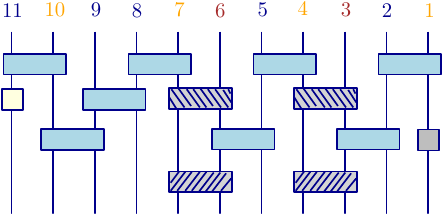}
        \caption{Quantum circuit for $\rho_+=2$, with $\rho$'s located at $\vec{n}^{\rho}=(6,3)$.}
        \label{fig:N11rho2}
    \end{subfigure}

    \medskip

    \begin{subfigure}[t]{0.45\textwidth}
        \centering
        \includegraphics[width=\linewidth]{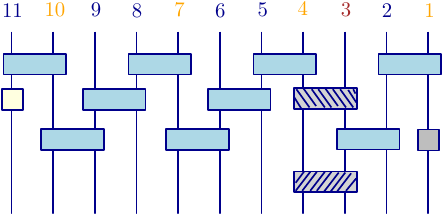}
        \caption{Quantum circuit for $\rho_+=1$ and $\vec{n}^{\rho}=(3)$.}
        \label{fig:N11rho1}
    \end{subfigure}

    \caption{Quantum circuits for $N=11$ ($m=4$), with $\kappa_-=4$ and $1\le \rho_+\le 3 $. The $-\kappa$'s are located in the following sites $\vec{n}=(10,7,4,1)$.}
    \label{fig:N11rho}
\end{figure}

\paragraph{Conjecture 3b: } For a system with $N=3m-1$ sites, $m\in \mathbb{N}_{\ge 2}$, $\kappa_-=m$ and $1\le \rho_+\le m -1$, the configuration that minimises the depth is 
\begin{align}
    &\vec{n}=(N-1,N-4,\cdots, 4, 1),\nonumber\\
    &\vec{n}^{\rho}=(3\rho_+,3\rho_+-3,\cdots,6,3).
\end{align}
Figure \ref{fig:N11rho} presents an example of all quantum circuits in this conjecture for $N=11$ (which corresponds to $m=4$).

\begin{figure}[H]
    \centering

    \begin{subfigure}[t]{0.45\textwidth}
        \centering
        \includegraphics[width=\linewidth]{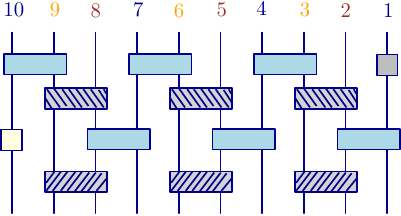}
        \caption{Quantum circuit for $\rho_+=3$, with $\rho$'s located at $\vec{n}^{\rho}=(8,5,2)$.}
        \label{fig:N10rho3}
    \end{subfigure}
    \hfill
    \begin{subfigure}[t]{0.45\textwidth}
        \centering
        \includegraphics[width=\linewidth]{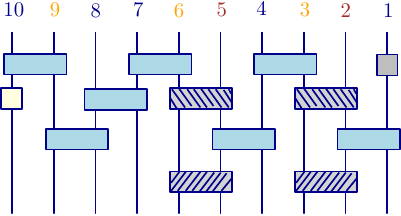}
        \caption{Quantum circuit for $\rho_+=2$, with $\rho$'s located at $\vec{n}^{\rho}=(5,2)$.}
        \label{fig:N10rho2}
    \end{subfigure}

    \medskip

    \begin{subfigure}[t]{0.45\textwidth}
        \centering
        \includegraphics[width=\linewidth]{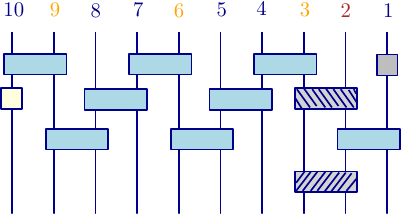}
        \caption{Quantum circuit for $\rho_+=1$, with $\rho$'s located at $\vec{n}^{\rho}=(2)$.}
        \label{fig:N10rho1}
    \end{subfigure}

    \caption{Quantum circuits for $N=10$ ($m=4$), with $\kappa_-=3$ and $1\le \rho_+\le 3 $. The $-\kappa$'s are located in the following sites $\vec{n}=(9,6,3)$.}
    \label{fig:N10rho}
\end{figure}

\paragraph{Conjecture 3c: } For a system with $N=3m-2$ sites, $m\in \mathbb{N}_{\ge 2}$, $\kappa_-=m-1$ and $1\le \rho_+\le m -1$, the configuration that minimises the depth is 
\begin{align}
    &\vec{n}=(N-1,N-4,\cdots, 6, 3),\nonumber\\
    &\vec{n}^{\rho}=(3\rho_+-1,3\rho_+-4,\cdots,5,2).
\end{align}
Figure \ref{fig:N10rho} presents an example of all quantum circuits in this conjecture for $N=10$ (which corresponds to $m=4$).

For all three conjectures, we can use Lemma 1, which, as mentioned above, is still valid here, to write $M=t(\kappa)$ as\vspace{-0.5cm}
\begin{align}
    M=\left(\mathop{\overset{\leftarrow}{\prod}}\limits_{1 \le r \le \kappa_-}U_{n_{r},n_{r}+1}\right)\left(\mathop{\overset{\leftarrow}{\prod}}\limits_{1 \le s \le \rho_+}V_{n^{\rho}_{s},\,n^{\rho}_{s}+1}\right)K_1^R(\kappa)\left(\mathop{\overset{\rightarrow}{\prod}}\limits_{\substack{
1 \le j \le N-1 \\[0.07cm]
j \notin \vec{n} \text{ and } j \notin \vec{n}^{\rho}}}U_{j,j+1}\right)\left(\mathop{\overset{\leftarrow}{\prod}}\limits_{1 \le s \le \rho_+}W_{n^{\rho}_{s},\,n^{\rho}_{s}+1}\right)\tilde{K}_N^{L}.\label{eq:Mfortwodiffthetas}
\end{align}\vspace{-0.5cm}

Notice that in Figures \ref{fig:N12rho4}-\ref{fig:N12rho1}, we have the inhomogeneities ordered as follows
\begin{align}
       & \{\theta_{12},\cdots,\theta_2,\theta_1\}_a=\{\kappa,-\kappa,\rho,\kappa,-\kappa,\rho,\kappa,-\kappa,\rho,\kappa,-\kappa,\rho\},\label{eq:N12a}\\
       &\{\theta_{12},\cdots,\theta_2,\theta_1\}_b=\{\kappa,-\kappa,\kappa,\kappa,-\kappa,\rho,\kappa,-\kappa,\rho,\kappa,-\kappa,\rho\},\label{eq:N12b}\\
       &\{\theta_{12},\cdots,\theta_2,\theta_1\}_c=\{\kappa,-\kappa,\kappa,\kappa,-\kappa,\kappa,\kappa,-\kappa,\rho,\kappa,-\kappa,\rho\},\label{eq:N12c}\\
       &\{\theta_{12},\cdots,\theta_2,\theta_1\}_d=\{\kappa,-\kappa,\kappa,\kappa,-\kappa,\kappa,\kappa,-\kappa,\kappa,\kappa,-\kappa,\rho\}.\label{eq:N12d}
\end{align}

In other words, from (a) to (b), we replaced one inhomogeneity $\rho$ by a $+\kappa$. As a result, one of the triplet becomes $\{\kappa,-\kappa, \kappa\}$, while the remaining triplet retain the form $\{\kappa,-\kappa, \rho\}$\footnote{We could equivalently have replaced the  $\rho$ at position (4), rather than the one at position (10), since permuting the positions of the inhomogeneities does not affect the spectrum. The resulting quantum circuits are equivalent, so we choose the arrangement that is more convenient for writing a general pattern}. From (b) to (c) we replace one more $\rho$ by $+\kappa$, and therefore, have one more triplet $\{\kappa,-\kappa, \kappa\}$ and one less $\{\kappa,-\kappa, \rho\}$. The same procedure is then repeated iteratively. 

A very similar effect is observed for the cases with $N=3m-1$ and $N=3m-2$. Comparing Figure \ref{fig:N12rho}, corresponding to $N=3m$, with Figures \ref{fig:N11rho} and \ref{fig:N10rho}, corresponding to  $N=3m-1$ and $N=3m-2$, respectively, reveals that the difference is confined to the right-hand side of the circuit. More precisely, for $N=3m$, the gate $K_1^R$ appears between $W_{12}$ and $V_{12}$;  for $N=3m-1$ it is applied before $U_{12}$; and for $N=3m-2$ it is applied after $U_{12}$ (recall that time flows from bottom to top). This distinction arises from the fact that the first site is always associated with an inhomogeneity $\rho$ for Conjecture 3a, with a $-\kappa$ for Conjecture 3b, and with a $+\kappa$ for Conjecture 3c.

Finally, in the case with only two types of inhomogeneities,  we defined the independent geometry by the value of $\kappa_-$. In particular, for each minimum depth, we associated one independent geometry. This is a consequence of Conjectures 1 and 2.

We now characterize an independent geometry by the number of inhomogeneities of type $-\kappa$ and the number of inhomogeneities of type $\rho$. In contrast to the previous setting, however, a fixed minimum possible depth may now correspond to several distinct independent geometries. Remarkably, for the minimum possible depth here, which is $d=4$, there are $m$ (Conjecture 3a) and $m-1$ (Conjectures 3b and 3c) different independent geometries. For $N=10,11,12$ they can be seen in Figures \ref{fig:N10rho}, \ref{fig:N11rho} and \ref{fig:N12rho}, respectively. Notice that we use again the fact that swapping the position of inhomogeneities does not affect the spectrum. 

\paragraph{What about $t(\rho)$?}

Since in this setting we have three types of inhomogeneity $\kappa$,$-\kappa$ and $\rho$, in addition to $M\propto t(\kappa)$, we can also define the operator $\widetilde{M}\propto t(\rho)$. A period can then be written as $M^{l_1}\tilde{M}^{l_2}$ for $l_1,l_2\in \mathbb{N}$. However, in the cases considered in Conjectures 3a–3c, $\widetilde{M}$ has depth significantly greater than four (while $M$ has always depth four). We therefore do not discuss $\widetilde{M}$ for these cases individually.

Instead, we write a general formula for $\tilde{M}$ (see Lemma 3 below) that includes all cases described in Conjectures 3a-3c.

\paragraph{Lemma 3:} For a circuit of length $N$ (given by $3m$, $3m-1$ and $3m-2$), assume that there is a $\rho$-inhomogeneity at position $N-(3j-1)$. Then for each value of $j\in \{1,...,m\}$, the quantum circuit $t(\rho)$ can be written in the following form
\begin{align}
    t(\rho)=&\left(\mathop{\overset{\leftarrow}{\prod}}\limits_{1 \le i \le N-3j}\check{R}_{i,i+1}(\rho,\theta_i)\right)K_1^R(\rho)\left(\mathop{\overset{\rightarrow}{\prod}}\limits_{1 \le i \le N-3j}\check{R}_{i,i+1}(\theta_i,-\rho)\right)\nonumber\\
    &\times \left(\mathop{\overset{\rightarrow}{\prod}}\limits_{N-3j+1 \le i \le N-1}\check{R}_{i,i+1}(\theta_{i+1},-\rho)\right)\tilde{K}_N^L(\rho)\left(\mathop{\overset{\leftarrow}{\prod}}\limits_{N-3j+1 \le i \le N-1}\check{R}_{i,i+1}(\rho,\theta_{i+1})\right).
    \label{eq:lemma3}
\end{align}
\vspace{0.2cm}

\begin{proof}
Let us start with the case $j=1$, and a transfer matrix with a $\rho$ at position $N-2$
\begin{align}
    t(u)=&\tr_a\left(K_a^L(u)R_{a,N}(u,\theta_N)R_{a,N-1}(u,\theta_{N-1})R_{a,N-2}(u,\rho)\left(\mathop{\overset{\leftarrow}{\prod}}\limits_{1 \le i \le N-3}R_{a,i}(u,\theta_i)\right)\right.\nonumber\\
    &\left. \times K_a^R(u)\left(\mathop{\overset{\rightarrow}{\prod}}\limits_{1 \le i \le N-3}R_{i,a}(\theta_i,-u)\right)R_{N-2,a}(\rho,-u)R_{N-1,a}(\theta_{N-1},-u)R_{N,a}(\theta_{N},-u)\right).
\end{align}

Now we compute $\tilde{M}=t(\rho)$ using the fact that $R(u,u)=g(u)P$ in the following way
\begin{align}
   & t(\rho)=g(\rho)\,\tr_a\left(K_a^L(\rho)R_{a,N}(\rho,\theta_N)R_{a,N-1}(\rho,\theta_{N-1})P_{a,N-2}\left(\mathop{\overset{\leftarrow}{\prod}}\limits_{1 \le i \le N-3}R_{a,i}(\rho,\theta_i)\right)\right.\nonumber\\
    &\hspace{1.3cm}\left. \times\, K_a^R(\rho)\left(\mathop{\overset{\rightarrow}{\prod}}\limits_{1 \le i \le N-3}R_{i,a}(\theta_i,-\rho)\right)R_{N-2,a}(\rho,-\rho)R_{N-1,a}(\theta_{N-1},-\rho)R_{N,a}(\theta_{N},-\rho)\right)\label{eq:lemma3a}\\[-0.3cm]
    &\hspace{0.8cm}= g(\rho)\,\tr_a\left(K_a^L(\rho)R_{a,N}(\rho,\theta_N)R_{a,N-1}(\rho,\theta_{N-1})\left(\mathop{\overset{\leftarrow}{\prod}}\limits_{1 \le i \le N-3}R_{N-2,i}(\rho,\theta_i)\right)K_{N-2}^R(\rho)\right.\nonumber\\
    &\hspace{0.8cm}\left. \times \,\left(\mathop{\overset{\rightarrow}{\prod}}\limits_{1 \le i \le N-3}R_{i,N-2}(\theta_i,-\rho)\right)P_{a,N-2}R_{N-2,a}(\rho,-\rho)R_{N-1,a}(\theta_{N-1},-\rho)R_{N,a}(\theta_{N},-\rho)\right)\label{eq:lemma3b}\\
    &\hspace{0.8cm}=g(\rho)\,\left(\mathop{\overset{\leftarrow}{\prod}}\limits_{1 \le i \le N-3}\check{R}_{i,i+1}(\rho,\theta_i)\right)K_{1}^R(\rho)\left(\mathop{\overset{\rightarrow}{\prod}}\limits_{1 \le i \le N-3}\check{R}_{i,i+1}(\theta_i,-\rho)\right)\tr_a\left(K_a^L(\rho)R_{a,N}(\rho,\theta_N)\right.\nonumber\\
    &\hspace{0.8cm}\times \left.R_{a,N-1}(\rho,\theta_{N-1})P_{a,N-2}R_{N-2,a}(\rho,-\rho)R_{N-1,a}(\theta_{N-1},-\rho)R_{N,a}(\theta_{N},-\rho)\right)\label{eq:lemma3c}\\
    &\hspace{0.8cm}=g(\rho)\,\left(\mathop{\overset{\leftarrow}{\prod}}\limits_{1 \le i \le N-3}\check{R}_{i,i+1}(\rho,\theta_i)\right)K_{1}^R(\rho)\left(\mathop{\overset{\rightarrow}{\prod}}\limits_{1 \le i \le N-3}\check{R}_{i,i+1}(\theta_i,-\rho)\right)R_{N-1,N-2}(\theta_{N-1},-\rho)\nonumber\\&\hspace{0.8cm}\times \tr_a\left(K_a^L(\rho)R_{a,N}(\rho,\theta_N)P_{a,N-2}R_{N-2,a}(\rho,-\rho)R_{N,a}(\theta_{N},-\rho)\right)R_{N-2,N-1}(\rho,\theta_{N-1})\label{eq:lemma3d}\\
    &\hspace{0.8cm}=g(\rho)\,\left(\mathop{\overset{\leftarrow}{\prod}}\limits_{1 \le i \le N-3}\check{R}_{i,i+1}(\rho,\theta_i)\right)K_{1}^R(\rho)\left(\mathop{\overset{\rightarrow}{\prod}}\limits_{1 \le i \le N-3}\check{R}_{i,i+1}(\theta_i,-\rho)\right)\nonumber\\&\hspace{0.8cm}\times R_{N-1,N-2}(\theta_{N-1},-\rho)R_{N,N-2}(\theta_{N},-\rho)\tilde{K}^L_{N-2}(\rho)R_{N-2,N}(\rho,\theta_N)R_{N-2,N-1}(\rho,\theta_{N-1})\label{eq:lemma3e}\\
    &\hspace{0.8cm}=g(\rho)\,\left(\mathop{\overset{\leftarrow}{\prod}}\limits_{1 \le i \le N-3}\check{R}_{i,i+1}(\rho,\theta_i)\right)K_{1}^R(\rho)\left(\mathop{\overset{\rightarrow}{\prod}}\limits_{1 \le i \le N-3}\check{R}_{i,i+1}(\theta_i,-\rho)\right)\nonumber\\
    &\hspace{0.8cm}\times \left(\mathop{\overset{\rightarrow}{\prod}}\limits_{N-2 \le i \le N-1}\check{R}_{i,i+1}(\theta_{N-1},-\rho)\right)\tilde{K}^L_{N}(\rho)\left(\mathop{\overset{\leftarrow}{\prod}}\limits_{N-2 \le i \le N-1}\check{R}_{i,i+1}(\rho,\theta_{N-1})\right).\label{eq:lemma3f}
\end{align}
From \eqref{eq:lemma3a} to \eqref{eq:lemma3b}, we move $P_{N-2,a}$ from its original position to the position immediately preceding $R_{N-2,a}(\rho,-\rho)$. This is achieved by swapping sites $a\leftrightarrow N-2$ in each operator that it passes through. In the next step, we observe that the entire block consisting of the two products with $K_{N-2}^R(\rho)$ inserted between them can be moved outside the trace to the left. Indeed, this block is independent of site $a$ and commutes with every operator it passes through.  In the same step, we use the relation $R=P\check{R}$ to rewrite the resulting factor outside the trace. 

From \eqref{eq:lemma3c} to \eqref{eq:lemma3d}, we proceed in three steps. First, we use $R_{a,N-1}P_{N-2,a}=P_{N-2,a}R_{N-2,N-1}$. Next, we apply the Yang–Baxter equation to reorder the relevant operators. Finally, we move outside the trace the operators that do not depend on $a$. They go to the left or to the right of the trace depending on which operators they commute with. 

To pass from \eqref{eq:lemma3d} to \eqref{eq:lemma3e}, we repeat the same procedure for the remaining operators.  In the final step, we again use $R=P\check{R}$ to rewrite the result in the desired form. The resulting expression \eqref{eq:lemma3f} is exactly \eqref{eq:lemma3} with $j=1$.

The computation for arbitrary $j$ follows immediately repeating steps \eqref{eq:lemma3a}-\eqref{eq:lemma3b}, and iterating the same sequence of steps \eqref{eq:lemma3c}-\eqref{eq:lemma3f} over the additional factors.
\end{proof}

The interested reader can easily generalise the computation in our Mathematica notebook \texttt{OpenQCforDiffGeom.nb}, to include this case. With that, one can write the analog of all the examples in Figures \ref{fig:N12rho}, \ref{fig:N11rho} and \ref{fig:N10rho}.

\subsubsection{Minimum ``effective'' depth}\label{sec:effectivedepth}

Achieving a complete classification for all geometries with minimum depth using three types of inhomogeneities is more challenging.
In this section, we propose an alternative, interesting choice that minimizes what we call an \textit{effective depth} (which we define below).

We recall that swapping inhomogeneities does not affect the spectrum. Therefore, for each choice of $\rho_+$ (number of $\rho$'s) and $\kappa_-$ (number of $-\kappa$'s), we are free to choose their positions. A convenient choice to achieve this \textit{effective} minimum depth, for instance, is to place all $\rho$'s at the beginning of the chain, while the remainder of the chain follows the configurations of Conjectures 1 or 2, shifted by $\rho_+$.

\begin{figure}[H]
    \centering
    \includegraphics[scale=0.9]{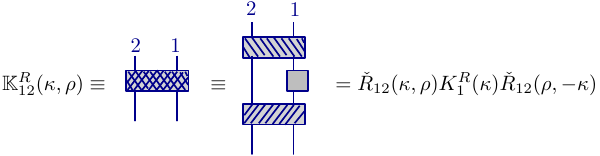}
    \caption{Effective boundary gates for $\theta_1=\rho$.}
    \label{fig:effKR12}
\end{figure}

\begin{figure}[H]
    \centering
    \includegraphics[scale=0.9]{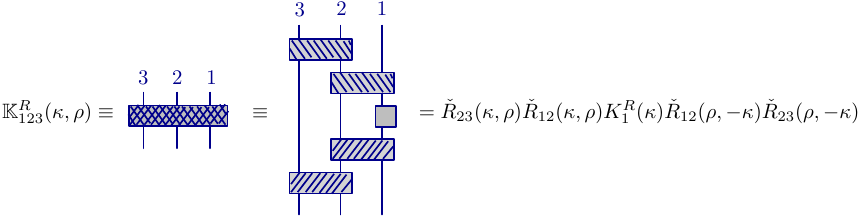}
    \caption{Effective boundary gates for $\theta_1=\theta_2=\rho$.}
    \label{fig:effKR123}
\end{figure}

\begin{figure}[H]
    \centering
    \includegraphics[scale=0.9]{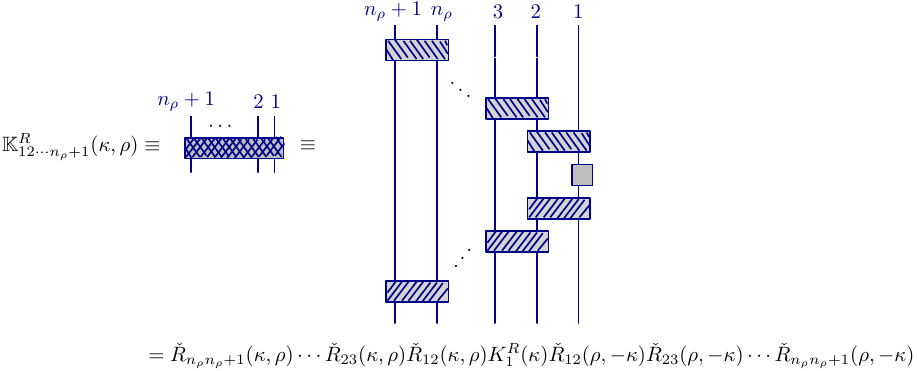}
    \caption{Effective boundary gates for $\theta_1=\theta_2=\cdots=\theta_n=\rho$.}
    \label{fig:effKR123n}
\end{figure}

To understand what we mean by \textit{effective} minimum depth, we need first to introduce the concept of \textit{effective} boundary gates\footnote{Placing all inhomogeneities of $\rho$ type in the end of the chain leads to the definition of a similar effective gate, but now for the left boundary.}.  For $n_\rho=1$ this is defined in Fig. \ref{fig:effKR12}
\noindent for $n_\rho=2$ in Fig. \ref{fig:effKR123},

etc, until $n_\rho=n$ in Fig. \ref{fig:effKR123n}

In other words, we view the $\rho$ inhomogeneities as effectively forming a larger boundary, extending over multiple sites.

A quantum circuit with effective minimum depth is one where the use of effective boundary gates, instead of fundamental gates, reduces the number of time steps required to implement $M$ to the minimum.

\begin{figure}[H]
    \centering
    \includegraphics[scale=0.9]{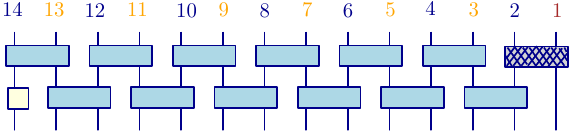}
    \caption{Quantum circuit with $N=14$ with $\theta_1=\rho$ and remaining $\theta_{\text{even}}=\kappa$ and $\theta_{\text{odd}}=-\kappa$. This is computed using $M=t(\kappa)$. If we use $t(\rho)$ instead, a completely different circuit is obtained.}
    \label{fig:N14rho}
\end{figure}

For instance, consider $N=14$ with $\theta_1=\rho$, and the remaining 13 sites alternating between $\kappa$ and $-\kappa$ as in Conjecture 1 but shifted by one site due to the initial $\rho$. This results in the quantum circuit shown in Fig. \ref{fig:N14rho}.

Thus, if we start from a brickwork-type circuit and add a new site at position one with inhomogeneity $\rho$, the resulting circuit has an effective depth of $d_e=2$. In other words, by employing the effective boundary gate defined in Fig. \ref{fig:effKR12}, the circuit retains a brickwork structure but with an enlarged boundary gate.

In general, using the definition in Fig. \ref{fig:effKR123n}, we consider a system with $n_\rho$ inhomogeneities $\theta_i=\rho$ at sites $i=1,2,\ldots,n_\rho$, while the remaining sub-chain follows Conjecture 1 or 2 shifted by $n_\rho$. Introducing the shifted position $\delta \equiv d_e + n_\rho$, the resulting general circuit configuration can be expressed as shown in Fig. \ref{fig:newgates}.
\begin{figure}[H]
    \centering
    \includegraphics[scale=0.9]{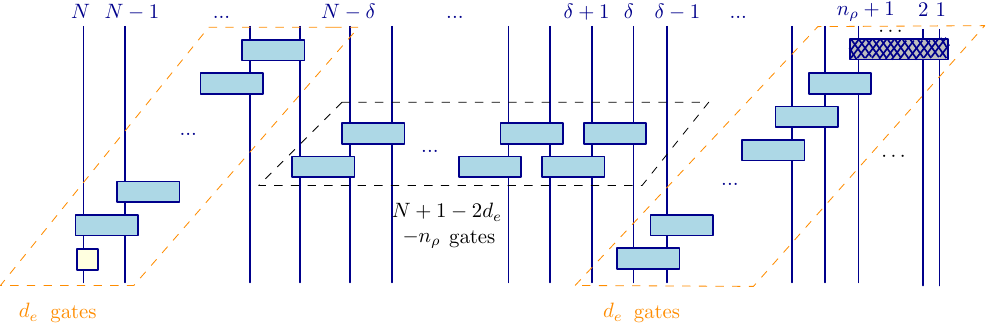}
    \caption{General circuit for $N$ sites and odd $N-\rho_+$. For even $N-\rho_+$, the only difference is that there is one less $U$ gate on the \textit{left} staircase part.}
    \label{fig:newgates}
\end{figure}

Placing all the $\rho$’s at the end of the chain produces the same effect, but now on the left boundary. Alternatively, one can distribute them by placing $n_\rho$ of the $\rho$’s at the beginning and the remaining $\tilde{n}_\rho$ at the end. This results in a left effective boundary gate of length $\tilde{n}_\rho + 1$ and a right effective boundary gate of length $n_\rho + 1$. Both choices minimise the effective depth, provided that the rest of the chain is described by Conjectures 1 and 2 with the appropriate shifts by $\tilde{n}_\rho$ and $n_\rho$. In order to apply the construction above, one needs $N-\rho_+\ge 2$.

Note, however, that while these choices minimize the effective depth, they do not necessarily minimize the actual depth. For instance, if $\rho_+=\kappa_{-}$, a better choice is to have inhomogeneities alternating $\{\kappa,-\kappa,\rho\}$ in the following way
\begin{equation}
    \{\theta_{N},\cdots,\theta_2,\theta_1\}=\{\kappa,-\kappa,\rho,\kappa,-\kappa,\rho,\kappa,-\kappa,\rho,\cdots\},\label{eq:alternativeconfig}
\end{equation}
like described in section \ref{sec:actualdepth}.

\subsubsection{More types of inhomogeneities?}

First, let us denote the number of different types of inhomogeneity by $n_{\theta}$. For simplicity, in the discussion below, we assume that the dynamical evolution operator is given by only $M=t(\kappa)$. 

When describing the case with $n_{\theta}=2$ (namely $\kappa$ and $-\kappa$), the minimum possible depth is $d=2 $. It is obtained by staggering the inhomogeneities. A natural separation between odd ($N=2m-1$) and even ($N=2m$) numbers of sites occurs in that case. For each $N$, there is only one configuration (and its complement) that generates $d=2$.

For $n_{\theta}=3$, we have seen that there is a natural choice obtained by alternating between $\kappa$, $-\kappa$ and $\rho$.
This generates circuits with depth $d=4$, and naturally splits into three cases: $N=3m, 3m-1, 3m-2$. Here, however, for every $N$ there are $m$ independent circuits for $N=3m$ (and $m-1$ for $N=3m-1,3m-2$) that can be generated with depth $d=4$.  See an example for $N=3m $ in figure \ref{fig:N12rho}, and for   $N=3m-1 $ and $N=3m-2 $ in Figures \ref{fig:N11rho} and \ref{fig:N10rho}, respectively.

The strategy is as follows: start with the fully three-fold alternating case (see equation \eqref{eq:N12a}) and  obtain the next one by replacing a $\rho$ by a $+\kappa$. To go to the next, replace one more $\rho$ by a $+\kappa$, and so on. 

Beyond $n_{\theta}=3$, we do not present the computations explicitly here, but from studying $n_{\theta}\le 8$ we conjecture the following:

\begin{itemize}
    \item For a system with $n_{\theta}$ different types of inhomogeneity, the minimum possible depth is $d=2n_{\theta}-2$.
    \item It splits naturally into $n_{\theta}$ parts, given by $N=n_{\theta}m, n_{\theta}m-1,\cdots, n_{\theta}m-(n_{\theta}-1)$;
    \item More and more independent circuits with a specific minimum possible depth can be obtained as we increase $n_{\theta}$. These are obtained by replacing each of the new types of inhomogeneities by $+\kappa$, one by one, by following an analogous strategy to Conjectures 3a-3b (see discussion immediately after Figure \ref{fig:N12rho} for example).
\end{itemize}

Moreover, an example for $n_\theta=4$ can be found in \texttt{OpenQCforDiffGeom.nb} stored in \cite{zenodo}.

\section{Boundary gates for a quantum circuit made of qubits }\label{Kmatrices}

Up to this point, all our constructions apply to any $R$-matrix and $K$-matrices solving the YBE (Eq.~\eqref{eq:ybe}) and BYBE (Eq.~\eqref{eq:bybe}), and are therefore valid for local Hilbert spaces of any finite dimension. In this section, we focus on the specific case of local Hilbert spaces given by $\mathbb{C}^2$. In other words, we restrict to regular $R$-matrices acting on $\mathrm{End}(\mathbb{C}^2 \otimes \mathbb{C}^2)$ and $K$-matrices acting on $\mathrm{End}(\mathbb{C}^2)$.  We note that these new boundary matrices are of independent interest and can also be used in other contexts beyond quantum circuits. 

The task of classifying all the $K$-matrices corresponding to a specific $R$-matrix can be considered an independent problem. Several groups have obtained different solutions for the boundary Yang-Baxter equation over the years. We refer to \cite{Sklyanin:1988yz} for the seminal work in this direction, but also to \cite{Cherednik:1984vvp,deVega:1993xi,Batchelor:1996np,Lima-Santos:2002mku,Lima-Santos:2003vnq,Malara:2004bi,Ghoshal:1993tm,Nepomechie:2018wzp} for applications to generalised Toda systems,  to \cite{Ahn:2010xa,deLeeuw:2012fy,Bielli:2024xuv,Bielli:2025abu,Gil:2026ost} for applications in AdS/CFT,  \cite{bajnok2006equivalences} to models with defects, to \cite{Crampe:2010cs,Frassek:2019isa} to stochastic models and to \cite{vanicat2018integrable,Frahm:2022gtk,Paletta:2024uzj} to open quantum systems. Additionally, the work in \cite{Gombor:2025qvk} finds integrable reflection matrices using an alternative approach.

Over the years, different groups have been involved in the classification of solutions to the Yang--Baxter equation. In the following, we build on the classification of regular $4\times 4$ $R$-matrices of up to 8-vertex type given in \cite{deLeeuw:2020ahe} and \cite{classificationybandboost}  for models of difference and non-difference form, and determine the corresponding solutions of the BYBE \eqref{eq:bybe} for the right $K$-matrices. We emphasise that we do not restrict to the case of regular $K$-matrices; hence, our solutions can also produce circuits with non-local conserved charges. The $R$-matrices we considered are of 6- and 8- vertex type. This assumption is more general than it may initially appear: in \cite{classificationybandboost}, we proved that every Hermitian $4\times 4$ integrable Hamiltonian can be mapped, through some of the integrability-preserving transformations described in Appendix \ref{App:transformations}, to a matrix of at most eight-vertex type, without requiring the transformed matrix to be Hermitian. We emphasize that for the integrable quantum circuit construction, we do not need to restrict to unitary solutions. We construct parametrized families of K-matrices; unitarity can be recovered for suitable choices of parameters, although we do not focus on this constraint here.

For each of the right $K$-matrices, we also computed the corresponding left $K$-matrix, guaranteeing that their transfer matrices commute for different values of the spectral parameter.

\subsection{Computing right reflection matrices $K^R(u)$: summary of the methods}
\label{methodsused}
We aim to solve the boundary Yang--Baxter equation, Eq.~\eqref{eq:bybe}, which is quadratic in the unknown matrix $K^R(u)$. We therefore substitute the most general form for $K^R(u)\in \text{End}(\mathbb{C}^2)$
\begin{equation}
    K^R(u)=\begin{pmatrix}
        k^R_{1,1}(u) & k^R_{1,2}(u)\\
        k^R_{2,1}(u) & k^R_{2,2}(u)
    \end{pmatrix}\label{eq:ansatzK}
\end{equation}
into the BYBE, and then differentiate with respect to one of the spectral parameters
\begin{equation}
    \frac{\partial}{\partial u}\left( R_{12}(u,v)K_1^R(u)R_{21}(v,-u)K_2^R(v)-K_2^R(v)R_{12}(u,-v)K_1^R(u)R_{21}(-v,-u)\right)\bigg|_{u=u_0}=0.
    \label{eq:d(bybe)/du}
\end{equation}

This produces a coupled system of differential equations, which we solve using two different well-known methods summarised below.

\subsubsection{Method 1}
\label{abelsmethod}
Consider the choice $u_0=v$. This reduces the problem to a system of coupled ordinary differential equations for the functions $k^R_{i,j}(v)$. When the model is sufficiently simple, this system can be solved directly. In particular, this approach is often effective for models of difference form. For more complicated cases, however, a direct treatment may become impractical. In such situations, we have used Abel's method, see \cite{Vieira:2017vnw} for a review. The method consists in treating $k^{R,\prime}_{i,j}(v)$ and $k^R_{i,j}(v)$ as independent variables, rendering the system linear in the set of unknown variables. One first solves a subset of the equations for the derivatives $k^{R,\prime}_{i,j}(v)$, substitutes the result into the remaining equations, and then solves for the functions $k^R_{i,j}(v)$. In most cases, this procedure leads to several distinct solutions.

After obtaining the solutions, one must substitute each resulting $K^R(v)$ back into the BYBE to verify consistency, since the temporary treatment of functions and derivatives as independent variables may introduce incompatible solutions. An additional complication arises from the structure of the equations treated, which admits multiple solutions. This requires analysing some cases separately, for instance, by imposing that specific matrix elements $k^R_{i,j}(v)$ vanish or by constraining the form of the $R$-matrix.

\subsubsection{Method 2}
Consider instead $u_0=0$ and impose the following boundary conditions on the $K$-matrix
\begin{equation}
    K^R(0)=\begin{pmatrix}
        \alpha_{1,1} & \alpha_{1,2}\\
        \alpha_{2,1} & \alpha_{2,2}
    \end{pmatrix},\quad 
    K^{R\,\prime}(0)=\begin{pmatrix}
        \beta_{1,1} & \beta_{1,2}\\
        \beta_{2,1} & \beta_{2,2}
    \end{pmatrix},
\end{equation}
with $\alpha_{i,j}$ and $\beta_{i,j}$ unspecified constant parameters\footnote{We remark that this choice also includes the case of regular $K$-matrices ($\alpha_{1,1}=\alpha_{2,2}=1$ and $\alpha_{1,2}=\alpha_{2,1}=0$).}.

One then solves the system of equations generated by \eqref{eq:d(bybe)/du} for $k^{R}_{i,j}(v)$, $\alpha_{i,j}$, and $\beta_{i,j}$. This procedure requires some care since, for certain solutions, some of the $\alpha$'s or $\beta$'s may be infinite.

\subsubsection{Comparison}

\begin{tabular}{p{2.5cm} p{6cm} p{6cm}}
\hline
 & \hspace{1.3cm}Advantages & \hspace{1.3cm}Disadvantages \\
\hline
\vspace{1.3\baselineskip}Method 1 &
\begin{itemize}
  \item Few unknowns (only four functions)
  \item Not many branches
\end{itemize}\vspace{0.01cm}
&
\begin{itemize}
  \item Coupled differential equations
\end{itemize}
\\
\hline
\vspace{1.0\baselineskip}Method 2 &
\begin{itemize}
  \item Algebraic equations
\end{itemize}
&
\begin{itemize}
  \item Many unknowns (8 constants and 4 functions)
  \item Many branches
\end{itemize}\vspace{0.01cm}
\\
\hline
\end{tabular}

\vspace{0.8cm}

For certain cases, after starting with Method 1 and solving the simplest differential equations in the system, one encounters one or two difficult differential equations. At this stage, it is sometimes useful to substitute the partial results into Eq. \eqref{eq:ansatzK} and then apply Method 2 to the partially solved ansatz to complete the solution. The reverse is also possible. This hybrid approach was useful whenever one of the methods alone led to a difficult system of equations. In particular, it was especially efficient for the 8-vertex-B, where the presence of elliptic Jacobi functions initially led to complicated ODEs.

\subsection{Known $K$-matrices: $K^R(u)$ for difference form $R$-matrix}
\label{knownk}
The general $K$-matrices for the XXX, XXZ and XYZ models were classified in \cite{deVega:1993xi,Inami:1994wn} (with some solutions found earlier in \cite{Sklyanin:1988yz,Cherednik:1984vvp,Destri:1991zm}); for completeness, they are explicitly presented below. Since these solutions are known, we do not detail the methods used to derive them, but one can easily apply the methods we provided in Sec. \ref{methodsused} for $R$-matrices of difference form. 

\paragraph{XXX}
For the XXX model, the $R$-matrix is 
\begin{equation}
    R_{XXX}(u)=\frac{1}{u+1}\begin{pmatrix}
        u+1 & 0 & 0 & 0\\
        0 & u & 1 & 0\\
        0 & 1 & u & 0\\
        0 & 0 & 0 & u+1
    \end{pmatrix},\label{eq:RXXX}
\end{equation}
and the most general $K$-matrix is
\begin{equation}
    K^{(XXX)}(u)=\begin{pmatrix}
        b_1(b_2+u)  & b_3 \,u\\
        b_4 \,u & b_1(b_2-u)) 
    \end{pmatrix},
\end{equation}
with $b_i$ constant parameters.

\paragraph{XXZ}

For the XXZ model, the $R$-matrix is 
\begin{equation}
    R_{XXZ}(u)=\frac{1}{\sin(u+\eta)}\begin{pmatrix}
        \sin(u+\eta) & 0 & 0 & 0\\
        0 & \sin u & \sin\eta & 0\\
        0 & \sin\eta & \sin u & 0\\
        0 & 0 & 0 & \sin(u+\eta)
    \end{pmatrix},\label{eq:RXXZ}
\end{equation}
where $\eta$ is a constant parameter.

The corresponding $K$-matrix is given by
\begin{equation}
    K^{(XXZ)}(u)=\begin{pmatrix}
        b_1\,\sin(b_2+u)  & b_3 \sin 2u\\
        b_4 \sin 2u & b_1\,\sin(b_2-u) 
    \end{pmatrix},
\end{equation}
where $b_i$ are constant parameters.
\paragraph{XYZ}
For the XYZ model, the $R$-matrix is
\begin{equation}
    R_{XYZ}(u)=\frac{1}{ \text{sn}(u+\eta) }\begin{pmatrix}
        \text{sn}(u+\eta) & 0 & 0 & k\,\text{sn}(u)\text{sn}(\eta)\text{sn}(u+\eta) \\
        0 & \text{sn}(u) & \text{sn}(\eta) & 0\\
        0 & \text{sn}(\eta) & \text{sn}(u) & 0\\
        k\,\text{sn}(u)\text{sn}(\eta)\text{sn}(u+\eta) & 0 & 0 & \text{sn}(u+\eta)
    \end{pmatrix}\label{eq:RXYZ},
\end{equation}
where $\text{sn}(u)=\text{JacobiSN}(u,k^2)$, $\eta$ and $k$ are constant parameters.

The corresponding $K$-matrices are given by
\begin{equation}
    K^{(XYZ,a)}(u)=\begin{pmatrix}
        1+\epsilon\, k\, \text{sn}^2u  & \epsilon\,b_1\,\text{sn}\,u\\
        b_1\,\text{sn}\,u & 1+\epsilon\, k\, \text{sn}^2 u
    \end{pmatrix},
\end{equation}
and
\begin{equation}
    \quad  K^{(XYZ,b)}(u)=\begin{pmatrix}
        1+\epsilon\, k\, \text{sn}^2u  & \epsilon \,b_1\, \text{cn} \,u\,\text{dn} u\\
       b_1\, \text{cn}\, u\,\text{dn}\, u  & -(1+\epsilon\, k\, \text{sn}^2u) 
    \end{pmatrix},
\end{equation}
with $\epsilon^2=1$ and $b_1$ an arbitrary constant.

\vspace{0.6cm}

For these three models, the corresponding solutions for the left boundary $K^L(u)$ are discussed in \cite{deVega:1993xi}. The spectra of these open quantum spin chains have been investigated in numerous works using various methods, including the Bethe ansatz, the off-diagonal Bethe ansatz, and the separation of variables, see for example  \cite{Pasquier:1989kd,Fan:1996jq,Nepomechie:2003vv,Frahm:2008qc,Faldella:2013qha}.

\subsection{New $K$-matrices: $K^R(u)$ for non-difference form $R$-matrix}
\label{newreflmat}
We proceed to classify the reflection matrices $K^R$ corresponding to the 6vB and 8vB $R$-matrices of \cite{deLeeuw:2020ahe, classificationybandboost}. We remark that this case also includes all possible 4x4 Hermitian Hamiltonians\footnote{Any 4x4 Hermitian Hamiltonian can be mapped to a model of up to eight-vertex type (not necessarily Hermitian) using the transformations in Section \ref{subsec:symmetries}. See \cite{classificationybandboost} for further details.}.

We adopt the following convention:
\begin{itemize}
    \item $a_i$ are free parameters in the $R$-matrix.
    \item $b_i$ are free parameters in the $K^R$-matrix.
    \item $F(u)$ and $G(u)$ are the two free functions in the $R$-matrix. Moreover, for readability, we will use the notation
        \begin{align}
            \Delta X = X(u)-X(v), \quad \Sigma X =X(u) + X(v), \quad \text{for} \; \, X\in \{F,G\}
        \end{align}
    \item $k(u)$ and $k_i(u)$ are free functions in the $K^R$-matrix.
\end{itemize}
For brevity, we omit the superscript $R$ from $K^R$ throughout the following list of solutions.

When solving the BYBE for $K^R$, we always find that we need to set the parameters $a_i$ of the $R$-matrix to some specific values or impose some constraints on the functions $F,G$. These constraints either fix the parity of the functions or establish relations between them. Importantly, each constraint leads to a different $R$-matrix, with its own set of $K^R$-matrices. This explains the large number of reflection matrices found for non-difference form models.

Now, let us suppose that $F$ is an odd function, the same argument applies if instead $G$ is odd. By performing a reparametrization~\eqref{eq:reparameterizationcondition} with $g(u)=F^{-1}(u)$, one can always map $F(u)$ to the identity function $u$,
\begin{align}
    F(u) &\mapsto \tilde{F}(u) = F(F^{-1}(u))=u, \\
    G(u) &\mapsto \tilde{G}(u) = G(F^{-1}(u)).
\end{align}
Moreover, since $F^{-1}$ is odd, the  equation~\eqref{eq:reparameterizationcondition} implies that the $K$-matrix for generic odd $F$ is preserved under the above reparametrization. Therefore, without loss of generality, if there exists at least one odd function $X(u)$, we can select it and set $X(u)=u$; while the remaining functions $Y(u)$ are reparametrized as $Y(X^{-1}(u))$. This feature will be used in the classification below.

Finally, if we consider the case where $F$ and $G$ are constant functions, for either the 6vB or 8vB, the $R$-matrix reduces to the permutation operator, and the reflection matrix is easily found. In fact, the BYBE~\eqref{eq:bybe} reduces to the equation
\begin{align}
    [K(u),K(v)]=0 \quad \forall \,u,v,
\end{align}
which is solved by
\begin{equation}
    K(u) = \left(
\begin{array}{cc}
1 &k(u) \\
 b_1 k(u) & 1+b_2 k(u) \\
\end{array}
\right).
\end{equation}

In what follows, we find the reflection matrices for both classes assuming that $F$ and $G$ are not simultaneously constant.

\subsubsection{6-vertex B}
For the 6-vertex B model of non-difference form, the $R$-matrix is
\begin{equation}
    R(u,v)=\begin{pmatrix}
        1+G(u) \Delta F  & 0 & 0 & 0\\
        0 & \Delta F & 1 & 0\\
        0 & 1 &  G(u) G(v)\Delta F+G(v)-G(u) & 0\\
        0 & 0 & 0 &1 -  G(v) \Delta F
    \end{pmatrix}.
\end{equation}
For this model, we obtained a full classification of the $K^R$-matrices.
\begin{itemize}
    \item \textbf{$\boldsymbol{F}$ constant}: The Hamiltonian corresponding to this $R$-matrix is $h\propto \sigma^+_j \sigma_{j+1}^-$. In this limit, we find three solutions of the boundary Yang-Baxter equation.
    
    \textbf{a), b)}
    If $G(u)$ is an arbitrary function, there exist the following two solutions
    \begin{equation}
    K^{(6\text{v},a)}(u) = \begin{pmatrix}
    1 & k(u) \\
    0 & 1
    \end{pmatrix}, \quad K^{(6\text{v},b)}(u) = \begin{pmatrix}
    G(u) & b_1\left(G(u)-G(-u)\right) \\
    0 & G(-u)
    \end{pmatrix}.
\end{equation}
    \textbf{c)} Moreover, if $G(u)=u$ there is the extra solution
    \begin{equation}
    K^{(6\text{v},c)}(u) = \left(
    \begin{array}{cc}
    1+ b_1 u& b_2 u \\
    b_3 u & 1-b_1 u \\
    \end{array}
    \right).
    \end{equation}
    
\item \textbf{$\boldsymbol{G}$ even}: 

\textbf{d)} If $F$ is also even, the $K$-matrix is the identity matrix,
\begin{equation}
    K^{(6\text{v},d)}(u) = \left(
\begin{array}{cc}
1 & 0 \\
 0 & 1 \\
\end{array}
\right).
\end{equation}
\textbf{e)} On the other hand, if $F$ satisfies the relation $ G(u) F(u)=-1$, there is a solution
\begin{equation}
    K^{(6\text{v},e)}(u) = \left(
\begin{array}{cc}
1 & 0 \\
 k(u) & 1 \\
\end{array}
\right).
\end{equation}

\textbf{f)} Finally, if $F$ and $G$ satisfy the relations
\begin{align}
    G(u) = - \frac{1}{ F(u)+F(-u)}, \quad F(-u)^2=a_1-F(u)^2,
\end{align}
we find
\begin{equation}
    K^{(6\text{v},f)}(u) = \left(
\begin{array}{cc}
b_1-F(-u) & b_2 \left( a_1-2F(u)^2\right) \\
 b_3 \left(F(u)-F(-u)\right) & b_1- F(u) \\
\end{array}
\right).
\end{equation}
\item \textbf{$\boldsymbol{G}$ odd}:

\textbf{g)} If $F(u)=u$, we have
\begin{equation}
    K^{(6\text{v},g)}(u) = \left(
\begin{array}{cc}
1 & -\frac{2 u}{b_1} \\
 b_1 G(u) & 1 \\
\end{array}
\right).
\end{equation}
\textbf{h)} Moreover, if $F(u)=u$ and $G(u)=\frac{ u}{a_1-u^2}$, there is an extra solution
\begin{equation}
    K^{(6\text{v},h)}(u) = \left(
\begin{array}{cc}
b_1-2  u & b_2 u \\
 \frac{b_3 u}{a_1-u^2} & b_1+2 u \\
\end{array}
\right).
\end{equation}

\item \textbf{$\boldsymbol{G}$ of arbitrary parity}:

\textbf{i)} If $F(u)=u$ and $G(u)=-\frac{1}{u+a_1}$, the $K$-matrix is
\begin{equation}
    K^{(6\text{v},i)}(u) = \left(
\begin{array}{cc}
u+a_1 & 0 \\
 k(u) & -u+a_1 \\
\end{array}
\right).
\end{equation}

\textbf{j)}
Finally, if the function $G$ satisfies the relation
\begin{align}
    G(u) = \frac{a_1-a_2F(-u)}{a_3-a_1F(u)-F(-u) \left(a_1-a_2 F(u)\right)},
\end{align}

we find the diagonal solution
\begin{equation}
    K^{(6\text{v},j)}(u) = \left(
\begin{array}{cc}
b_1-F(-u) & 0 \\
 0 & b_1- F(u) \\
\end{array}
\right).
\end{equation}
\end{itemize}

\subsubsection{8-vertex B} 
For the 8-vertex B model of non-difference form, the $R$-matrix is
\begin{equation}
    R(u,v)=\begin{pmatrix}
        r_{1}(u,v)  & 0 & 0 & r_{6}(u,v) \\
        0 & r_{2}(u,v) & r_{5}(u,v) & 0\\
        0 & r_{5}(u,v) &  r_{3}(u,v) & 0\\
        r_{6}(u,v) & 0 & 0 & r_{4}(u,v)
    \end{pmatrix},
\end{equation}
where we have defined
\begin{align}
    r_{1}(u,v)&= \text{cn} \sin\left(\frac{\Sigma G}{2}\right)-\text{dn} \:\text{sn} \cos\left(\frac{\Sigma G}{2}\right), \\
    r_{2}(u,v) &=\text{cn} \sin\left(\frac{\Delta G}{2}\right)+ \text{dn} \:\text{sn} \cos\left(\frac{\Delta G}{2}\right), \\
    r_3(u,v)&=-\text{cn} \sin\left(\frac{\Delta G}{2}\right)+\text{dn} \:\text{sn}  \cos\left(\frac{\Delta G}{2}\right),\\
    r_{4}(u,v)&= \text{cn} \sin\left(\frac{\Sigma G}{2}\right)+\text{dn} \:\text{sn} \cos\left(\frac{\Sigma G}{2}\right), \\
    r_{5}(u,v)&=\text{dn} \sqrt{\sin{G(u)}} \sqrt{\sin{G(v)}}, \\
    r_{6}(u,v)&=a_1 \:\text{cn} \:\text{sn} \sqrt{\sin{G(u)}} \sqrt{\sin{G(v)}}.
\end{align}
The functions cn, sn, dn are the Jacobi elliptic functions with argument $\Delta F$ and modulus $a_1$.

For this model, a complete classification of the reflection matrices is technically challenging because of the presence of elliptic functions. Hence, we provide a partial classification by restricting our analysis to the cases in which the parity of the functions $F$ and $G$ is fixed.

\begin{itemize}
\item  \textbf{$\boldsymbol{F}$ and $\boldsymbol{G}$ even}:

\textbf{a)} We find that the identity matrix is a solution of the boundary Yang--Baxter equation
\begin{equation}
    K^{(8\text{v},a)}(u) = \left(
\begin{array}{cc}
1 & 0 \\
 0 & 1 \\
\end{array}
\right).
\end{equation}
This solution is expected since in this case $R_{i,j}(u,v)=R_{i,j}(u,-v)=R_{i,j}(-u,v)$.

\textbf{b), c)} If we impose that $F$ is given by the following elliptic integral of the first kind 
\begin{align}
    F(u) = \frac{1}{2} \int_0^{G(u)} \frac{d \phi}{\sqrt{1-a_1^2 \sin^2 \phi}},
\end{align}

we find two additional solutions
\begin{equation}
    K^{(8\text{v},b)}(u) = \left(
\begin{array}{cc}
1 & 1-\sqrt{1-a_1^2 \sin ^2(G(u))} \: k(u) \\
 a_1 \sin (G(u)) \: k(u) & 1 \\
\end{array}
\right),
\end{equation}
\begin{equation}
    K^{(8\text{v},c)}(u) = \left(
\begin{array}{cc}
1 &-\sqrt{\frac{a_1^2\sin{(G(u))}}{1+\sqrt{1-a_1^2 \sin ^2(G(u))}}} k(u) \\
 \sqrt{\frac{1+\sqrt{1-a_1^2 \sin ^2(G(u))}}{\sin{(G(u))}}} k(u) & 2 \sqrt{a_1} k(u)+1 \\
\end{array}
\right).
\end{equation}

\item  \textbf{$\boldsymbol{F(u)=u}$ and $\boldsymbol{G}$ even}: For this constraint, we do not provide a complete classification of the solutions.

\textbf{d)} If we impose
\begin{align}
    G(u) = \arccos\left(\frac{a_2}{\text{dn}(2u)}\right),\label{Gdefd}
\end{align}
we find a solution that depends on Jacobi elliptic functions
\begin{equation}
    K^{(8\text{v},d)}(u) = \left(
\begin{array}{cc}
1 &\frac{l_1(b_1,b_2)}{l_1(b_2,b_1)} l_2(u) \\
 l_2(u) & \frac{\text{sn} \:\text{dn} \cos\left(\phi_+(u)\right)-\text{cn} \sin\left(\phi_+(u)\right) }{\text{sn} \: \text{dn} \cos\left(\phi_+(u)\right)+\text{cn} \sin\left(\phi_+(u)\right) } \\
\end{array}
\right),
\end{equation}
where we have defined 
\begin{align}
    l_1(b_1,b_2)&=a_1 b_2 \cos\left(\phi_-\right)\text{cn}\: \text{dn}\:\text{sn}^2 + a_1 \:\text{sn}^3\left(a_1 b_1 \cos\left(\phi_+\right)-b_2 \sin\left(\phi_-\right)\right)+ \nonumber \\
    &+\text{sn}\left(a_1 b_2 \sin\left(\phi_-\right)-b_1 \cos\left(\phi_+\right)\right)-b_1 \text{cn} \: \text{dn} \sin\left(\phi_+\right), \label{k1def}\\
    l_2(u)&=\frac{b_2 \:\text{dn}\left(1-a_1^2-\text{dn}^4\right)+a_1^2 b_2 \cos\left(2\phi_+\right)\text{dn}\left(1-a_1^2\text{sn}^4\right)+2 a_1^3 b_1 \text{cn}^2 \text{sn}^2 \: \text{dn} \: s(u)^2 }{2 a_1^2 \left(1-a_2^2\right)^{\frac{1}{4}}\left(-1+a_1^2\text{sn}^4\right) s(u) \left(\cos\left(\phi_+\right) \text{dn} \: \text{sn}+\sin\left(\phi_+\right) \text{cn}\right)},
    \label{k2def}
\end{align}
and
\begin{align}
    s(u)&=\left(1-\frac{a_1^4 a_2\left(1-a_1^2\text{sn}^4\right)^2}{\left(-1+a_1^2+\text{dn}^4\right)^2}\right)^{\frac{1}{4}}, \\
   2 \phi_{\pm}&=\arccos(a_2) \pm \arccos\left(\frac{a_1^2 a_2\left(1-a_1^2\text{sn}^4\right)}{-1+a_1^2+\text{dn}^4}\right). \label{thetadef}
\end{align}
Notice that at the values $a_2=\pm 1$ the function~\eqref{k2def} diverges. For this reason, we restrict the value of the parameter $a_2 \neq \pm1$.

\textbf{e), f), g), h)} Now, if $G$ is a constant function, $G=a_2$, the $R$-matrix is the one of the XYh model\footnote{In this point, in fact, the Hamiltonian becomes $h_{i,j}=\frac{1}{2}(1+a_1 \sin a_2) \sigma^x_i\sigma^x_j+\frac{1}{2}(1-a_1 \sin a_2) \sigma^y_i\sigma^y_j-\cos a_2 \sigma_i^z$.}. In this case, we find a complete classification of the $K$-matrices. Firstly, if $a_1$ is a generic constant, there exist two solutions that depend explicitly on the Jacobi elliptic functions
\begin{equation} 
    K^{(8\text{v},e)}(u) = \left(
\begin{array}{cc}
s_-(u,a_2) &l_-(u) \\
 l_-(u) & -s_+(u,a_2) \\
\end{array}
\right), \quad K^{(8\text{v},f)}(u) = \left(
\begin{array}{cc}
s_+(u,-a_2) &-l_+(u) \\
 l_+(u) & -s_-(u,-a_2) \\
\end{array}
\right), \label{kef}
\end{equation}
where we have defined
\begin{align}
    s_{\pm}(u,a_2) &= \text{dn}^2\pm a_1 \cos{(a_2)}\: \text{cn} \: \text{dn} \:\text{sn}-a_1 \sin{(a_2)} \: \text{cn}^2, \label{spm}\\
    l_{\pm}(u)&=\sqrt{2 a_1 \sin{(a_2)}\left(1\pm a_1 \sin{(a_2)}\right)} \:\text{cn} \: \text{dn}.\label{kpm}
\end{align}

On the other hand, if $a_1=1$ the Jacobi elliptic functions of modulus $a_1$ reduce to hyperbolic functions, and there are two extra $K$-matrices 
\begin{align}
    K^{(8\text{v},g)}(u) &= \left(
\begin{array}{cc}
s_-(u,-a_2)+b_1 & \sqrt{2\left(-1+\csc{(a_2)}\right)} \sinh{(2u)} \\
  \sqrt{2\left(-1+\csc{(a_2)}\right)} \sinh{(2u)} & s_+(u,-a_2)+b_1\\
\end{array}
\right), \label{kg} \\
 K^{(8\text{v},h)}(u) &= \left(
\begin{array}{cc}
s_+(u,a_2)+b_1 & -\sqrt{2\left(1+\csc{(a_2)}\right)} \sinh{(2u)} \\
  \sqrt{2\left(1+\csc{(a_2)}\right)} \sinh{(2u)} & s_-(u,a_2)+b_1\\
\end{array}
\right), \label{kh}
\end{align}
where we have defined
\begin{align}
    s_{\pm}(u,a_2)=\cot{(a_2)} \cosh{(2 u)}\pm(1+\csc{(a_2)})\sinh{(2u)}.\label{smpforgh}
\end{align}
Notice that, upon normalizing the entry $K_{11}(u)$ of~\eqref{kef}, \eqref{kg} and \eqref{kh} to one, the $K$-matrices~\eqref{kef} reduce, in the limit $a_1 \to 1$, to~\eqref{kg} and~\eqref{kh} with $b_1=\cot{a_2}$ and $b_1=-\cot{a_2}$, respectively.

Moreover, if $a_2$ is an integer multiple of $\pi$, the $K$ matrices \eqref{kg} and \eqref{kh} diverge. However, in this case $G = n \pi$, $n \in \mathbb{Z}$ and the $R$-matrix becomes diagonal, therefore non-regular.

\item $\boldsymbol{G(u)=u}$ \textbf{and F even}: In this case, there is no solution of the boundary Yang--Baxter equation.
\item $\boldsymbol{F(u)=u}$ \textbf{and G odd}: In this case, the boundary Yang--Baxter equation admits no solution either.
\end{itemize}

\subsection{Computing left reflection matrices $K^L(u)$: summary of the method}\label{sec:Kleft}

For models with open boundary conditions, integrability is fundamentally determined by the bulk $R$-matrix and the left and right $K$-matrices, which describe boundary scattering. We now outline the methodology for deriving the left reflection algebra specifically for $R$-matrices that are of non-difference form.

In order to obtain the condition on the $K^L(u)$, by following \cite{vanicat2018integrable}, we first define the dual reflection matrix $\bar{K}(u)$, that satisfies the dual equation\footnote{We remark that the steps involving $\bar{K}(u)$ work for any $R$-matrix solution of the Yang-Baxter equation and do not require any additional symmetry.} 

\begin{equation}
    R_{12}^{-1}(u,v)\bar{K}_1(u)R_{21}^{-1}(v,-u)\bar{K}_2(v)=\bar{K}_2(v)R_{12}^{-1}(u,-v)\bar{K}_1(u)R_{21}^{-1}(-v,-u).
    \label{eq:dualbybenondiff}
\end{equation}

The $K^L(u)$ is related to $\bar{K}(u)$ by the following automorphism
\begin{equation}
    K_1^L(u)=\text{tr}_0 \Big( \bar{K}_0(-u) \Big(\Big(\big(R_{01}(u,-u)\big)^{t_1}\Big)^{-1}\Big)^{t_1}P_{01}\Big),
    \label{KLandKbarnondiff}
\end{equation}
where $t_1$ denotes the transposition with respect to the first space.

Here, we sketch the general algorithm that we have used to compute the $K^L(u)$ matrices.

\begin{itemize}
    \item[1.] \textbf{Solve the Dual Reflection Equation.} We solve Eq. \eqref{eq:dualbybenondiff} for the dual reflection matrix $\bar{K}(u)$ for each of the $R$-matrices analysed in Secs. \ref{newreflmat}. While this can be done directly using either of the two methods described in Sec. \ref{methodsused}, symmetry allows for a more efficient approach:  if the $R$-matrix possesses specific symmetries, the dual equation can be mapped to the right reflection equation (Eq. \eqref{eq:bybe}) by substituting $R_{ij}$ with its inverse $R_{ij}^{-1}$. Consequently, $\bar{K}(u)$ can often be derived from the known $K^R(u)$. For the models under consideration, we were able to obtain $\bar{K}(u)$ starting from $K^R(u)$. However, since this method is not universal, it must be adapted to each case as outlined below.
    \item[2.]\textbf{Determine $K^L(u)$ via the automorphism.} We determine $K^L(u)$ using the automorphism given in Eq. \eqref{KLandKbarnondiff}. In some specific cases, we encounter two possible problems: either the automorphism becomes singular, or the inverse required in Eq. \eqref{eq:bybel} is not well-defined. For these cases, we obtained the $K^L(u)$ by directly solving the commutation relation $[t(u),t(v)] = 0$ for $L=4$. We checked the transfer matrix for $L=4$ for these problematic cases and we notice that it exhibits a trivial structure, which we analyse further in the subsequent sections.
    
    \item[3.] \textbf{Check commutativity $[t(u),t(v)] = 0$.} To validate our results, we explicitly verified the commutativity condition $[t(u), t(v)] = 0$ for system sizes up to $L=5$ using the $K^L(u)$ matrices.
\end{itemize}

In the following, we use the following conventions
\begin{itemize}
    \item Since the $K$-matrix always acts on a single spatial index, we lower the superscript $L$ to a subscript and write $K_{L}^{(i,j)}$, where $i$ labels the class and in case of multiple solutions $j=a,b,c,\dots$ enumerates the different solutions.
    \item $c_i$ are constants.
    \item $q_i(u)$ are free functions.
\end{itemize}

\subsubsection{6-vertex B}

For models in this class, we noticed that
\begin{align}
    & (A\otimes A)R^{-1}(u,v) (A\otimes A)^{-1}|_{F(u)=-F(u), G(u)=-G(u)}=R(u,v),
\end{align}
where $A=\left(
\begin{array}{cc}
 0 & 1 \\
 1 & 0 \\
\end{array}
\right)$.

This relation allowed us to obtain
\begin{align}
    &\bar K(u)= A K^R(u)A^{-1}|_{F(u)=-F(u), G(u)=-G(u)}.
    \label{Kbar}
\end{align}

We remark that since $K^R(u)$ and $K^L(u)$ independently satisfy the boundary Yang-Baxter equations \eqref{eq:bybe} and \eqref{eq:bybel}, one may relabel the constants $b_i \to c_i$ within $\bar{K}(u)$ without violating the commutativity condition $[t(u), t(v)] = 0$.

In what follows, we explicitly give the $K_L(u)$-matrix corresponding to the different cases.

\begin{itemize}
    \item \textbf{F constant:}
    
 \textbf{a) G arbitrary:} This is one of the problematic case that we mentioned. The automorphism \eqref{KLandKbarnondiff} is singular. We obtained the $K_L(u)$ by solving the commutativity condition $[t(u),t(v)]=0$ and we obtain
 \begin{align}
    K_L^{(6v,a)}(u)=\left(
\begin{array}{cc}
 1 & 0 \\
q_1(u) & q_2(u) \\
\end{array}
\right).
\end{align}
In this case, the transfer matrix has only one different eigenvalue.

 \textbf{b) G arbitrary:} The reflection equation \eqref{eq:bybel} is singular, hence we obtained the $K_L(u)$ by solving the commutativity condition $[t(u),t(v)]=0$ and we obtain
 \begin{align}
    K_L^{(6v,b)}(u)=\left(
\begin{array}{cc}
 1 & 0 \\
q_1(u) & q_2(u) \\
\end{array}
\right).
\end{align}
In this case, the transfer matrix has two different  eigenvalues with degeneracies $2^{L-1}$.

    \textbf{c) $\mathbf{G(u)=u}$:} Also in this case the reflection equation \eqref{eq:bybel} is singular and we obtain
 \begin{align}
    K_L^{(6v,c)}(u)=\left(
\begin{array}{cc}
 1 & c_1(1+q_4(u)) \\
q_3(u) & q_4(u) \\
\end{array}
\right).
\end{align}

    \item \textbf{G even:}

    \textbf{d) F even:} In this case, we perform the steps of Sec. \ref{sec:Kleft} and we obtain
    
 \begin{align}
    K_L^{(6v,d)}(u)=\left(
\begin{array}{cc}
 1 & 0 \\
 0 & -1 \\
\end{array}
\right).
\end{align}
In this case, the transfer matrix vanishes. We observed that, any $K^L(u)$ also solves the commutation relation $[t(u),t(v)]=0$, but the transfer matrix remains trivial (proportional to the identity matrix).

    \textbf{e) $\mathbf{G(u)F(u)=-1}$: } The reflection equation \eqref{eq:bybel} is singular, any general $K_L(u)$ satisfies the commutation condition $[t(u),t(v)]=0$. In this case, the resulting transfer matrix vanishes.

    \textbf{f) $\mathbf{G(u) = - \frac{1}{ F(u)+F(-u)}, \,\, F(-u)^2=a_1-F(u)^2}$:} For this case, we perform the steps illustrated in \ref{sec:Kleft}. First, we obtain the solution for $\bar{K}$, from Eq. \eqref{Kbar}.

We calculated $K^L$ from the automorphism Eq. \eqref{KLandKbarnondiff} and we obtain,
\begin{align}
    K_L^{(6v,f)}(u)=\left(
\begin{array}{cc}
 1 & \frac{c_3 \left(F(u)^2-F(-u)^2\right)}{F(u)-c_1} \\
 \frac{c_2 \left(2 F(u)^2-a_1\right)}{(F(-u)+F(u)) \left(F(u)-c_1\right)} & \frac{c_1-F(-u)}{F(u)-c_1} \\
\end{array}
\right).
\end{align}

\item \textbf{G odd}:

\textbf{g) $\mathbf{F(u)=u}$.} 
We perform the steps illustrated in \ref{sec:Kleft} and we obtain
\begin{align}
    K_L^{(6v,g)}(u)=\left(
\begin{array}{cc}
 1 & c_1 \\
 \frac{2 u G(u)+2}{c_1} & 1 \\
\end{array}
\right).
\end{align}

\textbf{h)}
We perform the steps illustrated in \ref{sec:Kleft} and we obtain
\begin{align}
    K_L^{(6v,h)}(u)=\left(
\begin{array}{cc}
 c_1 u-2 a_1 & c_3 u \\
 \frac{a_1 c_2 u}{u^2-a_1} & 2 a_1+c_1 u \\
\end{array}
\right)
\end{align}

\textbf{i)} In this case, the automorphism \eqref{KLandKbarnondiff} is singular, hence we computed the $K_L(u)$ by solving the commutator conditions $[t(u),t(v)]=0$ and we obtain
\begin{align}
    K_L^{(6v,i)}(u)=\left(
\begin{array}{cc}
 1 & q_1(u) \\
 0 & q_2(u) \\
\end{array}
\right).
\end{align}

This model is special since the transfer matrix does not depend on the inhomogeneities parameters.

\textbf{j)} We perform the steps illustrated in \ref{sec:Kleft} and we obtain
\begin{align}
    K_L^{(6v,j)}(u)=\left(
\begin{array}{cc}
 w(u) & 0 \\
 0 & -w(-u) \\
\end{array}
\right),
\end{align}
where $w(u)=a_1 c_1+a_3-\left(a_2 c_1+a_1\right) F(u)$.
\end{itemize}

\subsubsection{8-vertex B}

For models in this class, we noticed that
\begin{align}
    &R^{-1}(u,v) |_{F(u)=-F(u)}=R(u,v).
    \label{Rinv8v}
\end{align}
Hence, we obtain
\begin{align}
    &\bar K(u)= K^R(u)|_{F(u)=-F(u)}.
    \label{K2bar}
\end{align}

    \begin{itemize}
    \item \textbf{F and G even}:

    \textbf{a), b), c)} In this case, the automorphism \eqref{KLandKbarintro} is singular since, $R(u,-u)\propto P$ and $P^{t_1}$ is singular. For these three cases, we have verified that any arbitrary $K_L(u)$ satisfies the commutation relations $[t(u),t(v)]=0.$ In particular, we computed the homogeneous transfer matrix for model a) and we obtain that is proportional to the identity matrix, for model b) it has two eigenvalues with degeneracies $2^{L-1}$ and for model of class c) there is only one different eigenvalue.

    \item \textbf{$\boldsymbol{F(u)=u}$ and $\boldsymbol{G}$ even:} In this case, since the relation between $R(u,v)$ and $R(u,v)^{-1}$ given in \eqref{Rinv8v}, we obtained
    \begin{align}
        &\bar{K}^{(8v,i)}(u)=K^{(8v,i)}(-u),\,\,\,\, i=d,e,f,g,h.
    \end{align}
We calculated the corresponding $K_L(u)$ by using the automorphism \eqref{KLandKbarnondiff} and we obtained the following solutions

\textbf{d) }
\begin{align}
    K_L^{(8v,d)}=\left(
\begin{array}{cc}
 \frac{r _5 \left(1-\frac{2 \text{cn}(-u)}{\text{cn}(-u)+\text{dn}(-u) \text{sn}(-u) \cot \left(\phi _+(-u)\right)}\right)-r _4}{r _5^2-r_1 r _4} & \frac{r _2 l_1(-u)-r_6 l_2(-u)}{r_2 r_3-r_6^2} \\
 \frac{r_6 l_1(-u)-r_3 l_2(-u)}{r_6^2-r_2 r_3} & \frac{r_1 \left(1-\frac{2 \text{cn}(-u)}{\text{cn}(-u)+\text{dn}(-u) \text{sn}(-u) \cot \left(\phi_+(-u)\right)}\right)-r_5}{r_1 r_4-r_5^2} \\
\end{array}
\right),
\end{align}
where $l_1$, $l_2$ and $\phi_\pm$ are defined in \eqref{k1def}, \eqref{k2def} and \eqref{thetadef} and $r_i=r_i(u,-u)$ are the entries of the $R$-matrix with the restriction $F(u)=u$ and $G(u)$ in \eqref{Gdefd}. We remark that, the constants $b_i$ appearing in $l_i$ can now take different values as the one in $K^R$ since, as mentioned,  $K^R$ and $K^L$ solve separately the boundary Yang-Baxter equations.

\textbf{e), f)}
\begin{align}
&K_L^{(8v,e)}(u)=\left(
\begin{array}{cc}
 \frac{r_4 s_-\left(-u,a_2\right)+r_5 s_+\left(-u,a_2\right)}{r_1 r_4-r_5^2} & \frac{\left(r_2-r_6\right) l_-(-u)}{r_2 r_3-r_6^2} \\
 \frac{\left(r_3-r_6\right) l_-(-u)}{r_2 r_3-r_6^2} & \frac{r_5 s_-\left(-u,a_2\right)+r_1 s_+\left(-u,a_2\right)}{r_5^2-r_1 r_4} \\
\end{array}
\right),
\\&K_L^{(8v,f)}(u)=\left(
\begin{array}{cc}
 \frac{r_5 s_-\left(-u,-a_2\right)+r_4 s_+\left(-u,-a_2\right)}{r_1 r_4-r_5^2} & -\frac{\left(r_2+r_6\right) l_+(-u)}{r_2 r_3-r_6^2} \\
 \frac{\left(r_3+r_6\right) l_+(-u)}{r_2 r_3-r_6^2} & \frac{r_1 s_-\left(-u,-a_2\right)+r_5 s_+\left(-u,-a_2\right)}{r_5^2-r_1 r_4} \\
\end{array}
\right),
\end{align}
where $l_\pm$, $s_\pm$ are defined in \eqref{kpm} and \eqref{spm}; $r_i=r_i(u,-u)$ are the entries of the $R$-matrix with the restrictions $F(u)=u$ and $G(u)=a_2$.

\textbf{g), h)}

\begin{align}
    &K_L^{(8v,g)}(u)=\left(
\begin{array}{cc}
 \frac{r _4 \left(s_-\left(-u,-a_2\right)+b_1\right)-r _5 \left(s_+\left(-u,-a_2\right)+b_1\right)}{r _1 r _4-r _5^2} & \frac{\sqrt{2} \left(r_6-r_2\right) \sqrt{\csc \left(a_2\right)-1} \sinh (2 u)}{r_2 \rho _3-r_6^2} \\
 \frac{\sqrt{2} \left(r_6-r _3\right) \sqrt{\csc \left(a_2\right)-1} \sinh (2 u)}{r_2 r_3-r_6^2} & \frac{r_1 \left(s_+\left(-u,-a_2\right)+b_1\right)-r_5 \left(s_-\left(-u,-a_2\right)+b_1\right)}{r_1 r_4-r_5^2} \\
\end{array}
\right),
\\
&K_L^{(8v,h)}(u)=\left(
\begin{array}{cc}
 \frac{r_4 \left(s_+\left(-u,a_2\right)+b_1\right)-r_5 \left(s_-\left(-u,a_2\right)+b_1\right)}{r_1 r_4-r_5^2} & \frac{\sqrt{2} \left(r_2+r_6\right) \sqrt{\csc \left(a_2\right)+1} \sinh (2 u)}{r_2 r_3-r_6^2} \\
 -\frac{\sqrt{2} \left(r_3+r_6\right) \sqrt{\csc \left(a_2\right)+1} \sinh (2 u)}{r_2 r_3-r_6^2} & \frac{r_1 \left(s_-\left(-u,a_2\right)+b_1\right)-r_5 \left(s_+\left(-u,a_2\right)+b_1\right)}{r_1 r_4-r_5^2} \\
\end{array}
\right),
\end{align}
where $s_\pm(u)$ are defined in \eqref{smpforgh}; $r_i=r_i(u,-u)$ are the entries of the $R$-matrix with the restrictions $F(u)=u$, $G(u)=a_2$, $a_1=1$. 

\end{itemize}

\section{Conclusion and Outlook}\label{sec:conclusions}

We classify Yang–Baxter integrable quantum circuits with open boundary conditions and different geometries. We conjecture that time-periodic quantum circuits where the local bulk and boundary gates satisfy the Yang-Baxter equation and the same bulk gate is applied exactly once per period to every nearest-neighbor pair of spins are integrable. We further prove that all such circuits are isospectral. We separate these circuits into equivalence classes according to the number of $-\kappa$ inhomogeneities. The eigenvectors of circuits belonging to the same equivalence class are related by similarity transformations involving only bulk gates, whereas relating circuits from different equivalence classes necessarily requires boundary gates. For each class, we derive the minimal circuit depth required to implement one period of the time evolution. The geometry corresponding to this choice consists of a staircase part on the left, a staircase part on the right, and a brickwork part in the middle. For staggered inhomogeneities $(\kappa,-\kappa)$, the circuit is of brickwork type, but as the number of $-\kappa$'s decreases, more and more gates contribute to the staircase part on the sides.

Additionally, our construction provides an algorithm to determine Yang-Baxter integrability of a circuit with open boundary conditions. We also introduce multiple types of inhomogeneities and we obtain for each of them the configurations that lead to the minimum possible depth. For example, for three types of inhomogeneities, the minimum possible depth is $d=4$. 

As a practical example, we provide all qubit gates whose Hamiltonians are of six- and eight-vertex types. We did this by solving the Sklyanin reflection algebra for the $R$-matrices of non-difference form classified in \cite{classificationybandboost}.

Each of these results opens avenues for further exploration.

First, for two types of inhomogeneities, we conjecture that a circuit is Yang–Baxter integrable if: (1) the bulk gate is a solution of the Yang–Baxter equation; (2) the boundary gates satisfy the boundary Yang–Baxter equations; and (3) each bulk gate is applied exactly once to every pair of consecutive qubits. We have verified this conjecture numerically up to thirty sites. However, an analytic proof, even for the periodic case, remains an open problem.

A strong indication of the conjecture’s validity comes from the structure of the double-row transfer matrix. Each $R$-matrix with arguments $(u,\theta_i)$ carries a distinct spatial index, while the second string of $R$-matrices depends on the inhomogeneities as $(\theta_i,-u)$. This structure ensures that each gate is applied exactly once to each pair of spins. This is immediately clear in Theorems 1 and 2 where: any pair appearing in the first product does not appear in the second, and vice-versa. 

Furthermore, our classification provides a basis for further investigation into how the choice of circuit geometry in systems with open boundaries affects physical observables. In particular, it would be interesting to investigate hydrodynamic quantities, such as spin correlators, and determine their dependence on the underlying geometry.

In the spirit of \cite{maruyoshi2023conserved}, these circuits may also serve as benchmarks for quantum algorithms. A natural direction is to derive the conserved quantities associated with each geometry and analyse how noise propagation depends on the geometry under consideration.

We propose a numerical algorithm to test Yang–Baxter integrability of quantum circuits with open boundary conditions. The output of the algorithm should be interpreted with care. A positive result guarantees Yang–Baxter integrability, since one can explicitly construct the transfer matrix generating the conserved quantities. However, a negative result does not imply that the circuit is non-integrable. For example, for the circuits studied in \cite{popkov2025exact,Zhang:2026avw,paletta2026integrability}, our algorithm would return a negative result. Nevertheless, numerical constructions of conserved quantities and level-statistics analysis indicate signatures of integrability. This suggests that other algebraic mechanisms may exist for verifying integrability beyond the Yang–Baxter framework considered here. Identifying such mechanisms would be an interesting development for quantum circuits and, more broadly, for quantum integrable models.

In the Mathematica notebook \texttt{OpenQCforDiffGeom.nb} \cite{zenodo}, we provide a user-friendly algorithm for constructing circuits with different types of inhomogeneities.  As examples, we consider circuits with two, three, and four types of inhomogeneities, although the generalisation to more than four types is straightforward. 

For three types of inhomogeneities, the minimum possible depth is $d=4$. However, unlike the case with two types of inhomogeneities, several inequivalent circuits realize this depth (see for example figure \ref{fig:N12rho}). It would be interesting to understand how these different geometries with the same depth affect physical observables.

We also classified solutions of the boundary Yang–Baxter equation for the six- and eight-vertex models of non-difference form \cite{classificationybandboost}, which we report in the Mathematica notebook \texttt{KL.nb} \cite{zenodo}. The classification is complete for the six-vertex case, while for the eight-vertex case we focus on a restricted class of solutions. This classification provides explicit examples of boundary gates for the new classes of circuits constructed here. Furthermore, it is interesting in its own right, and further investigation on the spectrum and the symmetries of the quantum integrable spin chains built from them is a natural next step. 

We remark that the class of models considered includes, up to integrability-preserving transformations, all R-matrix corresponding to Hermitian Hamiltonians with up to sixteen-vertex structure, corresponding to the most general $4 \times 4$ Hamiltonian ansatz. A natural follow-up would be to complete the classification for the eight-vertex case. Moreover, some of the models we identify depend on several parameters, such as model d) in the 8vB. For certain parameter choices, the resulting dynamics is unitary, while for others it is dissipative. It would be interesting to characterize this transition by first determining the parameter regimes.

Another possible direction is to start from a non-unitary R-matrix and investigate whether there exists an integrability-preserving transformation that makes it unitary. One could then apply the same transformation at the level of the corresponding K-matrix and study whether the resulting transfer matrix becomes unitary.

We emphasize that our construction applies to both difference- and non-difference-form models, of any rank or spin. This is relevant because it opens the possibility of studying circuits whose gates are given by the Hubbard model. For instance, after classifying all $K$-matrices, and in the spirit of \cite{Paletta:2024uzj}, it would be interesting to analyse the possible connection between the solvability of the non-equilibrium steady states and the integrability of the full spectrum. In this regard, following \cite{vanicat2018integrable}, it would also be interesting to study the steady states associated to circuits of different geometries.

\section*{Acknowledgements}
We would like to thank M. de Leeuw, S. Driezen, D. Gregori, R. Nepomechie, D. Polvara, T. Prosen, E. Ragoucy, T. Skrzypek, L. Vinet  and M. Yamazaki for helpful discussions and M. de Leeuw, R. Nepomechie and T. Prosen for comments on the manuscript. 

ALR was supported by UKRI Future Leaders Fellowship (grant number MR/T018909/1) and by  ERC-2021-CoG - BrokenSymmetries 101044226. ALR thanks the organizers and participants of the  Workshop on higher-dimensional integrability (2025), in  Favignana (Italy), and of the Integrability, Dualities and Deformations Conference (IDD 2025) in Sweden, where part of this work was carried out, for enlightening discussions. ALR is grateful for the hospitality of Perimeter Institute where part of this work was carried out. Research at Perimeter Institute is supported in part by the Government of Canada through the Department of Innovation, Science and Economic Development and by the Province of Ontario through the Ministry of Colleges and Universities. This work was supported by a grant from the Simons Foundation (1034867, Dittrich). CP acknowledge funding from the European Union HORIZON-CL4-2022-QUANTUM-02-SGA through PASQuanS2.1 (Grant Agreement No. 101113690), European Research Council (ERC) through Advanced grant QUEST (Grant Agreement No. 101096208). The work of MGF was funded by Xunta de Galicia through the ``Programa de
axudas \'a etapa predoutoral da Xunta de Galicia'' (Consellería de Cultura, Educaci\'on e Universidade) with reference code ED481A-2024-096. MGF also acknowledges the grants 2023-PG083 (with reference code ED431F 2023/19 funded by Xunta de Galicia),  PID2023-152148NB-I00 (funded by AEI-Spain), the Mar\'ia de Maeztu grant CEX2023-001318-M (funded by MICIU/AEI /10.13039 / 501100011033), the CIGUS Network of Research Centres, and the European Union. 

\begin{appendix}
\numberwithin{equation}{section}
\section{BYBE symmetries}\label{App:transformations}

In this appendix, we consider transformations of the $R$-matrix that preserve the YBE, Eq.~\eqref{eq:ybe}, see \cite{classificationybandboost}. We then substitute them into the BYBE (Eq.~\eqref{eq:bybe}) and determine the corresponding transformation of the reflection matrix $K^R$ required to keep the BYBE invariant.

To improve readability, we first recall the BYBE \eqref{eq:bybe} here
\begin{equation}
        R_{12}(u,v)K_1^R(u)R_{21}(v,-u)K_2^R(v)=K_2^R(v)R_{12}(u,-v)K_1^R(u)R_{21}(-v,-u)\label{eq:bybeapp}.
\end{equation}

\subsection{Local basis transformations}\label{App:lbt}

It is well known that local basis transformations of the form 
\begin{equation}
    \tilde{R}_{12}(u,v)=V_1(u)V_2(v)R_{12}(u,v)V_1(u)^{-1}V_2(v)^{-1},
\end{equation}
preserve the Yang-Baxter equation. We now substitute $R$ in terms of $\tilde{R}$ into the BYBE and determine the resulting transformation of $K^R(u)$
\begin{align}
    &V_1(u)^{-1}V_2(v)^{-1}\tilde{R}_{12}(u,v)V_{1}(u)\cancel{V_2(v)}K_1^R(u)V_1(-u)^{-1}\cancel{V_2(v)}^{-1}\tilde{R}_{21}(v,-u)V_{1}(-u)V_2(v)K_2^R(v)=\nonumber\\
    &=K_2^R(v)V_1(u)^{-1}V_2(-v)^{-1}\tilde{R}_{12}(u,-v)V_{1}(u)\bcancel{V_2(-v)}K_1^R(u)V_1(-u)^{-1}\bcancel{V_2(-v)}^{-1}\tilde{R}_{21}(-v,-u)V_{1}(-u)V_2(-v),\label{eq:lbt1}\\
    &V_1(u)^{-1}V_2(v)^{-1}\tilde{R}_{12}(u,v)V_{1}(u)K_1^R(u)V_1(-u)^{-1}\tilde{R}_{21}(v,-u)V_2(v)K_2^R(v)V_{1}(-u)=\nonumber\\
    &=V_1(u)^{-1}K_2^R(v)V_2(-v)^{-1}\tilde{R}_{12}(u,-v)V_{1}(u)K_1^R(u)V_1(-u)^{-1}\tilde{R}_{21}(-v,-u)V_{1}(-u)V_2(-v)\label{eq:lbt2},\\[0.15cm]
     &\cancel{V_1(u)}^{-1}V_2(v)^{-1}\tilde{R}_{12}(u,v)V_{1}(u)K_1^R(u)V_1(-u)^{-1}\tilde{R}_{21}(v,-u)V_2(v)K_2^R(v)\bcancel{V_{1}(-u)}=\nonumber\\
    &=\cancel{V_1(u)}^{-1}K_2^R(v)V_2(-v)^{-1}\tilde{R}_{12}(u,-v)V_{1}(u)K_1^R(u)V_1(-u)^{-1}\tilde{R}_{21}(-v,-u)V_2(-v)\bcancel{V_{1}(-u)}\label{eq:lbt3},\\[0.15cm]
     &\tilde{R}_{12}(u,v)V_{1}(u)K_1^R(u)V_1(-u)^{-1}\tilde{R}_{21}(v,-u)V_2(v)K_2^R(v)V_2(-v)^{-1}=\nonumber\\
    &=V_2(v)K_2^R(v)V_2(-v)^{-1}\tilde{R}_{12}(u,-v)V_{1}(u)K_1^R(u)V_1(-u)^{-1}\tilde{R}_{21}(-v,-u),\label{eq:lbt4}\\[0.15cm]
    &\Rightarrow \quad\tilde{R}_{12}(u,v)\tilde{K}_1^R(u)\tilde{R}_{21}(v,-u)\tilde{K}_2^R(v)=\tilde{K}_2^R(v)\tilde{R}_{12}(u,-v)\tilde{K}_1^R(u)\tilde{R}_{21}(-v,-u),\label{eq:lbt5}
\end{align}
where 
\begin{equation}
    \tilde{K}^R(u)=V(u)K^R(u)V(-u)^{-1}\label{eq:Klbt},
\end{equation}
as mentioned in section \ref{eq:lbt1}.

In the first step from \eqref{eq:lbt1} to \eqref{eq:lbt2}, we used the fact that operators acting on site 1 and site 2 commute with each other and then we cancel terms like $V_i(u)V_i(u)^{-1}$.  In the next step, from \eqref{eq:lbt2} to \eqref{eq:lbt3}, we again use that operators acting on different sites commute to make the leftmost/rightmost term on both lhs and rhs the same and then cancel them. We then multiplied on the left by $V_2(v)$ and on the right by $V_2(-v)^{-1}$. Finally, we identify that if we define $\tilde{K}^R(u)$ as in \eqref{eq:Klbt}, we recover the same BYBE as in \eqref{eq:bybeapp}.

\subsection{Normalization}\label{App:normalization}

If we plug the following transformation

\begin{equation}
    \tilde{R}(u,v)=f(u,v)R(u,v),\label{eq:normalizationapp}
\end{equation}
in the BYBE \eqref{eq:bybeapp} we obtain 
\begin{align}
       f(u,v)^{-1}f(v,-u)^{-1}&\tilde{R}_{12}(u,v)K_1^R(u)\tilde{R}_{21}(v,-u)K_2^R(v)=\nonumber\\
       &f(u,-v)^{-1}f(-v,-u)^{-1}K_2^R(v)\tilde{R}_{12}(u,-v)K_1^R(u)\tilde{R}_{21}(-v,-u).
\end{align}
It is clear that we can rewrite it back as a BYBE and $\tilde{K}^R(u)=K^R(u)$, iff 
\begin{equation}
    f(u,v)f(v,-u)=f(u,-v)f(-v,-u),
\end{equation}
as mentioned in \eqref{eq:normalizationcondition}.

\subsection{Reparameterization}\label{App:reparameterization}

If we reparameterize the spectral parameters in $R(u,v)$ as
\begin{equation}
    \tilde{R}(u,v)=R(g(u),g(v)),
\end{equation}
it still satisfies the YBE. However, when we do the same in the BYBE we obtain the following
\begin{equation}
    R_{12}(g(u),g(v))\tilde{K}_1^R(u)R_{21}(g(v),g(-u))\tilde{K}_2^R(v)=\tilde{K}_2^R(v)R_{12}(g(u),g(-v))\tilde{K}_1^R(u)R_{21}(g(-v),g(-u)).
\end{equation}
If we take
\begin{equation}
    \tilde{K}^R(u)=K^R(g(u)),
\end{equation}
we almost obtain the correct BYBE. The only missing point is that we still want to interpret this as a reflection, and therefore need
\begin{equation}
    g(-u)=-g(u),
\end{equation}
namely, where $g(u)$ is an odd function of $u$.

This means that if we have a solution of the original BYBE and we transform $R(u,v)$ using a $g(u)$ that is not odd, we need to solve the BYBE again starting from the beginning with the transformed $R$-matrix.

\subsection{Twists}\label{App:twists}

Let us start with a twist of the type
\begin{equation}
    \tilde{R}_{12}(u,v)=W_1(u)R_{12}(u,v)W_2(v)^{-1}, \quad \text{ with } \quad [R_{12}(u,v),W_1(u)W_2(v)]=0.
\end{equation}
Substituting $R$ in terms of $\tilde{R}$ in the BYBE we obtain
\begin{align}
     W_2(v)^{-1}&\tilde{R}_{12}(u,v)W_1(u)K_1^R(u)W_1(-u)^{-1}\tilde{R}_{21}(v,-u)W_2(v)K_2^R(v)=\nonumber\\
     &K_2^R(v)W_2(-v)^{-1}\tilde{R}_{12}(u,-v)W_1(u)K_1^R(u)W_1(-u)^{-1}\tilde{R}_{21}(-v,-u)W_2(-v),\label{eq:twist1}\\[0.2cm]
     \tilde{R}_{12}(u,v)&W_1(u)K_1^R(u)W_1(-u)^{-1}\tilde{R}_{21}(v,-u)W_2(v)K_2^R(v))W_2(-v)^{-1}=\nonumber\\
     &W_2(v)K_2^R(v)W_2(-v)^{-1}\tilde{R}_{12}(u,-v)W_1(u)K_1^R(u)W_1(-u)^{-1}\tilde{R}_{21}(-v,-u),\label{eq:twist2}\\[0.2cm]
   &\Rightarrow \quad\tilde{R}_{12}(u,v)\tilde{K}_1^R(u)\tilde{R}_{21}(v,-u)\tilde{K}_2^R(v)=\tilde{K}_2^R(v)\tilde{R}_{12}(u,-v)\tilde{K}_1^R(u)\tilde{R}_{21}(-v,-u),\label{eq:twist3}     
\end{align}
where 
\begin{equation}
    \tilde{K}^R(u)=W(u)K^R(u)W(-u)^{-1}\label{eq:Ktwist}.
\end{equation}
From the \eqref{eq:twist1} to \eqref{eq:twist2} we multiply by $W_2(v)$ from the left and by $W_2(-v)^{-1}$ from the right. And then from \eqref{eq:twist2} to \eqref{eq:twist3} we notice that if we define $\tilde{K}^R(u)$ as in \eqref{eq:Ktwist} we recover the boundary Yang-Baxter equation.

\section{Proof that the spectrum remains invariant under swapping inhomogeneities }\label{App:swappinginhomogeneities}

If we build a transfer matrix with $\kappa_-$ inhomogeneities $-\kappa$ and $N-\kappa_-$ inhomogeneities $+\kappa$, the spectrum will be the same regardless of the positions at which we place the $-\kappa$'s. In models where the Bethe ansatz is known, it is clear from the Bethe equations that swapping inhomogeneities does not affect the spectrum. However, in this paper, we are considering open spin chains built from arbitrary $R$-matrices, including those of non-difference form.  To the best of our knowledge, an explicit proof that takes all these aspects into account is missing. Therefore, for completeness, we present a general proof here.

In \cite{Ferrando:2023lrx}, the authors present a simple and beautiful proof for periodic chains with $R$-matrices of difference-form. More specifically, they prove that a transfer matrix with two inhomogeneities swapped is related to the one before the swapping by a similarity transformation.

In this appendix, we generalise their proof to open spin chains with a bulk described by $R$-matrices of both difference and non-difference form.

Let us start by writing the YBE as the following
\begin{equation}
    R_{a_1,a_2}(u_1,u_2)R_{a_1,a_3}(u_1,u_3)R_{a_2,a_3}(u_2,u_3)=R_{a_2,a_3}(u_2,u_3)R_{a_1,a_3}(u_1,u_3)R_{a_1,a_2}(u_1,u_2).
    \label{eq:ybeapp}
\end{equation}

\paragraph{Step 1} 
Like in \cite{Ferrando:2023lrx}, we write 
\begin{equation}
    \check{R}_{i,i+1}(\theta_i,\theta_{i+1})R_{a,i+1}(u,\theta_{i+1})R_{a,i}(u,\theta_{i})=R_{a,i+1}(u,\theta_{i})R_{a,i}(u,\theta_{i+1})\check{R}_{i,i+1}(\theta_i,\theta_{i+1}).\label{eq:lemma4}
\end{equation}

This follows from a straightforward manipulation of the Yang--Baxter equation. To obtain it, in \eqref{eq:ybeapp}, plug $a_1=a$, $a_2=i$ and $a_3=i+1$, as well as $u_1=u$, $u_2=\theta_i$ and $u_3=\theta_{i+1}$. Then multiply on the left by $P_{i,i+1}$ and use the properties of $P$ to rewrite it in the form \eqref{eq:lemma4}, (using $\check{R}=PR$). 

\paragraph{Step 2} 
Similarly, we write
\begin{equation}
    \check{R}_{i,i+1}(\theta_i,\theta_{i+1})R_{i,a}(\theta_{i},-u)R_{i+1,a}(\theta_{i+1},-u)=R_{i,a}(\theta_{i+1},-u)R_{i+1,a}(\theta_{i},-u)\check{R}_{i,i+1}(\theta_i,\theta_{i+1})\label{eq:lemma5}.
\end{equation}
This again follows as a direct consequence of the Yang--Baxter equation. In \eqref{eq:ybeapp} substitute $a_1=i$, $a_2=i+1$, $a_3=a$, $u_1=\theta_i$, $u_2=\theta_{i+1}$ and $u_3=-u$. Then again multiply on the left by $P_{i,i+1}$ and use the properties of $P$ to rewrite the relation in the form \eqref{eq:lemma5}.

\paragraph{Step 3}
For $\theta_i$ and $\theta_{i+1}$, as long as $\text{det}(\check{R}(\theta_i,\theta_{i+1}))\neq 0$, one can write
\begin{equation}
    T_a(u,\sigma(\theta_i,\theta_{i+1}))=\check{R}_{i,i+1}(\theta_i,\theta_{i+1})T_a(u)\check{R}_{i,i+1}(\theta_i,\theta_{i+1})^{-1}.\label{eq:lemma6}
\end{equation}
To prove this we write
\begin{align}
    \check{R}_{i,i+1}&(\theta_i,\theta_{i+1})T_a(u)=\check{R}_{i,i+1}(\theta_i,\theta_{i+1})R_{aN}(u,\theta_N)\cdots R_{ai+1}(u,\theta_{i+1})R_{ai}(u,\theta_{i})\cdots R_{a,1}(u,\theta_{i})\label{eq:lemma6p1}\\[-0.2cm]
    &=R_{aN}(u,\theta_N)R_{aN-1}(u,\theta_{N-1})\cdots \check{R}_{i,i+1}(\theta_i,\theta_{i+1})R_{ai+1}(u,\theta_{i+1})R_{ai}(u,\theta_{i})\cdots R_{a,1}(u,\theta_{i})\label{eq:lemma6p2}\\[-0.2cm]
    &=R_{aN}(u,\theta_N)R_{aN-1}(u,\theta_{N-1})\cdots R_{ai+1}(u,\theta_{i})R_{ai}(u,\theta_{i+1})\check{R}_{i,i+1}(\theta_i,\theta_{i+1})\cdots R_{a,1}(u,\theta_{i})\label{eq:lemma6p3}\\[-0.2cm]
    &=R_{aN}(u,\theta_N)R_{aN-1}(u,\theta_{N-1})\cdots R_{ai+1}(u,\theta_{i})R_{ai}(u,\theta_{i+1})\cdots R_{a,1}(u,\theta_{i})\check{R}_{i,i+1}(\theta_i,\theta_{i+1})\label{eq:lemma6p4}\\[-0.2cm]
    &=T_a(u,\sigma(\theta_i,\theta_{i+1}))\check{R}_{i,i+1}(\theta_i,\theta_{i+1})\label{eq:lemma6p5},\\[0.2cm]
    &\quad \Rightarrow T_a(u,\sigma(\theta_i,\theta_{i+1}))=\check{R}_{i,i+1}(\theta_i,\theta_{i+1})T_a(u)\check{R}_{i,i+1}(\theta_i,\theta_{i+1})^{-1}.\label{eq:lemma6p6}
\end{align}
In equation \eqref{eq:lemma6p1}, we write the definition of $T_a(u)$ explicitly and multiply it on the left by $\check{R}(\theta_i,\theta_{i+1})$. It is important to remember that since $a$ is associated to the auxiliary space, we always have $a\neq 1,2, \cdots, i,i+1,\cdots , N$.  With this in mind, we can go from \eqref{eq:lemma6p1} to \eqref{eq:lemma6p2} by using the fact that $\check{R}(\theta_i,\theta_{i+1})$ commutes with every $R$ in $T_a(u)$ except  $R_{ai+1}(u,\theta_{i+1})$ and $R_{ai}(u,\theta_{i})$, allowing us to move it just before $R_{ai+1}(u,\theta_{i+1})$. From there to \eqref{eq:lemma6p3}, we use step 1 \eqref{eq:lemma4}. Next, we use the fact that operators acting on different spaces commute to move $\check{R}(\theta_i,\theta_{i+1})$ all the way to the end. Finally, we recognize that everything before $\check{R}(\theta_i,\theta_{i+1})$ is the monodromy matrix but with $\theta_i\leftrightarrow \theta_{i+1}$, i.e. $T_a(u,\sigma(\theta_i,\theta_{i+1}))$. As a final step we multiply both sides of the equation by $\check{R}(\theta_i,\theta_{i+1})^{-1}$ leading exactly to \eqref{eq:lemma6}. The later step is where the invertibility of $\check{R}(\theta_i,\theta_{i+1})$ is required. 

\paragraph{Step 4} 
For $\theta_i$ and $\theta_{i+1}$ such that $\text{det}(\check{R}(\theta_i,\theta_{i+1}))\neq 0$
\begin{equation}
    \hat{T}_a(u,\sigma(\theta_i,\theta_{i+1}))=\check{R}_{i,i+1}(\theta_i,\theta_{i+1})\hat{T}_a(u)\check{R}_{i,i+1}(\theta_i,\theta_{i+1})^{-1}\label{eq:lemma7}.
\end{equation}
 To prove this we write
\begin{align}
   &\check{R}_{i,i+1}(\theta_i,\theta_{i+1})\hat{T}_a(u)=\check{R}(\theta_i,\theta_{i+1})R_{1a}(\theta_1,-u)\cdots R_{ia}(\theta_i,-u)R_{i+1,a}(\theta_{i+1},-u)\cdots R_{Na}(\theta_N,-u)\label{eq:lemma7p1}\\[-0.2cm]
    &=R_{1a}(\theta_1,-u)R_{2a}(\theta_2,-u)\cdots \check{R}_{i,i+1}(\theta_i,\theta_{i+1})R_{ia}(\theta_i,-u)R_{i+1,a}(\theta_{i+1},-u)\cdots R_{Na}(\theta_N,-u)\label{eq:lemma7p2}\\[-0.2cm]
    &=R_{1a}(\theta_1,-u)R_{2a}(\theta_2,-u)\cdots R_{ia}(\theta_{i+1},-u)R_{i+1,a}(\theta_{i},-u)\check{R}_{i,i+1}(\theta_i,\theta_{i+1})\cdots R_{Na}(\theta_N,-u)\label{eq:lemma7p3}\\[-0.2cm]
    &=R_{1a}(\theta_1,-u)R_{2a}(\theta_2,-u)\cdots R_{ia}(\theta_{i+1},-u)R_{i+1,a}(\theta_{i},-u)\cdots R_{Na}(\theta_N,-u)\check{R}_{i,i+1}(\theta_i,\theta_{i+1})\label{eq:lemma7p4}\\[-0.2cm]
    &=\hat{T}_a(u,\sigma(\theta_i,\theta_{i+1}))\check{R}_{i,i+1}(\theta_i,\theta_{i+1}),\label{eq:lemma7p5}\\[0.2cm]
    &\Rightarrow \quad \hat{T}_a(u,\sigma(\theta_i,\theta_{i+1}))=\check{R}_{i,i+1}(\theta_i,\theta_{i+1})\hat{T}_a(u)\check{R}_{i,i+1}(\theta_i,\theta_{i+1})^{-1}\label{eq:lemma7p6}
\end{align}

This proof is very similar to the one for Step 3. In  \eqref{eq:lemma7p1} we write $\hat{T}_a(u)$ explicitly in terms of $R$-matrices and multiply them on the left by $\check{R}(\theta_i,\theta_{i+1})$. Next, we use the fact that $\check{R}(\theta_i,\theta_{i+1})$ commutes with all the $R$-matrices in $\hat{T}_a(u)$ except $R_{ia}(\theta_i,-u)$ and $R_{i+1,a}(\theta_{i+1},-u) $ to move it to just before $R_{ia}(\theta_i,-u)$. In the sequence, we use step 2 \eqref{eq:lemma5} to obtain \eqref{eq:lemma7p3}. Subsequently, we use the fact that $\check{R}(\theta_i,\theta_{i+1})$ commutes with all $R$-matrices after it, to move it to the end. In \eqref{eq:lemma7p4}, it is then clear that the product of $R$-matrices is again $\hat{T}_a(u)$ but with $\theta_i$ and $\theta_{i+1}$ swapped, which we write as $\hat{T}_a(u,\sigma(\theta_i,\theta_{i+1}))$. Finally, we multiplied both sides by $\check{R}(\theta_i,\theta_{i+1})^{-1}$ reaching the result mentioned in \eqref{eq:lemma7}.

\paragraph{Step 5}
For $\theta_i$ and $\theta_{i+1}$ such that $\text{det}(\check{R}(\theta_i,\theta_{i+1}))\neq 0$
\begin{equation}
    t(u,\sigma(\theta_i,\theta_{i+1}))=\check{R}_{i,i+1}(\theta_i,\theta_{i+1})t(u)\check{R}_{i,i+1}(\theta_i,\theta_{i+1})^{-1}.\label{eq:theorem3}
\end{equation}
 Let us start with $t(u,\sigma(\theta_i,\theta_{i+1}))$ and use steps 1-4 to prove the expression above
\begin{align}
    &t(u,\sigma(\theta_i,\theta_{i+1}))=\tr_a\left(K_a^{L}(u)T_a(u,\sigma(\theta_i,\theta_{i+1}))K_a^R(u)\hat{T}_a(u,\sigma(\theta_i,\theta_{i+1}))\right)\label{eq:theorem3p1}\\[0.2cm]
    & =\tr_a\left(K_a^{L}(u)\check{R}_{i,i+1}(\theta_i,\theta_{i+1})T_a(u)\cancel{\check{R}_{i,i+1}(\theta_i,\theta_{i+1})}^{-1}K_a^R(u)\cancel{\check{R}_{i,i+1}(\theta_i,\theta_{i+1})}\hat{T}_a(u)\check{R}_{i,i+1}(\theta_i,\theta_{i+1})^{-1}\right)\label{eq:theorem3p2}\\[-0.2cm]
    & =\check{R}_{i,i+1}(\theta_i,\theta_{i+1})\tr_a\left(K_a^{L}(u)T_a(u)K_a^R(u)\hat{T}_a(u)\right)\check{R}_{i,i+1}(\theta_i,\theta_{i+1})^{-1}\label{eq:theorem3p3}\\[0.2cm]
    &=\check{R}_{i,i+1}(\theta_i,\theta_{i+1})t(u)\check{R}_{i,i+1}(\theta_i,\theta_{i+1})^{-1}.\label{eq:theorem3p4}
\end{align}
For this proof we start with a transfer matrix with two inhomogeneities swapped and prove that this is the same as doing a similarity transformation in the original transfer matrix. In \eqref{eq:theorem3p1} we just wrote the definition of the transfer matrix with swapped inhomogeneities. Then used Step 3 and 4 to write the swapped monodromy matrices in terms of the original ones. In the same step we notice that $a\neq i,i+1$ so the $\check{R}_{i,i+1}(\theta_i,\theta_{i+1})$ commutes with $K_a^R(u)$ and can be then canceled by $\check{R}_{i,i+1}(\theta_i,\theta_{i+1})^{-1}$. From \eqref{eq:theorem3p2} to \eqref{eq:theorem3p3} we used the fact that $a\neq i,i+1$ to put $\check{R}_{i,i+1}(\theta_i,\theta_{i+1})$ out of the partial trace to left and $\check{R}_{i,i+1}(\theta_i,\theta_{i+1})^{-1}$ out of the trace to right. Finally we recognize that the term in the middle is in fact the original (not swapped) transfer matrix, therefore obtaining exactly equation \eqref{eq:theorem3}.

\section{Checking the BYBE numerically}\label{app:numericBYBE}

In section \ref{sec:gatespositions}, we conjecture that (analogously to the periodic setting \cite{Paletta:2025sap}), a quantum circuit where each gate $U$ acts exactly once at each pair of sites ${i,i+1}$, is integrable as long as $U$, $K^R$ and $\tilde{K}^L$ satisfy the Yang--Baxter and Boundary Yang--Baxter equations. 

In this appendix, we present a procedure to determine whether a set of gates satisfies the YBE and BYBE when the gates are known only numerically.

Let us start by writing the YBE for $\check{R}$
\begin{equation}
    \check{R}_{12}(u,v)\check{R}_{23}(u,w)\check{R}_{12}(v,w)=\check{R}_{23}(v,w)\check{R}_{12}(u,w)\check{R}_{23}(u,v).
    \label{eq:ybecheck}
\end{equation}
Now, we want to check the BYBE \eqref{eq:bybe}, for $R=P\check{R}$, numerically. This can be achieved with the following steps:

\paragraph{Step 1:} Write YBE \eqref{eq:ybecheck} for $u=\kappa$, $w=-\kappa$ and using that $\check{R}(\kappa,-\kappa)=U$ as
\begin{equation}
    \check{R}_{12}(\kappa,v)U_{23}\check{R}_{12}(v,-\kappa)=\check{R}_{23}(v,-\kappa)U_{12}\check{R}_{23}(\kappa,v).
    \label{eq:ybechecknumerically}
\end{equation}
Recall that the gate $U$ is known numerically.

\paragraph{Step 2a:} Substitute $U$ in \eqref{eq:ybechecknumerically} and solve the equation numerically for $\check{R}(\kappa,v)$ and $\check{R}(v,-\kappa)$. If the non-zero elements of these two operators are exactly in the same position as the ones in $U$, then you can claim that your gate $U$ is a solution of the YBE.  This step is usually used in the literature to check integrability in periodic quantum circuits.

\paragraph{Step 3a:} Assume that the $R$-matrix generating your gate is of difference form, which will then be of the form $\check{R}(\kappa-v)$ and $\check{R}(\kappa+v)$. For difference form, these are the only two operators that appear in the BYBE \eqref{eq:bybe} besides the $K$-matrix. So, plug them and the $K^R(\kappa)$ that you know numerically, in the BYBE and solve for $K^R(v)$. If the $K^R(v)$ has the non-zero elements in the exact same positions as the ones in $K^R(v)$, the procedure worked and you can go to step 4. If not, try  Steps 2b and 3b instead.

\paragraph{Step 2b:} There is still a chance that your gate $U$ is made of a non-difference form $R$-matrix. Write $\check{R}(\kappa,v)$ and $\check{R}(v,-\kappa)$ as numerical expansions in $v$, with same non-zero entries as the gate $U$, plug them in the YBE and solve the equation numerically for the coefficients to find $\check{R}(\kappa,v)$ and $\check{R}(v,-\kappa)$ for a few orders in $v$. If the non-zero elements of these two operators are exactly in the same position as the ones in $U$, then you can claim that your gate $U$ is a solution of the YBE up to certain order. Keep these $\check{R}(\kappa,v)$ and $\check{R}(v,-\kappa)$ aside, and  repeated the process from the beginning with  $v$ to $-v$, instead. Altogether, this will give you the four $R$-matrices \footnote{Recall that  $R=P\check{R}$.} that you need to plug in the BYBE \eqref{eq:bybe}. 

\paragraph{Step 3b} Now write $K^R(v)$ as an expansion in $v$ with same non-zero matrix elements as the numerical $K^R(\kappa)$ you have.  Plug both the $R$-matrices you found on step 2b, the expansion for $K^R(v)$  and the $K_1^R(v)$ that you know numerically, on the BYBE. Solve for the coefficients in $K_1^R(v)$. If the resulting matrix still has the same non-zero elements as your numerical $K_1^R(\kappa)$, for a few orders, it is a good indication that your quantum circuit is integrable but of non-difference form. If so, go to step 4. If not, your circuit is not integrable in the sense described in this paper.

\paragraph{Step 4:} If Steps 1-3a or 1-3b are satisfied, and your numerical gate $\tilde{K}^L(\kappa)$ has the same non-zero entries and symmetries of your numerical $K^R(\kappa)$, then  your quantum circuit is Yang-Baxter integrable.

\section{Almost a good choice}\label{app:almost}

\begin{figure}[H]
\centering

\begin{subfigure}[t]{0.48\textwidth}
    \vspace{-4cm}
    \includegraphics[scale=1.1]{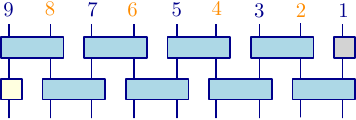}
    \label{fig:kappa_-=4b}
    \caption{$\kappa_-=4,$ depth $=2$ and $\Vec{n}=(8,6,4,2)$}
    \vspace{1cm}
    \includegraphics[scale=1.1]{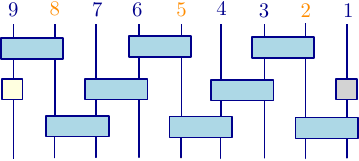}
    \label{fig:kappa_-=3b}
    \caption{$\kappa_-=3,$ depth $=3$ and $\Vec{n}=(8,5,2)$}
    \vspace{1cm}
    \includegraphics[scale=1.1]{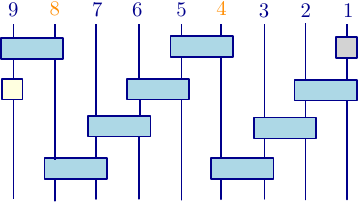}
    \label{fig:kappa_-=2b}
    \caption{$\kappa_-=2,$ depth $=4$ and $\Vec{n}=(8,4)$}
\end{subfigure}\hfill
\begin{subfigure}[t]{0.48\textwidth}
    \includegraphics[scale=1.1]{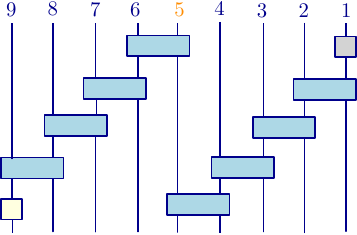}
    \label{fig:kappa_-=1b}
    \caption{$\kappa_-=1,$ depth $=5$ and $\Vec{n}=(5)$}
    \medskip
    \includegraphics[scale=1.1]{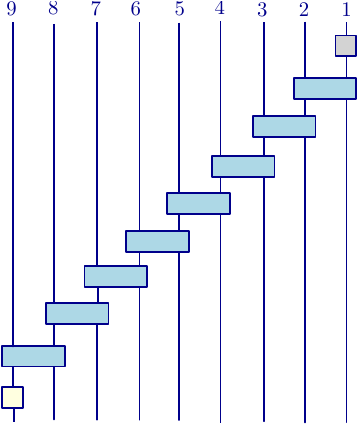}
    \label{fig:kappa_-=0b}
    \caption{$\kappa_-=0,$ depth $=10$ and $\Vec{n}=(\emptyset)$}
\end{subfigure}
\caption{Alternative open quantum circuits for length $N=9$. The orange colour indicates the position of the $-\kappa$'s.}
\label{fig:allgeometriesN=9}
\end{figure}

Given the fact that the  inhomogeneities configuration that minimises the depth is not unique, one could ask why did we choose the configurations in Conjectures 1 and 2, instead of the \textit{apparently} more natural one presented below (for $N=9$) in  Figure \ref{fig:allgeometriesN=9}. 

The reason is that although the choice in Figure \ref{fig:allgeometriesN=9} works for $N=9$, it does not generalise well for higher $N$ if we want to minimize the depth.

We can see that the case for $\kappa_-=0,1$ and $\kappa_-=\frac{N-1}{2}$ coincide with the ones in section \ref{sec:positiongatesoddN}, and the distribution of the gates looks much more like the choice made in the periodic case \cite{Paletta:2025sap}. However, in \cite{Paletta:2025sap}, minimising the depth was not one of their goals. It happens that simultaneously keeping the very regular gate distribution shown in Figure \ref{fig:allgeometriesN=9} and minimizing the depth is not possible beyond $N=11$. 

To understand the reason why not, let us focus in the case with depth $=3$. For achieving depth $=3$, for  $N=9$ we need three $U$ gates in the first line and an empty site in between them. To keep the regularity, for $N=11$ we will need four gates in the first line and for $N=13$ we will need five gates. So, every time we increase the number of sites by two we need one extra $U$ gate (that takes two sites) plus an extra site to have in between. So, to keep the regularity for depth $=3$ we need three new sites for each new odd number, but we get only two new ones. So, for $N=13$, we need five gates (that take 10 sites) plus four sites in between. This means we need $14$ sites for this configuration but we only have $13$ available. The problem gets worse as we increase the number of sites. This is the reason this gate distribution was not the one chosen in this paper.   The positions chosen in Conjectures 1 and 2 automatically avoid this issue and are valid for any number of sites. 

\end{appendix}

\bibliographystyle{SciPost_bibstyle}
\bibliography{references.bib}

\end{document}